\documentclass[pra,
    notitlepage,
    superscriptaddress,
    longbibliography,
    twocolumn]{revtex4-2}

\usepackage[utf8]{inputenc}
\usepackage{graphicx}
\usepackage{amsmath}
\usepackage{amsthm}
\usepackage{bm}
\usepackage{layout}
\usepackage{float}
\usepackage{amsfonts}
\usepackage{amssymb}%
\usepackage[margin=0.75in]{geometry}
\usepackage{color}
\usepackage{soul}
\usepackage{mathtools}
\usepackage{lineno}
\usepackage[caption=false]{subfig} 
\usepackage{ragged2e}              
\DeclareCaptionJustification{justified}{\justifying}
\usepackage{tabularx}
\usepackage{array}

\usepackage[colorlinks=true,citecolor=blue]{hyperref}
\theoremstyle{definition}

\usepackage[capitalise,compress]{cleveref}

\usepackage{physics}

\usepackage[normalem]{ulem}
\usepackage[subpreambles=false]{standalone}
\usepackage{tikz}
\usepackage[section]{placeins}

\usetikzlibrary{arrows}
\usetikzlibrary{shapes.geometric}

\usepackage{tabularx}
\usepackage[noend]{algpseudocode}

\makeatletter
\newcommand{\multiline}[1]{%
  \begin{tabularx}{\dimexpr\linewidth-\ALG@thistlm}[t]{@{}X@{}}
    #1
  \end{tabularx}
}
\makeatother
\newcolumntype{P}[1]{>{\centering\arraybackslash}p{#1}}

\renewcommand{\epsilon}{\varepsilon}

\renewcommand{\emph}[1]{\textit{#1}}

\newcounter{para}

\makeatletter
\newcommand*\bigcdot{\mathpalette\bigcdot@{.5}}
\newcommand*\bigcdot@[2]{\mathbin{\vcenter{\hbox{\scalebox{#2}{$\m@th#1\bullet$}}}}}

\newcommand{\llangle}[1][]{\savebox{\@brx}{\(\m@th{#1\langle}\)}%
  \mathopen{\copy\@brx\kern-0.5\wd\@brx\usebox{\@brx}}}
\newcommand{\rrangle}[1][]{\savebox{\@brx}{\(\m@th{#1\rangle}\)}%
  \mathclose{\copy\@brx\kern-0.5\wd\@brx\usebox{\@brx}}}

\makeatother

\newcommand{\ad}{\text{ad}}

\usepackage{array}
\newcolumntype{L}{>{$}l<{$}} 
\newcolumntype{C}{>{$}c<{$}} 
\newcolumntype{R}{>{$}r<{$}} 

\usepackage{xcolor}
\definecolor{darkgreen}{HTML}{009020}

\usepackage{pifont}
\newcommand{\cmark}{\ding{51}}%

\usepackage{mathtools}

\usepackage{xr}
\makeatletter
\newcommand*{\addFileDependency}[1]{
  \typeout{(#1)}
  \@addtofilelist{#1}
  \IfFileExists{#1}{}{\typeout{No file #1.}}
}
\makeatother

\usepackage{tikz}

\usepackage{mdframed}
\newmdenv[topline=false,rightline=false,bottomline=false,linewidth=2pt,linecolor=white!60!black,]{leftborder}

\newcommand{\Var}{\text{Var}}

\usepackage{physics}

\usepackage{array}

\usepackage{bbm}

\usepackage{booktabs}

\makeatletter
\DeclareFontFamily{OMX}{MnSymbolE}{}
\DeclareSymbolFont{MnLargeSymbols}{OMX}{MnSymbolE}{m}{n}
\SetSymbolFont{MnLargeSymbols}{bold}{OMX}{MnSymbolE}{b}{n}
\DeclareFontShape{OMX}{MnSymbolE}{m}{n}{
    <-6>  MnSymbolE5
   <6-7>  MnSymbolE6
   <7-8>  MnSymbolE7
   <8-9>  MnSymbolE8
   <9-10> MnSymbolE9
  <10-12> MnSymbolE10
  <12->   MnSymbolE12
}{}
\DeclareFontShape{OMX}{MnSymbolE}{b}{n}{
    <-6>  MnSymbolE-Bold5
   <6-7>  MnSymbolE-Bold6
   <7-8>  MnSymbolE-Bold7
   <8-9>  MnSymbolE-Bold8
   <9-10> MnSymbolE-Bold9
  <10-12> MnSymbolE-Bold10
  <12->   MnSymbolE-Bold12
}{}

\let\llangle\@undefined
\let\rrangle\@undefined
\DeclareMathDelimiter{\llangle}{\mathopen}%
                     {MnLargeSymbols}{'164}{MnLargeSymbols}{'164}
\DeclareMathDelimiter{\rrangle}{\mathclose}%
                     {MnLargeSymbols}{'171}{MnLargeSymbols}{'171}
\makeatother

\usepackage{float}
\makeatletter
\let\newfloat\newfloat@ltx
\makeatother
\usepackage{algpseudocode}
\usepackage{algorithm}
\makeatletter
\renewcommand{\ALG@name}{Algorithm }
\makeatother

\newcommand{\equref}[1]{Eq.~\eqref{#1}}

\newcommand{\figref}[1]{Fig.~\ref{#1}}

\newcommand{\secref}[1]{Sec.~\ref{#1}}
\newcommand{\refcite}[1]{Ref.~\cite{#1}}

\usepackage{siunitx}
\usepackage[acronym,nohypertypes={acronym},nolong,nosuper]{glossaries}
\makeglossaries
\newacronym{ole}{OLE}{Operator Loschmidt Echo}
\newacronym{otoc}{OTOC}{Out-of-Time-Order Correlator}
\newacronym{pec}{PEC}{Probabilistic Error Cancellation}
\newacronym{tnbp}{TN-BP}{Tensor Networks with Belief Propagation}
\newacronym{ppmc}{PP-MC}{Pauli Propagation with Monte Carlo}
\newacronym{bp}{BP}{Belief Propagation}
\newacronym{mps}{MPS}{Matrix Product State}
\newacronym{slc}{SLC}{Shaded Lightcone}
\newacronym{tls}{TLS}{Two-Level System}
\newacronym{svd}{SVD}{Singular Value Decomposition}
\newacronym{ttn}{TTN}{Tree Tensor Network}
\newacronym{dmrg}{DMRG}{Density Matrix Renormalization Group}
\newacronym{rcm}{RCM}{reverse Cuthill-McKee}

\newcommand{\beginsupplement}{%
  \onecolumngrid\clearpage
}

\usepackage{mathrsfs}
\usepackage{overpic}

\crefname{section}{App.}{Apps.}
\Crefname{section}{Appendix}{Appendices}
\crefname{subsection}{App.}{App.}
\Crefname{subsection}{Appendix}{Appendices}

\newif\ifneedcitehide
\newcommand{\needcite}[1][]{%
  \ifneedcitehide\else
    {\color{red}\textsuperscript{[cite\if\relax\detokenize{#1}\relax\else: #1\fi]}}%
  \fi}

\usepackage{multirow}

\begin{document}

\title{Observable Estimation in the Absence of Classical Verification}
\thanks{
\textbf{Author contributions:}
S. F. and A. K. led and co-ordinated the overall project. The Algorithmiq team led the model development, circuit design, and classical benchmarking efforts, with contributions from collaborators at Trinity College Dublin, EPFL, the University of Helsinki, the University of Ljubljana, the University of Padova, and the Flatiron Institute. The IBM team led the device calibration, the development and implementation of error-mitigation capabilities, and the experimental execution.}

\author{Samantha V. Barron}
\thanks{These authors contributed equally to this work.}
\author{Bradley Mitchell}
\thanks{These authors contributed equally to this work.}
\author{Vinay Tripathi}
\thanks{These authors contributed equally to this work.}
\affiliation{IBM Research}
\author{Francesco Grieco}
\thanks{These authors contributed equally to this work.}
\affiliation{Algorithmiq S.r.l., Via della Chiusa 15, 20123 Milan, Italy}
\author{Ilan Rosen}
\affiliation{IBM Research}

\author{Francesca Pietracaprina}
\affiliation{Algorithmiq S.r.l., Via della Chiusa 15, 20123 Milan, Italy}

\author{Davide Materia}
\affiliation{Algorithmiq S.r.l., Via della Chiusa 15, 20123 Milan, Italy}

\author{Alireza Seif}
\affiliation{IBM Research}

\author{Darvin Wanisch}
\affiliation{Dipartimento di Fisica e Astronomia "G. Galilei", via Marzolo 8, I-35131, Padova, Italy.}
\affiliation{Padua Quantum Technologies Research Center, Universit{\`a} degli Studi di Padova, Italy.}
\affiliation{INFN, Sezione di Padova, Italy.}

\author{Ramón L. Panadés-Barrueta}
\affiliation{Algorithmiq S.r.l., Via della Chiusa 15, 20123 Milan, Italy}
\author{Ewout van den Berg}
\author{Jay-U Chung}
\author{Andrew Eddins}
\author{Sam Ferracin}
\affiliation{IBM Research}

\author{Guillermo García-Pérez}
\affiliation{Algorithmiq S.r.l., Via della Chiusa 15, 20123 Milan, Italy}

\author{John Goold}
\affiliation{School of Physics, Trinity College Dublin, College Green, Dublin 2, Ireland}
\affiliation{Algorithmiq S.r.l., Via della Chiusa 15, 20123 Milan, Italy}

\author{Luke C. G. Govia}
\author{Holger Haas}
\author{Ian Hincks}
\author{Jesse C. Hoke}
\affiliation{IBM Research}

\author{Zoë Holmes}
\affiliation{Institute of Physics, École Polytechnique Fédérale de Lausanne (EPFL),
CH-1015 Lausanne, Switzerland}
\affiliation{Algorithmiq S.r.l., Via della Chiusa 15, 20123 Milan, Italy}
\affiliation{Centre for Quantum Science and Engineering, Ecole Polytechnique Fédérale de Lausanne (EPFL), CH-1015 Lausanne, Switzerland}

\author{Su-un Lee}
\affiliation{IBM Research}
\affiliation{Pritzker School of Molecular Engineering, The University of Chicago, Chicago, Illinois 60637, USA}

\author{Youngseok Kim}
\author{Swarnadeep Majumder}
\affiliation{IBM Research}

\author{Sabrina Maniscalco}
\affiliation{Algorithmiq S.r.l., Via della Chiusa 15, 20123 Milan, Italy}

\author{Simone Montangero}
\affiliation{Dipartimento di Fisica e Astronomia "G. Galilei", via Marzolo 8, I-35131, Padova, Italy.}
\affiliation{Padua Quantum Technologies Research Center, Universit{\`a} degli Studi di Padova, Italy.}
\affiliation{INFN, Sezione di Padova, Italy.}

\author{Daniel Puzzuoli}
\affiliation{IBM Research}

\author{Tomaž Prosen}
\affiliation{Faculty of Mathematics and Physics, University of Ljubljana,
Jadranska 19, SI-1000 Ljubljana, Slovenia}

\author{James Raftery}
\affiliation{IBM Research}

\author{Ricardo Rivera Cardoso}
\affiliation{Algorithmiq S.r.l., Via della Chiusa 15, 20123 Milan, Italy}

\author{Max Rossmannek}
\affiliation{IBM Research}

\author{Manuel Rudolph}
\affiliation{Institute of Physics, École Polytechnique Fédérale de Lausanne (EPFL),
CH-1015 Lausanne, Switzerland}

\author{Brendan Saxberg}
\author{Liran Shirizly}
\author{Karthik Siva}
\author{Joshua Skanes-Norman}
\affiliation{IBM Research}

\author{Ilaria Siloi}
\affiliation{Dipartimento di Fisica e Astronomia "G. Galilei", via Marzolo 8, I-35131, Padova, Italy.}
\affiliation{Padua Quantum Technologies Research Center, Universit{\`a} degli Studi di Padova, Italy.}
\affiliation{INFN, Sezione di Padova, Italy.}

\author{Kevin C Smith}
\affiliation{IBM Research}

\author{Boris Sokolov}
\affiliation{Algorithmiq S.r.l., Via della Chiusa 15, 20123 Milan, Italy}

\author{Maika Takita}
\affiliation{IBM Research}

\author{Yanting Teng}
\affiliation{Institute of Physics, École Polytechnique Fédérale de Lausanne (EPFL),
CH-1015 Lausanne, Switzerland}

\author{Mao Tian Tan}
\affiliation{Faculty of Mathematics and Physics, University of Ljubljana,
Jadranska 19, SI-1000 Ljubljana, Slovenia}

\author{Joseph Tindall}
\affiliation{Center for Computational Quantum Physics, Flatiron Institute,
162 Fifth Avenue, New York, New York 10010, USA}

\author{Zoltán Zimborás}
\affiliation{Department of Physics, University of Helsinki,
P.O. Box 43, FI-00014 Helsinki, Finland}
\affiliation{Algorithmiq S.r.l., Via della Chiusa 15, 20123 Milan, Italy}

\author{Matteo A.~C. Rossi}
\affiliation{Algorithmiq S.r.l., Via della Chiusa 15, 20123 Milan, Italy}

\author{Minh C. Tran}
\affiliation{IBM Research}

\author{Sergey N. Filippov}
\email{sergey.filippov@algorithmiq.fi}
\affiliation{Algorithmiq S.r.l., Via della Chiusa 15, 20123 Milan, Italy}

\author{Abhinav Kandala}
\email{akandala@us.ibm.com}
\affiliation{IBM Research}

\begin{abstract}
The predictive success of quantum mechanics underpins many areas of modern science, even as the exact simulation of large, interacting quantum systems remains beyond the reach of classical computation. This success has been enabled by the remarkable advancement of scalable numerical approximation methods, which often demonstrate practical accuracy despite the absence of formal guarantees. As quantum simulation pushes into regimes where these approximations struggle, a fundamental challenge arises: How can quantum outcomes be trusted when reliable classical benchmarks are unavailable? Here, we establish a framework for the independent validation of quantum estimates in this setting and present evidence that they provide the most credible result among several considered methods, in the absence of an immediately accessible ground-truth solution. We apply our framework to the semi-scrambling dynamics of a physical model that strains several leading classical simulation methods yet remains experimentally accessible, in part through our introduction of the \textit{operator Loschmidt echo}. We systematically design a series of experiments using quantum heuristics that, taken together, test the underlying assumptions and provide strong confidence in the observable estimates obtained from the quantum computer. 
We then show how this framework can be extended to place accuracy bounds on quantum estimates via careful characterization and manipulation of the device noise, transforming the problem of validating the observable estimation to validating the noise model. These results establish a route towards trusted quantum computation for scientific discovery, independent of classical verification.

\end{abstract}

\maketitle

The ultimate goal of quantum computing is to solve problems in regimes where classical methods break down.
Yet, benchmarking progress in quantum hardware has so far relied heavily on verifying that complex quantum computations match exact classical simulations~\cite{blatt2012quantum,kandala_hardware-efficient_2017,bluvstein_quantum_2022} at small scales and the best available classical approximations at larger ones~\cite{kim_evidence_2023,haghshenas_digital_2026}.
While this strategy has been effective at enabling quantitative validation of quantum processors, the resulting standard becomes circular: a quantum computation is considered trustworthy when it agrees with a classical calculation, but the purpose of the quantum computation is to go beyond what classical calculation can reliably provide.

This conundrum is not unique to quantum computing. 
Classical simulation methods for quantum many-body systems face the same predicament: 
techniques such as Density Functional Theory~\cite{RevModPhys.87.897} and Density Matrix Renormalization Group~\cite{RevModPhys.77.259} are routinely applied with great success in regimes where no exact solution or rigorous error bounds are available.
Their reliability is established not by comparison to a ground truth, but by evidence such as agreement among independent methods, validation against smaller instances where exact solutions exist, convergence with respect to controllable parameters, and recovery of known analytical limits.
Even though these classical methods are never certified, they ultimately earn our confidence through the accumulation of such evidence.

Here, we apply this validation principle to quantum computing. Concretely, we study the semi-scrambling dynamics of a heterogeneous Floquet Ising model using the operator Loschmidt echo (OLE), a generalization of the state Loschmidt echo~\cite{gorin_dynamics_2006} as a probe of operator spreading that produces a measurable signal on noisy quantum hardware.
The resulting dynamics generate entanglement growth and operator backflow that strain leading classical approximation methods, which in our benchmarks produce sharply disparate predictions. 
We consider two methods for estimating the OLE signal on superconducting quantum processors and demonstrate a distinct validation strategy for each.

The first is a heuristic method that mitigates the effect of noise by making a reasonable assumption about how the OLE signal decays under noise.
Previous attempts at validating such methods have relied on circuit variants where classical benchmarks are available (such as small-scale tests~\cite{Czarnik_2021,abanin_observation_2025}, Cliffordization~\cite{kim_evidence_2023}, analytical solutions~\cite{fischer2026dynamical}, or regimes where classical approximations converge~\cite{anand2023classical,farrell2024scalable}).
These variants, however, are not representative of the target circuit performance. 
Rather than relying solely on classical benchmarks, we directly test
the underlying assumption on the target circuit by probing its stability under controlled and uncontrolled changes to the noise, including reproducibility across multiple quantum processors.
Taken together, these tests provide the accumulated evidence that the quantum method, though heuristic, produces the most credible results among the considered computational methods.

\begin{figure*}[t]
\includegraphics[width=1.0\textwidth]{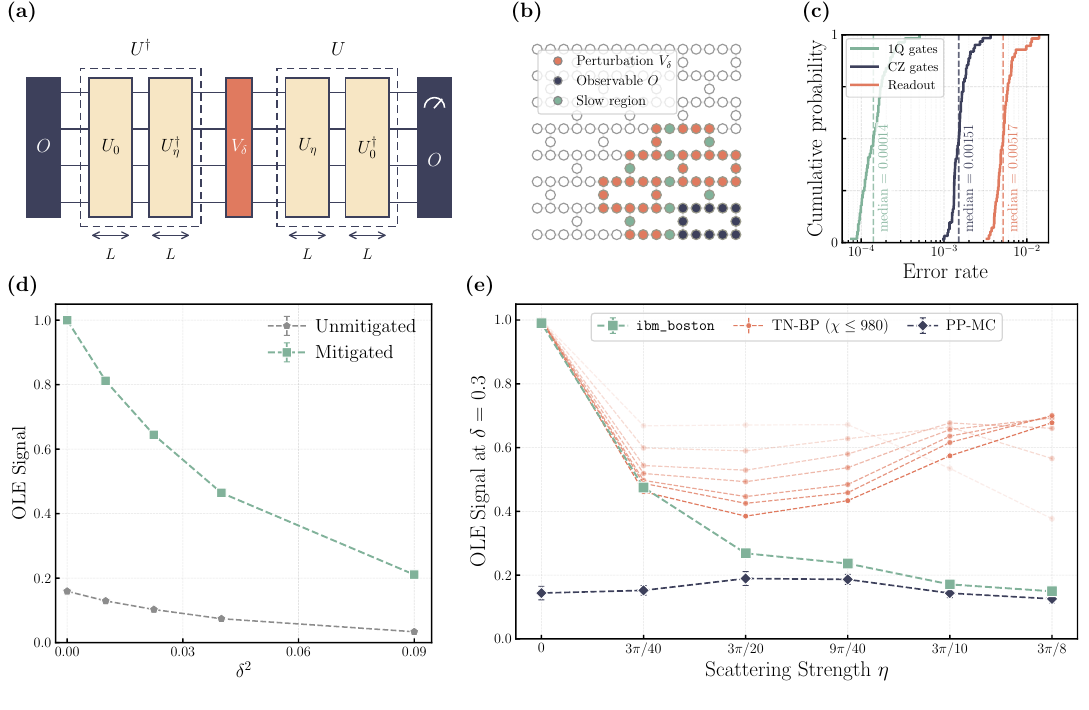}
\caption{
(a) Quantum circuits to compute the OLE signal, i.e. the overlap between an observable and its echoed version. 
The initial operator is approximated by an ensemble of eigenstates of $O$.
It is evolved under $U^\dag$, perturbed by $V_\delta$, and evolved back under $U$ before measuring in the eigenbasis of $O$.
(b) The quantum circuits are applied on 56 qubits (colored) of $\texttt{ibm\_boston}$.
Orange circles indicate qubits where the perturbation $V_\delta$ is applied.
Blue circles are the 12 qubits in the support of the representative observable $O$ for which the OLE signal is computed in later panels. 
The speed of operator propagation through each qubit can be controlled by tuning the angle of a single-qubit rotation on the qubit.
The spatial heterogeneity of $U$ is highlighted by the contrast between fast propagation through orange and blue qubits and the slow  through the green ones.
(c) Cumulative distribution of the one-qubit-gate, two-qubit-gate, and readout errors across the 56 qubits used in the experiment. 
The vertical dashed lines indicate the median error rates.
(d) Error mitigation of the OLE signal at $L = 6$ and $\eta = 9\pi/40$ via global rescaling.
The unmitigated signals (gray circles) at $\delta \neq 0$ are rescaled by the unmitigated signal at $\delta = 0$, where the ideal signal is exactly one, to produce the mitigated estimates (green squares). 
We collect $3.2$ million shots, with a quantum-processor runtime of roughly 45 minutes, for each $\delta$. 
The error bars for the experiment signals only account for the standard shot noise. 
(e) Estimates for the OLE signal at $L = 6$ and $\delta = 0.3$ from $\texttt{ibm\_boston}$ (green squares), TN-BP (orange circles), and PP-MC (blue diamonds) at different values of the scattering strength $\eta$.
The TN-BP estimates use increasing bond dimensions 64, 128, 256, 512, 640, 768, and 980, corresponding to the orange circles of light to bold shades.
At $\chi = 980$, the TN-BP simulation for each $\eta$ took about $160$ GPU-hours, distributed over eight NVIDIA H200 GPUs.
The PP-MC algorithm has an approximate limit of $M = 500$ million Pauli paths during the simulation, taking $\sim 290$\, CPU-hours distributed on AMD EPYC 9R14 nodes for each $\eta$.
The error bar is the standard deviation over 800 independent PP-MC samples. 
}
\label{fig:main}
\end{figure*}

While the confidence of the heuristic method is established through consistency checks, the second method equips its observable estimates with stand-alone error bounds.
These error bounds enable validation independent of classical verification, albeit at a higher computational cost than the heuristic.
In particular, we implement Probabilistic Error Cancellation (PEC)~\cite{temme_error_2017,van_den_berg_probabilistic_2023}, which effectively applies inverse noise channels to produce unbiased expectation values, given an accurately characterized noise model.
The central question is therefore no longer whether the quantum result agrees with a classical answer, but whether the noise model used to invert the hardware errors is sufficiently accurate. 
To this end, we carry out extensive tests to benchmark the noise model's assumptions and validity.
Building on these tests, we then implement PEC to obtain error-bounded estimates from the OLE circuits containing approximately 1000 two-qubit gates,
enabled by order-of-magnitude improvements in two-qubit error rates of Heron R3 devices relative to the previous fixed-frequency architecture~\cite{kim_evidence_2023}.

\textbf{Operator Loschmidt echo:}
In this work, we test our validation principle on circuits that measure the operator Loschmidt echo (OLE), which quantifies the overlap between an evolved operator and its perturbed echo.
Given an observable $O$, a unitary $U$ on $n$ qubits, and a perturbation $V_\delta$ of strength $\delta$, the OLE signal is
\begin{align}
    S_\delta = \frac{1}{2^n}\Tr(U^\dag O U \cdot V_\delta^\dag \cdot U^\dag O U \cdot V_\delta). \label{eq:ole_def}
\end{align}
Here, $V_\delta$ is the product of single-qubit rotations by an angle $\delta$ on a subset of qubits. \Cref{fig:main}(a) illustrates a circuit for computing the OLE signal of an observable $O$.
Experimentally, the initial operator $O$ is prepared through its ensemble representation---the qubits are initialized in random eigenstates of $O$.
An estimate for $S_\delta$ is obtained by measuring $O$ after the evolution and averaging over these initial states, weighted by the corresponding eigenvalues.

The OLE is the operator analog of the more familiar \emph{state} Loschmidt echo---a standard probe of quantum chaos, decoherence, and the stability of many-body dynamics under small perturbations~\cite{gorin_dynamics_2006}.
At small $\delta$, it has a linear dependence on the out-of-time-order correlator (OTOC) $C$~\cite{swingle_unscrambling_2018} between $O$ and the generator of $V_\delta$: $S_\delta \approx 1 - \frac{\delta^2}{2} C$.
Unlike the state Loschmidt echo, whose signal decays exponentially in system size and is therefore experimentally inaccessible at scale, the OLE at small $\delta$ decays only polynomially with the support of the perturbation $V_\delta$ and remains resolvable on noisy hardware even for complex operator dynamics.
However, we do not restrict to this small-$\delta$ limit.
At the values of $\delta$ studied here, $S_\delta$ carries information about higher-order correlators beyond the OTOC (see \cref{app:model} for dependence of OLE on the perturbation and its connection to higher-order correlators).

\textbf{Semi-scrambling Dynamics:}
The unitary $U$ is a Floquet, Ising-like circuit acting on $56$ qubits of a heavy-hex lattice.
It factorizes as $U^\dag = U_\eta^\dag U_0$, where $U_\eta = (U_{\rm FL}^{(\eta)})^L$ and $U_0 = U_{\eta = 0}$ each apply $L$ Floquet layers.
Each Floquet layer $U_{\rm FL}^{(\eta)}$ can be further decomposed into single-qubit gates sandwiched between three layers of CZ.
The difference between $U_0$ and $U_\eta$ is limited to only a few \emph{scattering centers} in $U_\eta$ where additional single-qubit rotations with an angle $\eta$ are applied. 
In the absence of scattering ($\eta = 0$), the two halves cancel exactly. 
The residual signal is therefore generated entirely by these scattering centers.
Each layer $U_{\rm FL}^{(\eta)}$ is spatially heterogeneous with fast and slow information propagation pathways set by the angles of single-qubit rotations, producing \emph{semi-scrambling} dynamics in which the operator support neither localizes nor grows ballistically.
This construction sits precisely at the intersection where leading classical heuristics diverge yet the experiment remains feasible: the semi-scrambling dynamics and the interference between $U_0$ and $U_\eta^\dag$ push different approximation methods into disagreement, while the partial recovery of the operator under echo keeps the signal experimentally resolvable (see \cref{app:model} for the full circuit construction).

\textbf{Quantum experiments:}
We estimate the OLE signal on a superconducting quantum processor, primarily $\texttt{ibm\_boston}$, with 156 fixed-frequency transmon qubits coupled with nearest neighbor tunable couplers in the heavy-hex topology. We focus on a 56-qubit section of the device [\cref{fig:main}(b)], chosen and optimized to maximize the overall fidelity. 

To improve signal quality, we employ several calibration and error-suppression strategies. First, we optimize the CZ layers using a batched calibration protocol that matches the layer structure of the target circuits, with a uniform CZ gate time of $\tau=\SI{96}{ns}$. This simultaneous calibration suppresses crosstalk during parallel two-qubit gate execution and substantially improves the gate fidelity. 
At the time of data acquisition for Fig.~\ref{fig:main}, the selected layout achieved a median CZ error rate of $0.15\%$, with the worst-case error rates below $0.40\%$ [see \cref{fig:main}(c)]. 
Second, to reduce the impact of noise drift from defect two-level systems (TLS), we modulate individual qubit bias electrodes at frequencies between $0.1-1$ Hz~\cite{kim2025error}. Third, we suppress the accumulation of coherent errors by Pauli twirling the CZ layers, thereby shaping the effective noise into stochastic Pauli noise~\cite{Bennett1996,Hashim_2021randomized}. 
Finally, we develop a scheme to detect and post-select on leakage outside the computational subspace, enhancing the experimental signal, see \cref{app:postselect}.
In \cref{fig:main}(d), we show the bare OLE signal measured at $L = 6$ (depth 72, 1488 CZ gates) for a representative observable $O = Z^{\otimes 12}$ supported on 12 qubits [\cref{fig:main}(b)] as a function of the perturbation strength $\delta$.

\begin{figure*}[t]
\includegraphics[width=\textwidth]{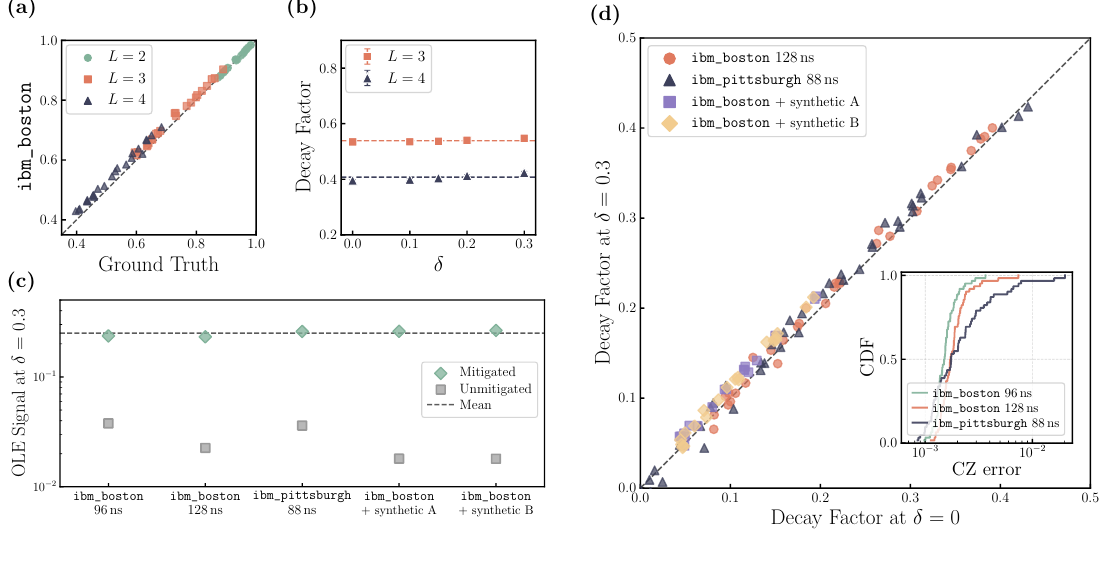}
\caption{
\textbf{Building trust in the quantum estimate.}
(a) Classical verification at small depth. The rescaled experimental OLE signal on $\texttt{ibm\_boston}$ is compared against the TN-BP ground truth at $\delta = 0.3$ and scattering strength $\eta = 9\pi/40$ for $L = 2$ (green circles), $L = 3$ (orange squares), and $L = 4$ (purple triangles), across several representative observables of Pauli weights 3 to 12. At these depths the TN-BP simulation at $\chi = 128$ has converged (fidelity more than $0.97$) and serves as a proxy for the ground truth.
The points lie along the $y = x$ line, showing that the rescaled quantum estimate tracks the exactly computable answer, albeit with a small positive bias.
(b) The decay factor of the unmitigated signal of the weight-12 observable in \cref{fig:main}(b) as a function of $\delta$ at $L = 3$ and $L = 4$. The decay is essentially independent of $\delta$, supporting the application of global rescaling for error mitigation.
(c) Reproducibility across device settings and controlled noise manipulation at $L = 6$. For the weight-12 observable considered in \cref{fig:main}(b), the OLE signal at $\delta = 0.3$ is reproduced across different settings: on the same device at different gate durations, different devices, and two synthetic noise models.
The maximum difference between the five mitigated signals is $0.03$.
The unmitigated signals (gray squares) vary by a factor of two across the settings, whereas the globally rescaled signals (green diamonds) collapse onto a common value, demonstrating that global rescaling is stable under both hardware and controlled modifications to the noise.
(d) Cross-device consistency at $L = 6$, where exact classical verification is unavailable. For each observable, we plot the decay factor at $\delta = 0.3$ against the decay factor at $\delta = 0$, computed using the rescaled $\texttt{ibm\_boston}$ signal at gate duration $\tau = 96$\,ns as the reference, for different device settings and synthetic noise models.
The points fall along the $y = x$ line, consistent with the assumption that the decay factor is $\delta$-independent across profiles with distinct noise. Inset: cumulative distribution of two-qubit (CZ) gate errors for the three device and gate-duration settings, highlighting their different noise levels.
}
\label{fig:building-trust}
\end{figure*}

Despite our circuit-performance optimizations, noise reduces the ideal OLE signal $S_\delta$ to a smaller value $\tilde S_\delta$.  A simple heuristic for mitigating this decay is global rescaling, which has been considered previously in experiments that share such a circuit structure~\cite{Swingle2018resilience,Mi2021information}. The underlying assumption is that noise primarily produces an overall attenuation of the OLE signal that is approximately independent of the perturbation strength $\delta$: $\tilde S_\delta \approx \alpha S_\delta$ for a $\delta$-independent factor $\alpha$. 
At $\delta=0$, the perturbation is absent and the ideal echo $S_0$ is exactly $1$.
The measured echo fidelity $\tilde S_0$ then estimates $\alpha$. Dividing by this value gives the global rescaling estimate $S_\delta \approx \frac{\tilde S_\delta}{\tilde S_0}$. 
We treat global rescaling as a heuristic mitigation method and test its central assumption, namely the approximate $\delta$-independence of the attenuation factor, in the following sections.

\textbf{Classical benchmarking:}
In the absence of exact verification, we attempt to benchmark the mitigated OLE signal from \texttt{ibm\_boston} 
with state-of-the-art classical heuristics: tensor network with belief propagation (TN-BP)~\cite{Tindall2023,Tindall2024efficient} and a bespoke Monte Carlo sampling version of Pauli propagation (PP-MC)~\cite{rudolph_pauli_2026,abanin_observation_2025}, as we sweep the scattering strength $\eta$ in \Cref{fig:main}(e). These classical heuristics match the quantum experiment in opposite limits, but disagree in the intermediate regime. 
In \cref{app:classical_heuristics,app:pauli_propagation,app:summary}, we test other classical heuristics, including matrix product states, tree tensor networks and other new variants of Pauli propagation, which broadly disagree with the considered methods including the quantum experiment, and propose future directions for classical simulation.
Nevertheless, an exhaustive list of classical heuristics is beyond the scope of this manuscript; we refer readers to the quantum advantage tracker~\cite{advantage_tracker} for ongoing efforts to improve both quantum and classical methods for estimating the OLE signals.

TN-BP represents an evolving quantum state as a tensor network, with the bond dimension $\chi$ controlling accuracy, and uses belief propagation to approximate the otherwise intractable contractions. 
At small $\eta$, the simulations converge with increasing bond-dimension $\chi$ to values that agree with the experiment. There, the operator is only weakly scattered and the entanglement generated by $U_0$ is nearly canceled out by $U_\eta^\dag$. As $\eta$ increases, TN-BP fails to converge even for the largest bond dimensions simulated, $\chi=980$, and diverges from the quantum experiment (see \cref{app:classical_heuristics} for several heuristic extrapolations of the TN-BP signal based on their fidelity and bond dimensions). 
This represents, to the best of our knowledge, the largest $\chi$ simulation of a finite system using TN-BP reported in the literature, requiring a multi-GPU implementation of the algorithm.

\begin{figure*}[t]
\includegraphics[width=\textwidth]{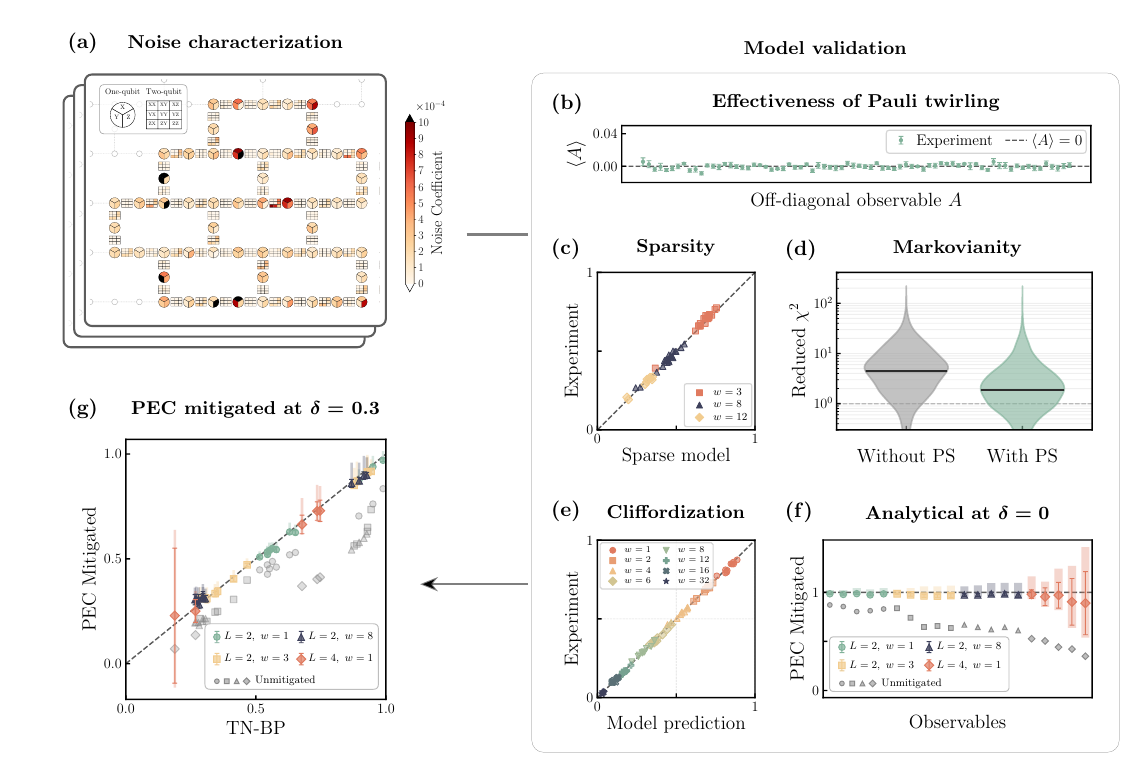}
\caption{
\textbf{Unbiased estimation with error bounds via Probabilistic Error Cancellation.}
PEC is applied to the OLE circuits at shallow depths ($L = 2$ and $4$), where converged TN-BP provides an exact reference for benchmarking the protocol.
(a) The noise model learned across the 56-qubit patch of $\texttt{ibm\_boston}$: each qubit and coupler is visualized with its learned Pauli error coefficients (color scale). PEC aims to invert this noise model to recover the ideal signal.
(b)--(f) Model validation: a series of tests that the learned sparse Pauli--Lindblad noise model is an accurate and predictive description of the device.
(b) Measured expectation values of off-diagonal observables that are predicted to be zero (dashed line) by a Pauli noise model. 
The values are within the statistical shot noise from zero, confirming that the noise has been twirled into incoherent Pauli channels.  
(c) Measured signal versus the prediction of the sparse local noise model for representative observables of Pauli-weight $w$. 
The agreement along the $y=x$ line indicates the effect of noise channels generated by high-weight Pauli strings is negligible, validating the sparse-model approximation.
(d) Distribution of the reduced $\chi^2$ of the exponential fits in noise learning with and without post-selection (PS) for samples that remain in the computational subspace. 
Post-selection alleviates the non-Markovianity of the noise model, shifting the distribution toward the ideal $\chi^2 \approx 1$.
(e) Measured versus exact signal from a Cliffordized version of the OLE circuit at $L = 4$ for representative observables of various weights $w$ between 1 and 32. 
The scatter points align with $y = x$, confirming the learned model captures the dominant device errors.
(f) PEC-mitigated signal at $\delta = 0$, where the ideal echo is exactly one (dashed line), for several observables at $L = 2$ ($w = 1,3,8$) and $L = 4$ ($w = 1$). 
The observables are selected based on their overhead $\gamma^2$, ranging from the smallest to the largest among those of the same $L$ and $w$. While the unmitigated values (gray) decay under noise, PEC recovers the signal after mitigation. The error bars are the PEC standard error $2\gamma/\sqrt{N}$. 
The shaded regions bound the distance from the estimates to the ground truth. 
The bound includes the PEC standard error, propagated errors from the noise characterization, and worst-case biases introduced by the classical computation of the shaded lightcones in PEC.
(g) PEC-mitigated signal versus the converged TN-BP estimates for observables of different weights at $L = 2$ and $L = 4$. 
The displayed observables are selected to include both the smallest and the largest sampling overhead $\gamma^2$ at each $L$ and weight.
The mitigation at $L = 2\ (4)$ for each $\delta$ used $N \approx 2.9 \times 10^{5}\ (7.3\times 10^{5})$ PEC randomizations, requiring approximately 1.5 (4) hours of quantum-processor runtime.
All mitigated estimates (colored) match the ground truth within the derived bounds.
}
\label{fig:pec_results}
\end{figure*}

Meanwhile, PP-MC exhibits the opposite dependence on $\eta$: it deviates from the experiment at small $\eta$, but shows closer agreement at larger $\eta$. This trend is consistent with the approximation made by the method. 
Our PP-MC implementation is tailored to the circuit as it propagates the observable $U^{\dag} O U$ in the Heisenberg picture through half of the circuit and stores at most $M$ Pauli strings and their coefficients. 
Once the evolved operator exceeds this memory budget, the method uses Monte Carlo \emph{2-norm} sampling to probabilistically compress the expansion, representing the evolution by a stochastic subset of its Pauli paths. 
Because this compression suppresses the interference between Pauli paths,
it is only reliable 
when the evolved operator scrambles into many effectively distinct, non-interfering paths, as in random circuits and in the large-$\eta$ regime studied here. When storing up to $M = 500$ million Pauli strings, PP-MC seemingly agrees with the quantum experiment for $\eta \geq 3\pi/10$, but diverges at smaller $\eta$, where coherent interference among the operator paths remains important (see \cref{app:pauli_propagation}). 
Since PP-MC drops a number of interference effects between Pauli paths, it is unclear how much this method should be trusted, even in the regime where it agrees with the experiment.

\Cref{fig:main}(e) exemplifies the central challenge we face as quantum devices simulate increasingly complex systems: no single method is reliable across the full parameter range.
The estimates agree at the extremes but diverge throughout the intermediate regime, precisely where no exact classical check exists. Which estimate, if any, should we trust? 
The remaining sections show that the quantum experiment is the most credible of the methods tested.

\textbf{Building trust in the quantum estimate:}
We now stress-test the global rescaling heuristic for estimating the OLE signal. 
Following standard practice, we first benchmark the method in regimes where classical verification is available. Here, we compare the experimental values at shallower depths $L = 2, 3$, and $4$, where TN-BP converges at a modest bond dimension $\chi = 128$ (fidelity above $0.97$) and provides a trustworthy ground truth. Across several observables of increasing Pauli weight, \cref{fig:building-trust}(a) shows that the globally rescaled \texttt{ibm\_boston} signal tracks the ground truth, albeit with a small overestimation. 
At these shallow depths, we test the underlying assumption for global rescaling: the ratio of the noisy signal to the ideal one (the decay factor) is set by the noise alone and is independent of $\delta$. Indeed, for $L = 3$ and $4$, we see that the measured decay factor is compatible with a constant independent of $\delta$ [\cref{fig:building-trust}(b)].
These smaller-scale tests support the heuristic, but they do not directly validate its assumptions on the target circuits, where exact classical verification is unavailable.

To address this gap, we use noise manipulation as a validation tool for the target circuits. By deliberately varying the noise affecting the target circuit, we can test whether the observed signal decays as assumed by global rescaling, without relying on access to a classically computed ground truth. 
We use two complementary forms of noise manipulation. First, we employ hardware-level perturbations of the noise, increasing the two-qubit gate duration (to $\tau=\SI{128}{ns}$), and repeating the experiment on a different quantum processor (\texttt{ibm\_pittsburgh}), to study noise profiles with distinct distributions for the CZ error rates. 
Second, we use a controlled mode of perturbations: insertion of predetermined Pauli noise channels of different strengths to form synthetic noise models. 
The globally rescaled estimate is reproducible under both types of noise manipulations, even as the unmitigated values vary due to the different noise profiles [\cref{fig:building-trust}(c)].
To validate the underlying assumption of global rescaling, now at $L = 6$, we perform a cross-device consistency check to test the $\delta$ independence of the decay factors across these different noise profiles. Treating the estimates from \texttt{ibm\_boston} at CZ gate duration $\tau=\SI{96}{ns}$ as the reference at $L=6$, we see that the decay factors for the additional noise profiles show broad agreement for $\delta=0,\, 0.3$ across several observables, in \cref{fig:building-trust}(d).
These hardware-level and controlled noise manipulations provide evidence that the global rescaling captures the dominant effect of noise on the target circuits.

\textbf{Towards estimation with error bounds:}
The validation of global rescaling's assumptions provides confidence in its estimate, which is now the most credible among the methods considered. 
Yet, this estimate lacks a quantitative statement of its accuracy, e.g., a bound on how far it could be from the ground truth. 
Producing such a bound is a growing priority~\cite{kraft_bounded-error_2025} for quantum computation when benchmarks against reliable classical heuristics are not available. Probabilistic error cancellation (PEC) provides a route to such an estimate~\cite{temme_error_2017,van_den_berg_probabilistic_2023}. Given a characterized model of the device noise, PEC probabilistically applies operations that invert the physical noise channels on average, producing an unbiased estimate of the ideal observable. Its remaining uncertainty is then statistical, a standard error $\sigma_{\text{PEC}} = 2\gamma / \sqrt{N}$, where $N$ is the number of circuit randomizations and the sampling overhead $\gamma^2$ grows with the total noise in the circuit (see \cref{app:pec} for details of our PEC implementation). In this sense, PEC converts the signal validation problem into a noise-model validation problem: the confidence interval is meaningful only to the extent that the learned noise model~[\cref{fig:pec_results}(a)] accurately represents the errors occurring during the experiment.

We model the device noise by a sparse Pauli-Lindblad model~\cite{van_den_berg_probabilistic_2023} and learn its coefficients using a protocol based on cycle benchmarking~\cite{Erhard2019characterizing}, which was previously demonstrated at scale on fixed-frequency processors~\cite{kim_evidence_2023}. In addition, our learning protocol incorporates Clifford circuits with the same structure as the target non-Clifford circuits, using a formalism similar to average circuit eigenvalue sampling~\cite{flammia2022aces}.
We systematically test the validity of several assumptions that underpin the learned noise model. First, in \cref{fig:pec_results}(b), we validate the
{\it Pauli-ness} of the noise, or the effectiveness of Pauli twirling in shaping the device noise, by confirming that off-diagonal observables vanish. Next, in \cref{fig:pec_results}(c), we test the {\it sparsity} assumptions of the noise model by verifying that the experimentally learned Pauli fidelities match the sparse model's predictions. We then test the {\it Markovianity} of our noise in \cref{fig:pec_results}(d). The noise learning traces reveal deviations from exponential decays that are often attributed to non-Markovian device errors~\cite{Fogarty2015,fong2017randomizedbenchmarkingcorrelatednoise}.
Leakage out of the computational subspace is a common source of non-Markovianity, which can arise from the weak anharmonicity of transmon qubits, dispersive interactions
to TLS, or other couplings on the device.
We show that post-selecting on these errors improves the quality of the exponential fit, implying a suppression of non-Markovianity, that we then use throughout our learning and mitigation experiments.

We then perform end-to-end validation of our noise model by comparing its predictions against the ground truth in variants of the target circuit where analytical solutions are possible.
\Cref{fig:pec_results}(e) validates the model in a Cliffordization---a highly entangled circuit obtained from the target circuit at $L = 4$ by replacing single-qubit rotations with single-qubit Clifford gates~(see \cref{app:pec}). 
Remarkably, the model provides accurate predictions across several high-weight stabilizers, even in the limit of strong damping.
In \cref{fig:pec_results}(f), we test the performance of PEC in mitigating the OLE signal at $\delta = 0$ where the signal is exactly one.
Our PEC implementation leverages the shaded lightcone (SLC)~\cite{eddins_lightcone_2024}, a post-processing technique that uses classical simulations to bound the influence of error channels on the observable, allowing selective cancellation of only important errors. 
Thereby, SLC significantly reduces the sampling overhead $\gamma^2$, at the cost of a small, bounded bias. 
The error bounds shown on \cref{fig:pec_results}(f) includes this bias, in addition to the PEC standard error and the propagated uncertainty in noise characterization (see \cref{app:pec} for details).
In both cases, the model's predictions and PEC experiments, across several randomly sampled observables, track the expected values, providing strong evidence that the learned model is representative of the device noise.
Despite temporal fluctuations in the relaxation times of superconducting qubits, the success of PEC at $\delta=0$ also stress-tests the {\it stability} of the noise over the duration of each mitigation run, with the noise re-characterized approximately every 45 minutes.
We further detail our noise model validation in \cref{app:exp}.

With this confidence in our noise model, we now apply PEC to the target circuit at $\delta = 0.3$.
The mitigated signal, again across several observables, agrees with the converged TN-BP estimates within the derived error bounds [\cref{fig:pec_results}(g)], confirming that PEC yields a reliable, error-bounded estimate in the regime where classical verification is still available. We further show in \cref{app:pec} that these methods can be extended to regimes where classical verification is unavailable, at $L=6$, by reducing the sampling overhead via modest reductions to hardware error rates.

\textbf{Discussion:} 
Validation is essential to the usefulness of any computational method, and especially to credible demonstrations of quantum advantage~\cite{lanes2025framework}. As quantum processors move into regimes where exact classical verification is unavailable, trust cannot rest solely on agreement with any single classical reference calculation: the effect of noise can be qualitatively different on the referenced circuits compared to the target circuits. Heuristic quantum estimates can be made credible through a hierarchy of tests that directly interrogate their assumptions on the target circuits, while more formal approaches such as probabilistic error cancellation can attach quantitative error bars when the underlying noise model is experimentally validated, albeit at increased sampling cost. The overheads of error-bounded estimation will continue to decrease with improvements in hardware error rates, faster sampling, and the use of quantum and classical resources in quantum-centric supercomputing architectures~\cite{seelam2026reference}. As these capabilities mature and quantum simulation can be performed at scale, real-world experimental measurements will provide a complementary path to validation~\cite{lee2026benchmarking}.
Indeed, the logic could also run in reverse: novel approximate classical methods will also require similarly careful validation, and well-characterized quantum experiments could themselves provide a benchmark within that pipeline.

The validation of both methods considered in this manuscript relies on how well we can characterize and control the device's noise. Our work therefore places noise-model validation at the center of trusted quantum computation: accurate modeling of the hardware errors under the same conditions as the target experiment enables meaningful error bars for quantum estimates. This perspective applies beyond near-term quantum hardware and is directly relevant to the transition from error mitigation to quantum error correction~\cite{eisert2025mind, zimboras2025myths}, where accurate models of noise will be essential for decoding and reliable logical quantum computation.

\textbf{Acknowledgments:}
A. K. thanks the entire IBM Quantum team for all the contributions that made this experiment possible.
The authors would like to thank 
Sajant Anand, 
Abhinav Deshpande,
Jay Gambetta, 
James Garrison, 
William Kirby,
Lixiang Luo,
Javier Robledo Moreno, 
Kunal Sharma, 
Oles Shtanko,  
and Kristan Temme
for their helpful comments and suggestions throughout the development of this manuscript. 
DM acknowledges EuroHPC for access to the supercomputers
Jupiter (JSC) through grant EHPC-DEV-2026D01-085 and Discoverer/Discoverer+
(SofiaTech) through grant EHPC-DEV-2026D02-279. FP acknowledges EuroHPC for access
to the supercomputer MareNostrum 5 (BSC) through grant EHPC-DEV-2026D01-027.
ZH acknowledges support from the Sandoz Family Foundation-Monique de Meuron program for Academic Promotion.
MSR acknowledges funding from the 2024 Google PhD Fellowship and the Swiss National Science Foundation [grant number 200021-219329]
YT acknowledges support from NCCR spin, a National Centre of Competence in Research, funded by the Swiss National Science Foundation (grant number 565785).
ZZ acknowledges support from the  Horizon Europe project 101113946 OpenSuperQPlus100 and the AI4QT project Nr. 2020-1.2.3-EUREKA-2022-00029.

\bibliography{main3}

\onecolumngrid
\newpage
\appendix
\beginsupplement
\begin{center}
  {\large\bfseries Appendices for\\[0.3em]
   ``Observable Estimation in the Absence of Classical Verification''}\\[1em]
\end{center}
\vspace{1em}

\tableofcontents

\section{Operator Loschmidt echo and semi-scrambling dynamics}\label{app:model}
\subsection{Operator Loschmidt echo (OLE)} \label{sec:OLE}

\begin{figure}
\includegraphics[width=0.75\textwidth]{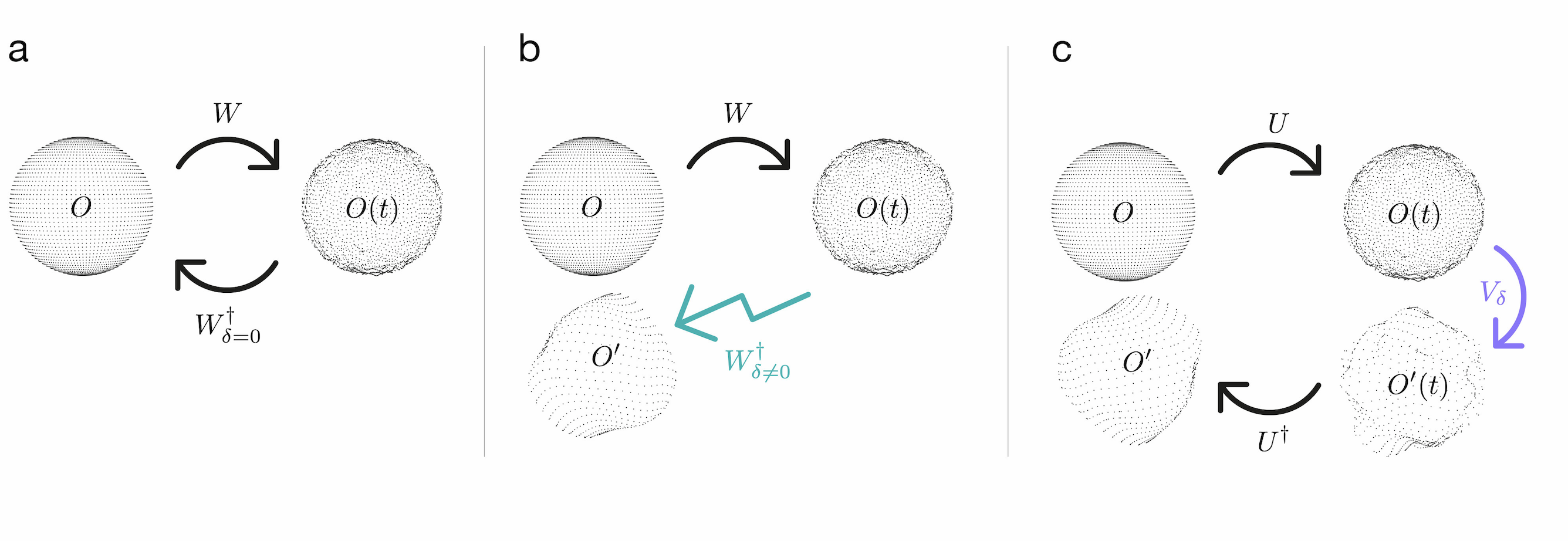}
\caption{\textbf{Loschmidt echo}: The Loschmidt echo can be visualized as evolving from an initial object (state or observable) under a certain unitary $W$, and back propagation with a perturbed $W_\delta$. (a) In the absence of perturbation, the initial condition is exactly recovered. (b) Perturbation leads to a mismatch between the initial and final states or observables. (c) In this experiment $W_{\delta} = U V_{\delta}$, so the perturbation is localized in time, in the middle of the echo dynamics.}
\label{fig:losch echo}
\end{figure}

Given a time-evolution unitary $W$ and its perturbed time-reversed version $W_\delta$, the probability $\big\vert \bra{\psi} W_{\delta}^{\dag} W \ket{\psi} \big\vert ^2$ of a pure state $\ket{\psi}$ under the unitary evolution $W_{\delta}^{\dag} W$, in \cref{fig:losch echo}, is the Loschmidt echo signal~\cite{gorin_dynamics_2006,prosen2003theory,Campisi2017}. This quantity exhibits high sensitivity to the perturbation strength $\delta$ and, therefore, has been proposed as a tool in identifying chaotic dynamics, probing dynamical phase transitions, and enhancing quantum metrology.
However, the echo signal in many-body systems typically decays exponentially in the system size, thus hindering its experimental observation unless scanning over smaller subsystems is done~\cite{karch2025probingquantummanybodydynamics}. Implementation of the conventional Loschmidt echo for states on an ideal quantum computer would imply repeated preparation of the multiqubit state $\ket{0}^{\otimes N}$, its evolution in the gate-based manner according to the echo unitary operator $W_{\delta}^{\dag} W$, and measurement of the resulting state in the computational basis $\{0,1\}^N$ with the aim to quantify the frequency of events in which the all-zero bitstring $0^{\times N}$ is observed as a measurement outcome. The current generation of quantum computers struggles with conducting such an experiment at scale because of the prohibitive requirements with respect to the initialization- and measurement-repetition rates as well as the gate and readout fidelities. 

In contrast to the return probability for states, the \acrfull{ole} quantifies the normalised overlap (the Hilbert-Schmidt scalar product) of the original operator $O$ with its perturbed time-reversed version~\cite{Sahu2022,Sreeram2025}, 
\begin{equation} \label{ole}
    S_{\delta} = \frac{{\rm tr}\left(O^{\dag} \; W_{\delta} W^{\dag} O W W_{\delta}^{\dag} \right)}{{\rm tr}\left( O^{\dag} O \right)} .
\end{equation}
\acrshort{ole} can also be seen as the overlap between the Heisenberg-picture observable $W^{\dag} O W$ and its perturbed counterpart $W_{\delta}^{\dag} O W_{\delta}$, namely, $S_{\delta} = {\rm tr} \big( W_{\delta}^{\dag} O^{\dag} W_{\delta} W^{\dag} O W \big) / {\rm tr}( O^{\dag} O )$. The denominator of Eq.~\eqref{ole} takes into account that the Frobenius norm is preserved by the unitary dynamics, i.e., $\| O \|_2 = \| W^{\dag} O W \|_2 = \| W_{\delta}^{\dag} O^{\dag} W_{\delta} \|_2 = \sqrt{{\rm tr}\left( O^{\dag} O \right)}$. Upon vectorisation of the $2^N \times 2^N$-dimensional operator space, Eq.~\eqref{ole} resembles the return amplitude in the echo signal for $4^N$-dimensional ``states'', $S_{\delta} =  \langle\langle W_{\delta}^{\dag} O W_{\delta} | W^{\dag} O W \rangle\rangle / \langle\langle O | O \rangle\rangle$, though it is drastically disparate from the Loschmidt echo for states from the viewpoint of experimental feasibility. The novelty of our approach is in the fact that we avoid dealing with $N$ ququarts ($4$-dimensional systems), which would be mathematically equivalent to the Loschmidt echo for $2N$-qubit pure states, but rather implement the measurement procedure for Eq.~\eqref{ole} at the level of $N$ qubits by exploiting the ensemble representation of mixed states and conventional measurements. 

\subsubsection{Measurement of OLE} \label{subsection-measurement-of-ole}

Let the operator $O$ be a Hermitian traceless operator with eigenvalues $\pm 1$, for instance, a Pauli string on $N$ qubits. Then $\| O \|_2^2 = {\rm tr}( O^2 ) = 2^N$ and Eq.~\eqref{ole} reduces to
\begin{equation} \label{ole-simplified}
   S_{\delta} = \frac{1}{2^N} {\rm tr}\left( W_{\delta}^{\dag} O W_{\delta} W^{\dag} O W \right).
\end{equation}
Let ${\cal V}^{\pm} := \{ \ket{\psi_i^{\pm}} \}_i$ be a set of orthonormal eigenvectors of $O$ corresponding to the eigenvalue $\pm 1$. Then the cardinality $|{\cal V}^{\pm}| = 2^{N-1}$, and
\begin{equation} \label{rho-O}
    \varrho_{\pm O} = \frac{I^{\otimes N} \pm O}{2^N} = \frac{1}{2^{N-1}} \sum_i \ket{\psi_i^{\pm}} \bra{\psi_i^{\pm}}
\end{equation}
are genuine density operators because they both have unit trace (${\rm tr}(\varrho_{\pm O}) = 1$) and nonnegative eigenvalues $0$ and $2^{-(N-1)}$ each of degeneracy $2^{N-1}$ ($\varrho_{\pm O} \geqslant 0$). Despite the fact that the state \eqref{rho-O} is generally mixed, its preparation on quantum hardware is straightforward thanks to the ensemble representation in terms of the eigenvectors of $O$. 

Similarly, one can effectively prepare on a quantum computer the operator $O$ itself by sampling uniformly $M$ eigenvectors $\ket{\phi_m} \in {\cal V}^{+} \cup {\cal V}^{-}$ and assigning the corresponding eigenvalues $ \bra{\phi_m} O \ket{\phi_m} = \pm 1$ to them, namely,
\begin{equation} \label{O-ensemble}
    \frac{1}{M} \sum_{m = 1}^M \underbrace{ \bra{\phi_m} O \ket{\phi_m} }_{\pm 1} \, \ket{\phi_m}\bra{\phi_m} \longrightarrow \frac{1}{2^N} O \quad \text{if} \quad M \longrightarrow \infty.
\end{equation}
There is no extra sampling cost for the implementation of Eq.~\eqref{O-ensemble} even though roughly half of the coefficients are negative. This is because $\frac{1}{2^N} O = \frac{1}{2} (\varrho_{+O} - \varrho_{-O})$ is a weighted difference of two density operators, with absolute weights summing to unity.

Let $\ket{\phi_m(\delta)} := W_{\delta} W^{\dag} \ket{\phi_m}$ denote the time-reversed perturbed evolution of eigenstates $\ket{\phi_m}$ of $O$. Substituting the representation \eqref{O-ensemble} for $O$ in $W^{\dag} O W$ from Eq.~\eqref{ole-simplified}, we obtain the operator Loschmidt echo
\begin{equation}
    S_{\delta} = \lim_{M \rightarrow \infty} S_{\delta}(M)
\end{equation} 
through its estimator
\begin{equation}
S_{\delta}(M) = \frac{1}{M} \sum_{m = 1}^M \underbrace{ \bra{\phi_m} O \ket{\phi_m} }_{\pm 1} \times \bra{\phi_m(\delta)} O \ket{\phi_m(\delta)}. 
\end{equation}
The \acrshort{ole} estimator $S_{\delta}(M)$ is experimentally accessible by means of measuring $O$ in the evolved states $\ket{\phi_m(\delta)}$, and its estimation error quickly decreases as $\propto 1/{\sqrt{M}}$. It is the correlation between $\bra{\phi_m} O \ket{\phi_m}$ and $\bra{\phi_m(\delta)} O \ket{\phi_m(\delta)}$ that constitutes the echo experiment and should be perfect if the perturbation is absent ($\delta = 0$). For non-zero perturbations, \acrshort{ole} provides valuable information about the nature of operator spreading.

\subsubsection{OLE estimation protocol}

\begin{figure}[t]
\includegraphics[width=0.60\textwidth]{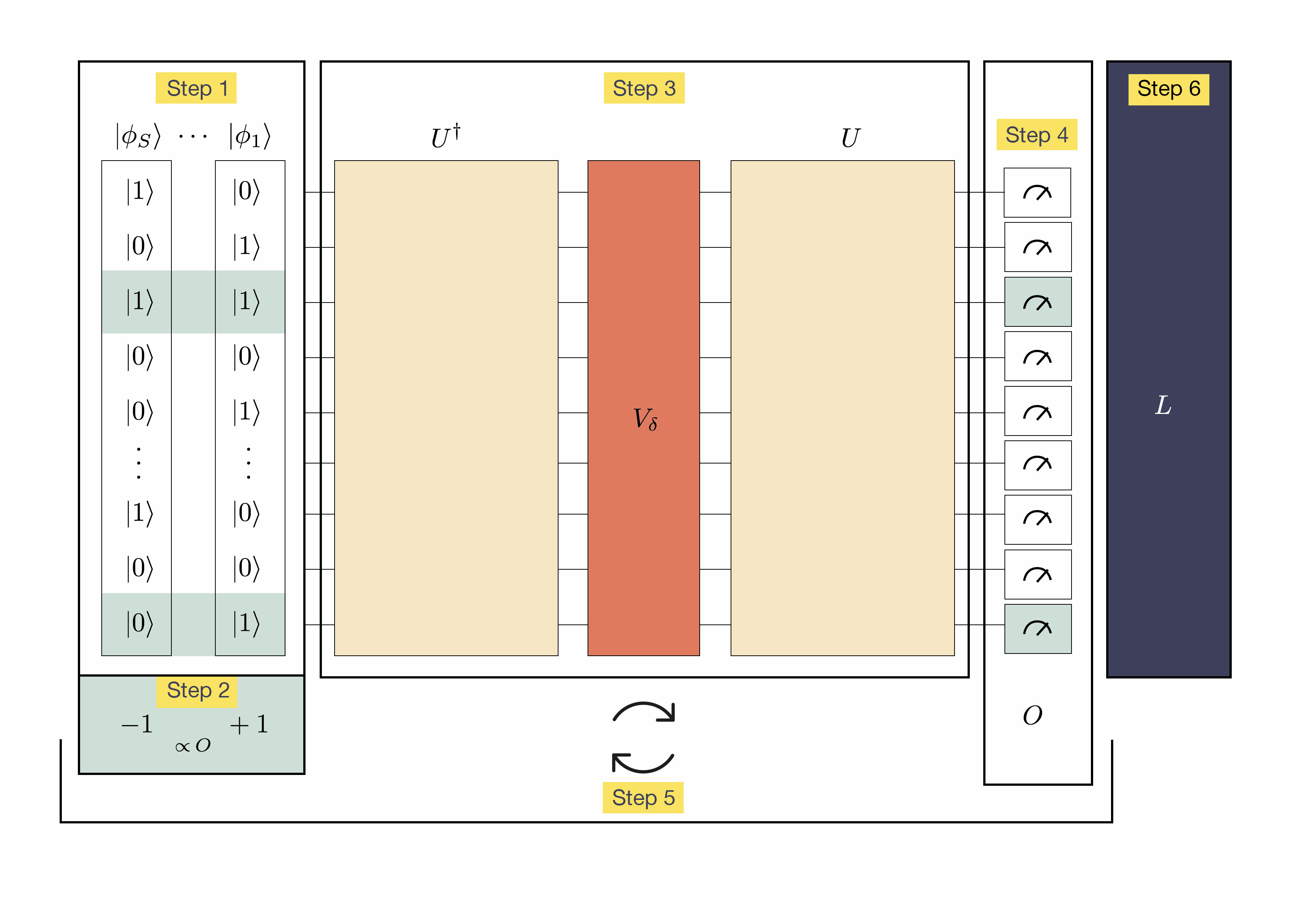}
\caption{
\textbf{\acrlong{ole} protocol} The physical question is to probe how much of the operator density is located in the (crosshatched) geometric area $\mathscr{V}_{\rm P}$ as a result of the eventual operator spreading $U^{\dag} O U$. The figure of merit is the \acrfull{otoc} $\propto {\rm tr}\{ [G, U^{\dag} O U]^{\dag}  [G, U^{\dag} O U] \}$ for the generator $G$ acting nontrivially in the area $\mathscr{V}_{\rm P}$. \acrshort{otoc} is probed with the \acrfull{ole} signal $\propto {\rm tr}[U V_{\delta} U^{\dag} O U V_{\delta}^{\dag} U^{\dag} O]$ with $V_{\delta} = e^{-i \delta G}$. The experimental setup for measuring the operator Loschmidt echo for the observable $O = Z^{\otimes \mathscr{V}_O}$ consists of the preparation of randomly chosen computational-basis-states $\ket{\phi_m} \in \{ \ket{0 \ldots 0}, \ldots, \ket{1 \ldots 1} \}$. Evolution of the state $\ket{\phi_m}$ under $U V_{\delta} U^{\dag}$ results in $\ket{\phi_m}_{\rm out}$. Measurement of $\ket{\phi_m}_{\rm out}$ is performed in the computational basis. Averaging the parity $\bra{\phi_m} O \ket{\phi_m} \in \{ +1, -1\}$ times $\bra{\phi_m}_{\rm out} O \ket{\phi_m}_{\rm out}$ over $S$ realizations provides the \acrshort{ole} estimation and its statistical error.}
\label{fig:ole-otoc}
\end{figure}

Suppose $O$ is a Pauli string comprising only $I$- and $Z$-operators on a few qubits ${\cal Q} \equiv \mathscr{V}_{O} \subset \{0, \ldots, N-1 \}$, $|{\cal Q}| \ll N$, i.e., $O = Z_{\cal Q} \equiv  \otimes_{ q \in {\cal Q} } Z_q$. Then the experiment design is as follows~\cite{algorithmiq2025} (see \cref{fig:ole-otoc}):
\begin{enumerate}
\item \label{item-preparation} prepare a randomly chosen bitstring-state $\ket{\phi_m} \in \{ \ket{0 \ldots 0}, \ldots, \ket{1 \ldots 1} \}$;
\item \label{item-parity} calculate the state parity $\bra{\phi_m} Z_{\cal Q} \ket{\phi_m} \in \{ +1, -1\}$ on the subset ${\cal Q}$ of qubits;
\item \label{item-evolution} evolve the state $\ket{\phi_m}$ with the unitary  $W_{\delta} W^{\dag}$;
\item \label{item-measurement} measure the evolved state $\ket{\phi_m(\delta)}$ in the computational basis so as to estimate $\bra{\phi_m(\delta)} Z_{\cal Q} \ket{\phi_m(\delta)}$;
\item \label{item-repeat} repeat the steps \ref{item-preparation} to \ref{item-measurement} $M$ times;
\item calculate the estimator $S_{\delta}(M)$ for the operator Loschmidt echo and its estimation error $\Delta S_{\delta}(M)$ as the average of products $\bra{\phi_m} Z_{\cal Q} \ket{\phi_m} \times \bra{\phi_m(\delta)} Z_{\cal Q} \ket{\phi_m(\delta)}$ and its standard error. 
\end{enumerate}

In stark contrast to the Loschmidt echo for states, the operator Loschmidt echo does not require any measurements to be performed on the qubits beyond the subset ${\cal Q}$. This results in the higher frequency of events, in which the values $\bra{\phi_m} Z_{\cal Q} \ket{\phi_m}$ and $\bra{\phi_m(\delta)} Z_{\cal Q} \ket{\phi_m(\delta)}$ correlate, as compared to the frequency of completely identical bitstrings in the standard Loschmidt echo for states.

Should all $N$ qubits be measured in the standard basis as an output of quantum computation anyway, this opens an option to specify the target subset of qubits ${\cal Q}$, and consequently the observable $Z_{\cal Q}$, \emph{a posteriori}. This feature of quantum computation unlocks its potential of estimating multiple \acrshort{ole}s at once, which is prohibitive to many classical simulation methods, such as Pauli propagation and tensor networks in the Heisenberg picture. The a posteriori estimation of many observables has indeed been performed to obtain the results in \cref{fig:building-trust,fig:pec_results}.

\subsubsection{OLE and OTOC}\label{sec:ole_otoc}

In this work, we consider the case when the $\delta$-perturbation is time-localised in the middle of echo dynamics, i.e.,  $W_{\delta} = U V_{\delta}$ and $W=U$, where now $U$ is the unperturbed time-evolution. In this case, the \acrshort{ole} is directly related to the \acrfull{otoc}, a concept of particular interest in operator dynamics extensively studied both theoretically and experimentally~\cite{Larkin1969QuasiclassicalMI,kitaev2015simple,kitaev2015hidden,roberts2015localized,roberts2015diagnosing,maldacena2016remarks,maldacena_bound_2016,Campisi2017,landsman2019,Yan2020,Mi2021information,Sanchez2022,Liu2025,Yan2020b,Harris2022,abanin_observation_2025,zhang2025molecular}. The \acrshort{ole} in Eq.~\eqref{ole} takes the form
\begin{equation} \label{ole-V}
    S_{\delta} = \frac{{\rm tr}\big( U^{\dag} O^{\dag} U V_{\delta} U^{\dag} O U V_{\delta}^{\dag} \big)}{{\rm tr}\left( O^{\dag} O \right)}.
\end{equation}
Let $G=G^{\dag}$ be a generator of the perturbation $V_{\delta}$, i.e., $V_{\delta} = e^{-i \delta G}$. Then we have
\begin{eqnarray} \label{VOVdag}
V_{\delta} U^{\dag} O U V_{\delta}^{\dag} & = & e^{-i \delta G} U^{\dag} O U e^{i \delta G} \nonumber\\
& = & U^{\dag} O U - i \delta [G,U^{\dag} O U] - \frac{\delta^2}{2} \left[ G, [G,U^{\dag} O U] \right] \nonumber\\ 
&& + \ldots + \frac{(- i \delta)^k}{k!} \underbrace{\left[ G, [G, [ \ldots [G,U^{\dag} O U] \ldots ] ] \right]}_{k\text{th-order commutator}} + \ldots
\end{eqnarray}
Substituting \eqref{VOVdag} in \eqref{ole-V} and taking into account the relation ${\rm tr}(A[B,C]) = {\rm tr}([A,B]C)$, we obtain 
\begin{equation} \label{ole-delta-expansion}
    S_{\delta} = 1 + i\delta \times \frac{{\rm tr}\big( [U^{\dag} O^{\dag} U, U^{\dag} O U] G \big)}{{\rm tr}\left(O^{\dag} O \right)} - \frac{\delta^2}{2} \times \frac{{\rm tr}\big( [G, U^{\dag} O U]^{\dag} [G, U^{\dag} O U] \big)}{{\rm tr}\left(O^{\dag} O \right)} + \ldots
\end{equation}
For a Hermitian operator $O=O^{\dag}$, the first-order term $\propto \delta$ vanishes. So do all odd-order terms because
\begin{equation*}
    {\rm tr}\bigg( U^{\dag} O^{\dag} U \underbrace{\left[ G, [G, [ \ldots [G,U^{\dag} O U] \ldots ] ] \right]}_{k\text{th-order commutator}} \bigg) = (-1)^k {\rm tr}\bigg( U^{\dag} O U \underbrace{\left[ G, [G, [ \ldots [G,U^{\dag} O^{\dag} U] \ldots ] ] \right]}_{k\text{th-order commutator}} \bigg).
\end{equation*}
Therefore, the \acrshort{ole} for a Hermitian traceless operator $O$ with eigenvalues $\pm 1$ reads
\begin{eqnarray} \label{ole-delta-expansion-2}
    S_{\delta} & = & 1 - \frac{\delta^2}{2} \times \underbrace{\frac{1}{2^N} {\rm tr}\big( [G, U^{\dag} O U]^{\dag} [G, U^{\dag} O U] \big)}_{\text{\acrshort{otoc}}} \nonumber\\ 
      &   & + \frac{\delta^4}{4!} \times \underbrace{\frac{1}{2^N} {\rm tr}\big( [G,[G,U^{\dag} O U]]^{\dag} [G,[G,U^{\dag} O U]] \big)}_{\geqslant 0} + \ldots \nonumber\\
      & \approx & \exp \left( - \frac{C \delta^2}{2} \right).
\end{eqnarray}
The terms in the second line of Eq.~\eqref{ole-delta-expansion-2} are negligible provided $\delta \ll 1$. The validity of omitting those terms is probed experimentally: the assumption is justified as long as the dependence of \acrshort{ole} on $\delta^2$ remains linear in the vicinity of $\delta = 0$ and does not bend up due to the positive fourth-order correction. The slope in the linear dependence $S_{\delta} \approx 1 - \frac{\delta^2}{2} C$ enables one to reveal the \acrshort{otoc} in the maximally mixed (infinite temperature) state $\varrho_{\infty} = (\frac{1}{2}I)^{\otimes N}$, 
\begin{equation}
\label{otoc}
    C = \langle [G, U^{\dag} O U]^{\dag} [G, U^{\dag} O U] \rangle_{\varrho_{\infty}}.
\end{equation}
The \acrshort{otoc} can also be seen as an ensemble average over the Loschmidt echoes for states~\cite{Yan2020,zhou2024generalizedloschmidtechoinformation}. However, the \acrshort{ole} provides a more practical way to estimate the \acrshort{otoc} in the presence of noise.

\subsection{Framework for implementing and probing the effect of scattering in semiscrambling dynamics}

\emph{Semi-scrambling} dynamics occupies a regime that is favorable for pre-fault-tolerant quantum computation: the dynamics is entangling enough that its interference effects strain state-of-the-art classical simulation methods, while the growth of the operator support is slow enough that the experimental signal survives the accumulation of hardware noise. Natural hosts for such dynamics are heterogeneous structures, in which interaction forms and strengths are distributed non-uniformly yet periodically, akin to metamaterials~\cite{zheludev2012metamaterials} and Kitaev materials~\cite{TREBST20221}. 

In this appendix we define a model realizing these conditions on the heavy-hex topology of IBM quantum processors. As mentioned, the full Loschmidt-echo evolution factorizes as $W_\delta W^\dag = U V_\delta U^\dag$, with the probing perturbation $V_\delta = e^{-i\delta G}$ localized in time in the middle of the echo. The dynamics $U$ simulates a scattering-like process with $U = U_0^{\dag} U_\eta$, where $U_0$ consists of $L$ identical Floquet layers and $U_\eta$ consists of the same layers modified on a small set of \emph{scattering} qubits $\mathscr{V}_{\rm S}$, with the scattering coefficients on those qubits rescaled by a factor set by $\eta$. Two distinct perturbations therefore enter the experiment and must not be confused: the \emph{scattering strength} $\eta$, a static modification localized on $\mathscr{V}_{\rm S}$ that breaks the exact cancellation $U_\eta^{\dag} U_0 = I$ and thereby generates the physical signal under study; and the \emph{probe strength} $\delta$, which is scanned to extract this signal through the \acrshort{ole}--\acrshort{otoc} relation of Eq.~\eqref{ole-delta-expansion-2}.

\begin{figure*}[t]
\centering
\centering
\includegraphics[width=\linewidth]{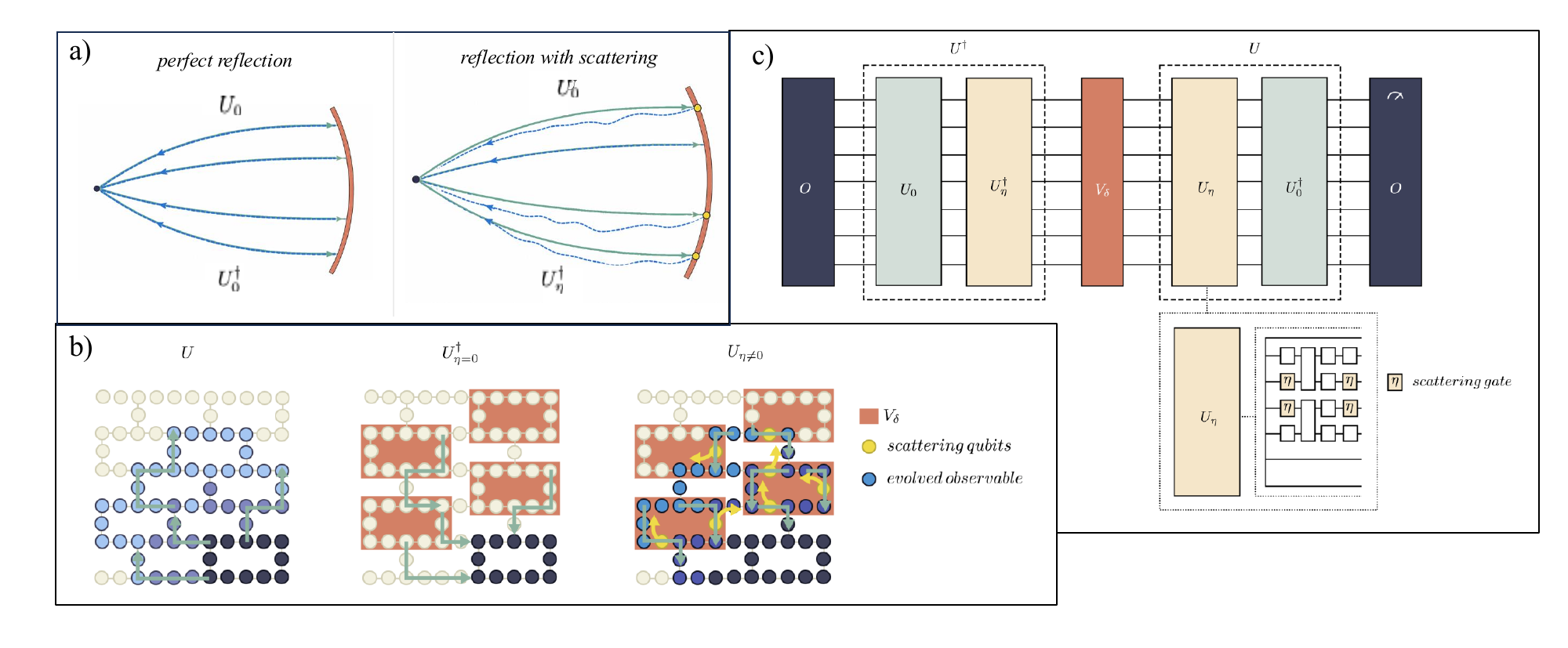}
\caption{
\textbf{\acrshort{ole} in scattering dynamics.} a)~Intuition for the scattering dynamics: starting from the observable $O$, the effective support spreads under the echo evolution $U^{\dag} = U_{\eta}^{\dag} U_0$. In the absence of scattering ($\eta = 0$) the forward and backward Floquet evolutions cancel exactly and the support returns to $O$; the few scattering centers $\mathscr{V}_{\rm S}$ break this cancellation and generate the residual signal probed by $V_\delta$ (perturbation region $\mathscr{V}_{\rm P}$, observable region $\mathscr{V}_O$). For $\eta \neq 0$, the refocusing of the observable is not possible anymore, and the observable continues to partially spread due to the presence of the scattering qubits. b) Intuition for the scattering dynamics as applied within the heavy hex topology. c)~Effective circuit for the \acrshort{ole}. The observable $O$ is prepared through its eigenstate ensemble (\cref{subsection-measurement-of-ole}) and measured at the end; in between, the state evolves under $U^{\dag} = U_\eta^{\dag} U_0$, the probe $V_\delta$, and $U = U_0^{\dag} U_\eta$ [Eq.~\eqref{U0-Ueta-U}]. The inset expands a single $U_\eta$ layer, highlighting the localized scattering gates ($\eta$) applied on the qubits $\mathscr{V}_{\rm S}$ that break the exact cancellation $U_\eta^{\dag} U_0 = I$.}
\label{fig:scattering}
\end{figure*}

For readers interested in reproducing the circuit, $U_0$ and $U_{\eta}$ are defined in Eq.~\eqref{U0-Ueta-U}, with the constituent kicked-Ising Floquet layer is given explicitly in Eq.~\eqref{UFL}. The model is composed of a combination of $ZZ$ type couplings, local $Z$ fields and a kicked local transversal $X$  field that controls the speed of information spread. The 56-qubit topology, together with the support $\mathscr{V}_{\rm P}$ of the perturbation generator $G = \sum_{q \in \mathscr{V}_{\rm P}} X_q$, is shown in Fig.~\ref{fig:scattering} and Fig.~\ref{fig:heterogeneity}. The circuit is also available in explicit form in the Quantum Advantage Tracker~\cite{advantage_tracker}.

The semi-scrambling behavior arises from the combination of two ingredients, described in the following subsections: (i)~\emph{scattering dynamics} (\cref{app:scattering}), the near-cancellation of the forward and backward Floquet evolutions everywhere except at the scattering centers, so that the entire signal originates from a few qubits while the circuit remains deep and highly entangling; and (ii)~a \emph{heterogeneous structure} (\cref{sec:specific-dynamics}), a spatial modulation of the transverse-field angles that partitions the lattice into fast and slow information-propagation channels and thus retards the growth of the operator support. The two ingredients are independent in principle and are introduced separately below; \cref{sec:graph} then shows how they are combined in a single native Floquet layer on hardware.

\subsubsection{Scattering dynamics}
\label{app:scattering}

By \emph{scattering dynamics} we mean the way the time-evolved observable $U^{\dag} O U$ is shaped by the near-cancellation of a forward and a backward Floquet evolution that differ only at a few scattering centers. The dynamics is built from $L$ forward and $L$ backward Floquet layers,
\begin{equation} \label{U0-Ueta-U}
    U_0 = (U_{\rm FL})^L, \qquad U_{\eta} = \big(U_{\rm FL}^{(\eta)}\big)^L, \qquad U^{\dag} = U_{\eta}^{\dag} U_0 = \big(U_{\rm FL}^{(\eta)\dag}\big)^L (U_{\rm FL})^L,
\end{equation}
where $U_{\rm FL}$ is a single Floquet layer of the unperturbed dynamics, $U_{\rm FL}^{(\eta)}$ is the same layer in the presence of scattering of strength $\eta$ [defined at the gate level below, Eq.~\eqref{UFL}], and $U_0 = U_{\eta = 0}$. In the absence of scattering ($\eta = 0$) the two halves cancel exactly, $U^{\dag} = U_{\eta}^{\dag} U_0 = I$, and the time-evolved observable is left unchanged. The entire signal is therefore generated by the few scattering centers $\mathscr{V}_{\rm S}$ alone, even though the circuit itself remains deep and highly entangling. In the weak-scattering regime $\eta \approx 0$ (equivalently $U_0 \approx U_\eta$), the lowest-order expansion of the signal in the scattering strength $\eta$ is derived in \cref{sec:physics-scattering}.

We probe the effect of scattering by measuring \acrshort{ole} for an observable $O$ and a probing perturbation $V_\delta$ whose geometric supports are disjoint. The target observable $O$ acts non-trivially on a subset of qubits $\mathscr{V}_O$; in the experiment it is a Pauli operator $O = \bigotimes_{q \in \mathscr{V}_O} Z_q$, so that the measurements are performed in the computational basis. The perturbation $V_{\delta} = e^{-i\delta G}$ probes the effect of scattering and is generated by a Hermitian operator $G$ acting non-trivially on a subset of qubits $\mathscr{V}_{\rm P}$; in the experiment $G = \sum_{q \in \mathscr{V}_{\rm P}} X_q$, implying $V_{\delta} = \bigotimes_{q \in \mathscr{V}_{\rm P}} e^{-i \delta X_q}$.

The figure of merit for the effect of scattering is the \acrfull{otoc} $\propto {\rm tr}\{ [G, U^{\dag} O U]^{\dag}  [G, U^{\dag} O U] \}$ for the generator $G$ acting nontrivially in the area $\mathscr{V}_{\rm P}$. The \acrshort{otoc} is probed with the \acrfull{ole} signal $\propto {\rm tr}[U V_{\delta} U^{\dag} O U V_{\delta}^{\dag} U^{\dag} O]$ with $V_{\delta} = e^{-i \delta G}$, see \cref{sec:OLE}. If no scattering takes place, the \acrshort{otoc} vanishes for non-overlapping observable and perturbation, $\mathscr{V}_O \cap \mathscr{V}_{\rm P} = \varnothing$.

\subsubsection{Heterogeneous structure} \label{sec:specific-dynamics}

\begin{figure*}[t]
\centering
\begin{minipage}[b]{0.35\textwidth}
\centering
\includegraphics[width=\linewidth]{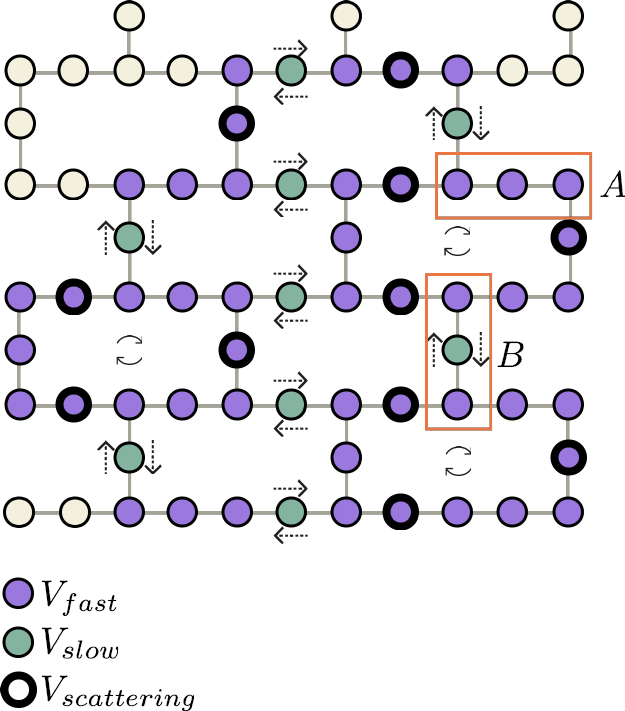}\\[2pt]
\textbf{(a)}
\end{minipage}\hfill
\begin{minipage}[b]{0.60\textwidth}
\centering
\includegraphics[width=\linewidth]{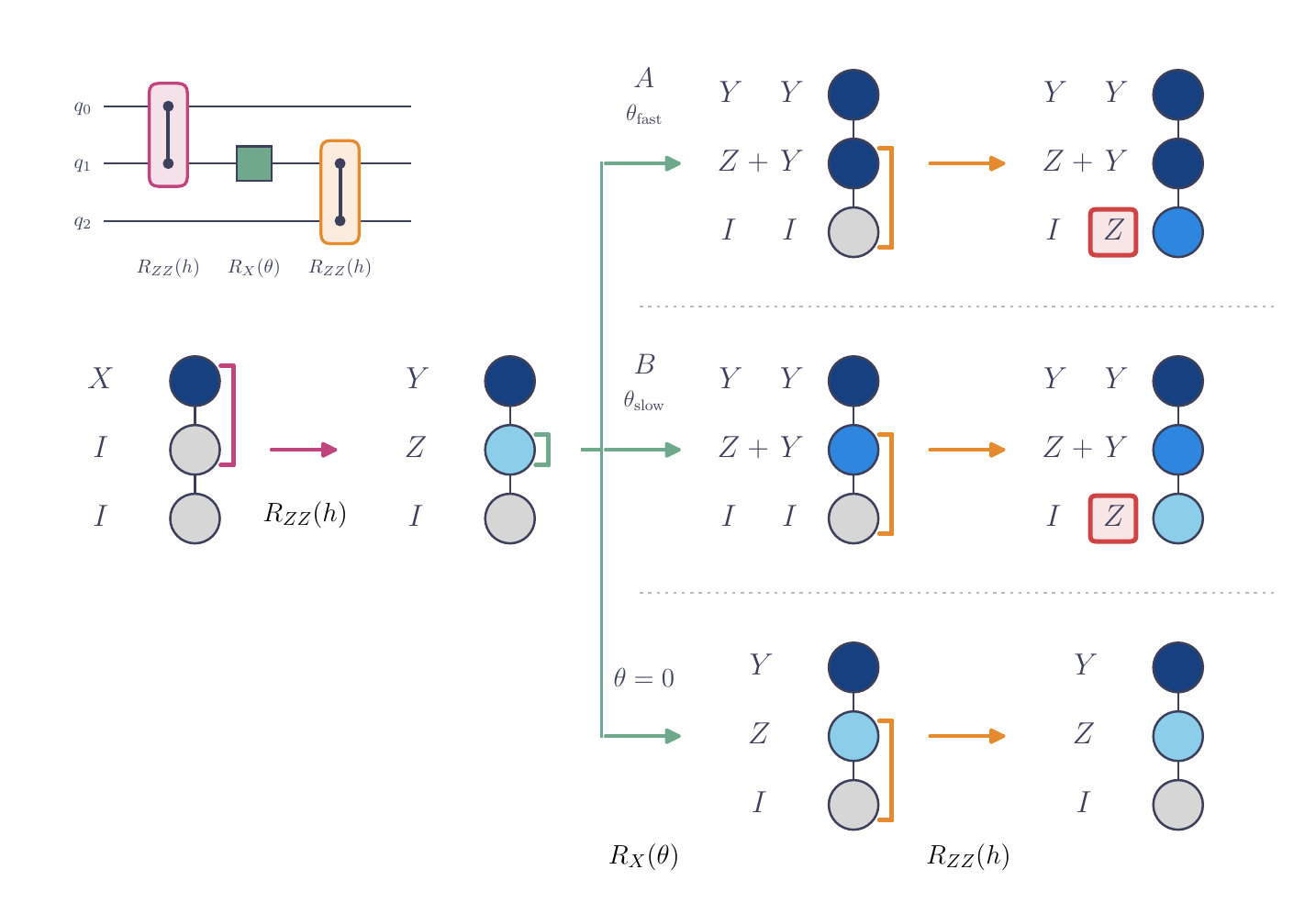}\\[2pt]
\textbf{(b)}
\end{minipage}
\caption{
\textbf{Fast and slow propagation in the heterogeneous structure.} (a)~The $56$-qubit heavy-hex layout with the vertices coloured by class: fast vertices $\mathscr{V}_{\rm fast}$, slow vertices $\mathscr{V}_{\rm slow}$, and the scattering centers $\mathscr{V}_{\rm S}$ (ringed). The dashed arrows indicate the direction and relative speed of the operator-support growth, while the highlighted rectangles single out example regions: a \emph{fast loop} composed only of fast vertices (label $A$) and a mixed fast/slow region (label $B$), corresponding to the two spreading scenarios detailed in~(b), where a section of 3 qubits is considered. (b)~Single-Floquet-layer mechanism of operator spreading, in a simplified 3-qubit example, but same logic as the experiment. A $Z$ operator landing on a qubit can propagate to a neighbour only after the local transverse-field rotation $R_X(\theta)$ converts part of it into a $Y$ component of amplitude $\sin(2\theta)$, which the subsequent $R_{ZZ}(h)$ gate turns into a $Z$ on the neighbouring qubit. For a fast vertex (large field angle $\theta = b_1$, case $A$) the conversion is efficient and the support grows, as depicted by the blue circles; for a slow vertex (small angle $\theta = b_2$, case $B$) it is weak and the growth is retarded; for $\theta = 0$ (or a multiple of $\tfrac{\pi}{2}$) the $Z$ is preserved and propagation through the qubit is blocked entirely.
}
\label{fig:heterogeneity}
\end{figure*}

The second ingredient is a spatial \emph{heterogeneity} that sets how fast the observable support grows under each Floquet layer, independently of the scattering discussed above. The lattice vertices are partitioned into two classes according to the intensity of their local transverse ($X$) field, $\mathscr{V} = \mathscr{V}_{\rm fast} \cup \mathscr{V}_{\rm slow}$: \emph{fast} vertices $\mathscr{V}_{\rm fast}$ and \emph{slow} vertices $\mathscr{V}_{\rm slow}$. This partition is what makes the dynamics \emph{semi-scrambling}: the operator support expands neither too fast, which would make the signal fragile to hardware noise, nor too slow which would keep it classically easy to simulate.  
It is the transverse field that controls the speed of information spreading, and it can block it entirely through a qubit $q$ whenever its field angle is a multiple of $\tfrac{\pi}{2}$. For $b_q \ll 1$, information instead \emph{leaks} through that qubit in a retarded way: the $Z$ operator on qubit $q$ is rotated by $e^{-i b_q X}$ into a small component $\sin(2 b_q)\,Y$, which the entangling gate then propagates onwards. Assigning a larger field angle to the fast vertices $\mathscr{V}_{\rm fast}$ than to the slow vertices $\mathscr{V}_{\rm slow}$ produces the imbalance between fast and slow propagation channels that defines the heterogeneous structure. The resulting effect on the operator support is illustrated in Fig.~\ref{fig:heterogeneity}: the support grows with larger weight along $\mathscr{V}_{\rm fast}$ and is retarded along $\mathscr{V}_{\rm slow}$.
This heterogeneous structure is not the only medium for reaching semi-scrambling dynamics, but it has the nice feature to create highly entangled regions, from which the operator can continue to spread in a controlled way.

\subsubsection{Implementation on real hardware topology}
\label{sec:graph}

On real hardware the two ingredients above, scattering (\cref{app:scattering}) and heterogeneity (\cref{sec:specific-dynamics}), are realized \emph{simultaneously}, within the same native Floquet layer: the heterogeneity is encoded in the choice of transverse-field angles across the lattice, while the scattering is a localized rescaling of those same angles on the qubits $\mathscr{V}_{\rm S}$.

A general framework for the description and design of such heterogeneous structures is based on a graph representation. The graph vertices $\mathscr{V}$ correspond to the lattice sites, whereas the graph edges $\mathscr{E}$ encode the interaction pattern among the sites; in a digital simulation of a spin-$\tfrac{1}{2}$ lattice each vertex is associated with a qubit. The closer the pattern of graph edges $\mathscr{E}$ to the actual connectivity of two-qubit operations in the device, the shallower the transpiled circuit, so we focus on graphs compatible with the topology of native gates in IBM Heron quantum computers. Heterogeneity is attained through a non-uniform distribution of parameters over the vertices $\mathscr{V}$, the edges $\mathscr{E}$, or both, at the level of native gates, a heterogeneity in the single-qubit gates ($\mathscr{V}$), the two-qubit gates ($\mathscr{E}$), or both. Following Ref.~\cite{algorithmiq2025}, where several mechanisms relate the difference in vertex or edge gates to the speed of operator-support growth, the main idea is to separate the whole set of vertices into fast and slow information-transmission sets, $\mathscr{V} = \mathscr{V}_{\rm fast} \cup \mathscr{V}_{\rm slow}$ (or to perform the analogous separation at the level of edges, $\mathscr{E} = \mathscr{E}_1 \cup \mathscr{E}_2$). Taking the hardware implementation into account, we split the set of edges $\mathscr{E} = \cup_j E_j$ into disjoint subsets of commuting gates that are applied simultaneously and form a single circuit layer.

Concretely, the heterogeneity is controlled by the intensity of the $X$ field in the Floquet layer
\begin{equation} \label{UFL}
U_{\rm FL}(h,b_1,b_2) = \prod_{E \subset \mathscr{E}} \bigotimes_{(q_1,q_2) \in E} e^{-i (b_{q_1} X_{q_1} + b_{q_2} X_{q_2})} e^{-i \frac{\pi}{4} Z_{q_1} Z_{q_2}} e^{-i h (Z_{q_1} + Z_{q_2})} e^{-i (b_{q_1} X_{q_1} + b_{q_2} X_{q_2})}.
\end{equation}
The scattering is then implemented as a rescaling of the $X$-field angle on the scattering qubits, $b_q \rightarrow b_q - \eta$ for $q \in \mathscr{V}_{\rm S}$, with $\eta$ the scattering strength; in effect, scattering changes only the angles of the $R_X$ gates acting on the qubits $q \in \mathscr{V}_{\rm S}$. The perturbed layer $U_{\rm FL}^{(\eta)}(h,b_1,b_2)$ entering Eq.~\eqref{U0-Ueta-U} thus has the same structure as Eq.~\eqref{UFL}, and it is here that the two ingredients combine: the field angle of each qubit $q$ is fixed jointly by its heterogeneity class ($\mathscr{V}_{\rm fast}$ or $\mathscr{V}_{\rm slow}$) and by whether it is a scattering center, according to the region of \cref{fig:heterogeneity} it belongs to,
\begin{enumerate}
    \item $b_q = b_1$ if $q \in \mathscr{V}_{\rm fast} \backslash \mathscr{V}_{\rm S}$ (fast vertices);
    \item $b_q = b_2$ if $q \in \mathscr{V}_{\rm slow} \backslash \mathscr{V}_{\rm S}$ (slow vertices);
    \item $b_q = b_1 - \eta  $ if $q \in \mathscr{V}_{\rm fast} \cap \mathscr{V}_{\rm S}$ and $b_q = b_2 - \eta $ if $q \in \mathscr{V}_{\rm slow} \cap \mathscr{V}_{\rm S}$ (scattering centers).
\end{enumerate}
If $\eta = 0$, then $U_{\rm FL}^{(0)} = U_{\rm FL}$, consistently with $U_0 = U_{\eta = 0}$ in Eq.~\eqref{U0-Ueta-U}.

Define a Pauli rotation gate as $R_Q(\theta) = e^{-i\theta Q/2}$. Note that $R_{ZZ}(\tfrac{\pi}{2}) = e^{-i \frac{\pi}{4} Z Z} = e^{-i \frac{\pi}{4}} \cdot S \otimes S \cdot CZ$, so the entangling gate $R_{ZZ}(\tfrac{\pi}{2})$ is implemented in hardware via the $CZ$ Clifford gate, while the extra $S$ gates are merged with neighbouring single-qubit operations and retranspiled in the conventional manner.

\begin{figure*}[t]
\centering
\includegraphics[width=0.95\textwidth]{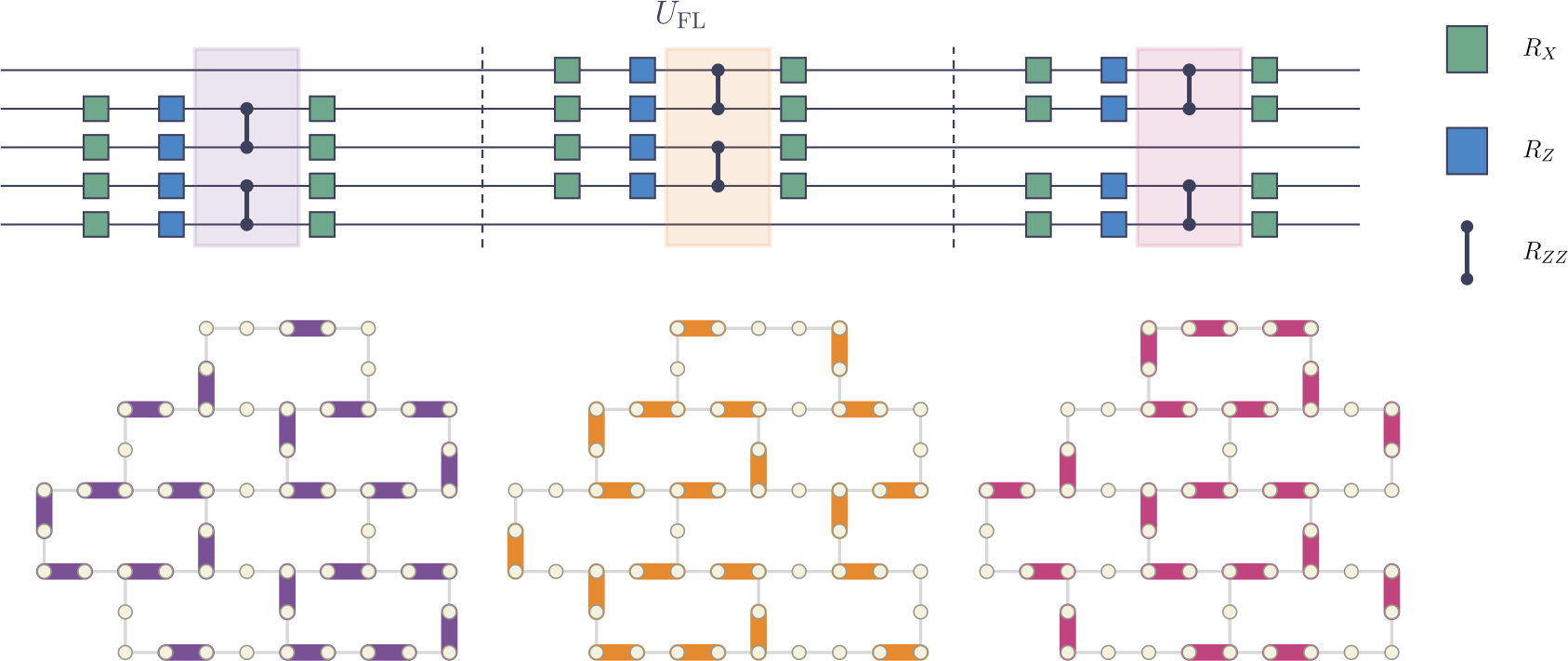}
\caption{
\textbf{Native Floquet layer and its scheduling on the heavy-hex topology.} \emph{Top:} gate structure of a single Floquet layer $U_{\rm FL}$ [Eq.~\eqref{UFL}], built from single-qubit transverse-field rotations $R_X$ and $R_Z$ and two-qubit entangling gates $R_{ZZ}$. \emph{Bottom:} the edges $\mathscr{E}$ of the $56$-qubit heavy-hex graph are partitioned into disjoint subsets $E_1, E_2, E_3$ (coloured), each a set of two-qubit gates acting on disjoint qubit pairs that therefore commute and are executed simultaneously as one circuit layer. Applying the three subsets in sequence realizes one full Floquet layer $U_{\rm FL}$, and, since the graph coincides with the device connectivity, no extra transpilation is incurred.
}
\label{fig:model}
\end{figure*}

The native Floquet layer and its edge scheduling are shown in \cref{fig:model}, while the vertex classes and the key regions of the graph ($\mathscr{V}_{\rm fast}, \mathscr{V}_{\rm slow}, \mathscr{V}_{\rm S}, \mathscr{V}_{\rm P}, \mathscr{V}_O$) are indicated in \cref{fig:heterogeneity}. The evolution $U$ is given by Eqs.~\eqref{U0-Ueta-U} and \eqref{UFL} with the parameters $h = \frac{\pi}{8}$, $b_1 = \frac{3\pi}{16}$, $b_2 = 0.125$, and $L=1,2,3,4,5,6$. Scattering takes place on $|\mathscr{V}_{\rm S}| = 11$ qubits with different scattering strengths $\eta \in \{0, 3\pi/40, 3\pi/20, 9\pi/40, 3\pi/10, 3\pi/8\}$. The generator of the probing perturbation $V_\delta = e^{-i \delta G}$ is $G = \sum_{q \in \mathscr{V}_{\rm P}} X_q$ and acts nontrivially on $|\mathscr{V}_{\rm P}| = 35$ qubits. The observable $O = \bigotimes_{q \in \mathscr{V}_O} Z_q$, which corresponds to the results depicted in \cref{fig:main}, has nontrivial support on $|\mathscr{V}_O| = 12$ qubits within a fast loop, i.e.\ a rectangle composed only of fast vertices.

\subsection{Fluctuation-dissipation relations for OLE} \label{sec:physics-scattering}

The scattering dynamics $U^{\dag} = U_\eta^{\dag}U_0$ is composed of $L$ forth- and $L$ backpropagation Floquet layers that do not fully cancel each other only due to a few scattering centers, see \cref{fig:scattering}. Eq.~\eqref{otoc} then takes the form
\begin{equation}
\label{otoc-lc}
    C = \underbrace{ \langle [U_\eta G U_\eta^{\dag}, U_0 O U_0^{\dag}]^{\dag} [U_\eta G U_\eta^{\dag}, U_0 O U_0^{\dag}] \rangle_{\varrho_{\infty}} }_{\text{Loschmidt correlator}}.
\end{equation}
We refer to the right-hand side of Eq.~\eqref{otoc-lc} as the \emph{Loschmidt correlator} as it quantifies how two initially commuting operators ($O$ and $G$) lose their commutation property due to the presence of differences in their individual evolutions $U_0$ and $U_\eta$. Were $U_0$ and $U_\eta$ identical, the operators $U_0 O U_0^{\dag}$ and $U_\eta G U_\eta^{\dag}$ would still commute despite the fact that their supports overlap as a result of operator spreading. This is in stark contrast to the conventional \acrshort{otoc}, which is sensitive to the operator spreading in the first place. Therefore, this is the scattering process $U^{\dag} = U_\eta^{\dag}U_0$ with $U_0 \approx U_\eta$ and $U_0 \neq U_\eta$ in which the original spirit of the Loschmidt paradox~\cite{prosen2003theory,gorin_dynamics_2006} extends to the realm of operators.

The operators $U_0 O U_0^{\dag}$ and $U_\eta G U_\eta^{\dag}$ cease to commute exclusively due to the difference in the evolutions $U_0$ and $U_\eta$. The greater the mismatch the greater the Loschmidt commutator. Consider a particular scattering generator $K$, with $\eta$ controlling its strength, e.g., $U_0 = \prod_{l=1}^L U_{\rm FL}$ and $U_\eta = \prod_{l=1}^L e^{- i \eta K} U_{\rm FL}$. Denote $\widetilde{K}_l := (U_{\rm FL}^{\dag})^l K U_{\rm FL}^l$ the scattering generator in the interaction picture. Then $U_\eta^{\dag}U_0 = I + i \eta \sum_l \widetilde{K}_l - \frac{\eta^2}{2} \sum_{l} \widetilde{K}_l^2 - \eta^2 \sum_{l' > l} \widetilde{K}_l \widetilde{K}_{l'} + O(\eta^3)$. Since $[G,O] = 0$, the Loschmidt correlator takes the form 
\begin{equation} \label{C-scattering}
C = \eta^2 \sum_{l,l'} C_{ll'}  + O(\eta^3), \qquad C_{ll'} = \frac{1}{2^N} {\rm tr}\left( \Big[ G, \big[ \widetilde{K}_l, O \big] \Big]^{\dag} \Big[ G, \big[ \widetilde{K}_{l'}, O \big] \Big] \right),
\end{equation}
with $C$ exhibiting quadratic growth with respect to the scattering strength $\eta$ in the leading term. Eq.~\eqref{C-scattering} has an extra interpretation in terms of the fluctuation-dissipation theorem in analogy with the Loschmidt echo for states \cite{prosen2003theory}. $C_{ll'}$ is the 2-point time-correlation function of the scattering generator $K$ with allowance for the observable $O$ and the perturbation generator $G$ (the first order contributions in the whole circuit vanish). The diagonal contribution $C_{ll} \geq 0$. If all three operators $O, G, K$ are mutually commuting, then $C_{l0} = C_{0l'} = 0$. In the relation $S \approx 1 - \frac{\delta^2 \eta^2}{2} \sum_{l,l'} C_{ll'}$, the \acrshort{ole} signal $S$ can be treated as the dissipation of quantum information and $\sum_{l,l'} C_{ll'}$ is an integrated time-correlation function (fluctuation).

\section{Tensor network simulation methods} \label{app:classical_heuristics}

\subsection{Belief propagation} \label{app:bp}

Tensor-network methods approximate a many-body wavefunction by restricting the amount of entanglement retained across the bonds of a network~\cite{Verstraete2008,Schollwock2011,Orus2014}. Their application to two-dimensional quantum circuits is challenging because evaluating observables and determining optimal bond truncations require contracting a loopy tensor network, a task whose exact cost generally grows prohibitively with the bond dimension. In the following sections, we address this contraction problem using \acrfull{bp}~\cite{Tindall2023}, which has been shown to be a powerful and reliable method for performing tensor-network simulations on the heavy-hex topology \cite{Tindall2024efficient,Rudolph2025,Tindall2026}. 

We represent the circuit wavefunction as a tensor-network state whose connectivity follows the heavy-hex graph of the quantum processor. In \acrshort{bp}, the environment surrounding a tensor or bond is approximated by a collection of self-consistent messages passed along the edges of this graph. Each message encodes the cavity environment of the branch it originates
from, i.e.\ the contraction of the network on one side of an edge once the
opposite vertex is removed. The environment of a vertex is then taken to
factorize into a product of the incoming messages, an approximation that is
exact on trees but neglects the correlations mediated by loops. Its accuracy therefore depends on the strength of the correlations carried around those loops, rather than directly on the total system size.

The converged \acrshort{bp} messages serve two purposes in our simulations. First, they provide approximate environments for evaluating norms and expectation values. Second, they are used to place the tensor-network state in an approximate Vidal gauge and to construct the local environment used when truncating a bond after the application of a gate~\cite{Tindall2023}. The resulting gate-update procedure is equivalent to a Simple Update performed in this approximate gauge~\cite{Jiang2008}. It avoids the cost of computing the full two-dimensional environment while retaining information about the surrounding network through the \acrshort{bp} messages.

Two distinct approximations must consequently be monitored. The first is the finite bond dimension $\chi$, which limits the entanglement retained during the evolution. We quantify the accumulated effect of these truncations using the product of the local gate fidelities defined below. The second is the \acrshort{bp} environment approximation itself. We assess it using loop-correlation diagnostics, which measure the correlations neglected when a closed loop is replaced by the corresponding \acrshort{bp} messages~\cite{Rudolph2025}. Small loop corrections indicate that the message-based environment is reliable, whereas convergence of the messages alone does not provide such a guarantee. For the circuits studied here, the loop corrections remain small ($\lesssim 10^{-5}$) at moderate truncation (bond dimensions $\approx 128$) and decrease as the maximum allowed bond dimension increases. This allows us to attribute the dominant loss of accuracy in the regimes that are difficult to classically simulate primarily to the finite-$\chi$ truncation, subject to the limitations of the diagnostics.

We update the \acrshort{bp} messages sequentially without damping and truncate every bond to a maximum dimension $\chi$. To reach the largest bond dimensions, the tensors and messages of a single simulation are distributed across multiple GPUs. 

We first present the resulting observable estimates and their convergence with $\chi$ in \cref{bp-tn-results}.
We then describe the distributed implementation in \cref{tn-bp-implementation}, followed by its computational scaling and resource requirements in \cref{tn-bp-computational-cost}.

\subsubsection{Results}
\label{bp-tn-results}
We simulate the model described in \cref{app:model} using \acrshort{bp}, adopting the same parameters as the quantum hardware experiments: $b=0.125$, $\delta \in [0, 0.3]$, and scattering strength $\eta \in [0, 3\pi/8]$. We consider circuits with different numbers $L \in \{2, \dots, 6\}$ of Floquet layers. For these circuits, we compute the \acrshort{ole} signal in the Schr\"odinger evolution and estimate the fidelity of the \acrshort{bp} state relative to the ideal state, where the latter is defined as the average gate fidelity
\begin{equation}
\label{eq:fidelity}
f = \prod_{i=1}^n f_i
\end{equation}
where the product runs on the gates and 
$$f_i = \left| \frac{\langle \psi_i|U_i|\psi_{i-1}\rangle }{\sqrt{\langle \psi_i|\psi_i\rangle \langle \psi_{i-1}|\psi_{i-1}\rangle} }\right|^2$$
is the overlap between the approximate state and the ideal state obtained by applying the gate exactly \cite{Tindall2023}. 
We note that Eq.~\eqref{eq:fidelity} provides only a heuristic estimate of the fidelity of the \acrshort{bp} state: the per-gate overlaps $f_i$ are evaluated using the rank-one \acrshort{bp} environments, which are exact only on loop-free networks, and the multiplicative accumulation over gates assumes that successive truncation errors contribute independently. Consequently, this estimator is neither exact nor a guaranteed lower bound on the true fidelity~\cite{haghshenas_digital_2026}, although the large loops of the heavy-hex geometry are expected to make it a reliable proxy for the circuits considered here.

We ran simulations across a range of bond dimensions \(\chi\), using at least 20 initial states per circuit setting to gather statistics. For the largest bond dimension tested, \(\chi\) = 980, we reduced this to 10 initial states; we highlight that the simulations shown here are currently the largest ones attempted in the literature for these types of circuits. We argue that the BP algorithm converges correctly for the circuits that have been considered. We checked the loop convergence via the first-order approximation of the BP error, as done in Eq.~3 of~\cite{Rudolph2025}. At L=3 and with increasing bond dimension, the error decreases up to $\varepsilon\sim10^{-5}$ already at $\chi\approx 320$, and the quality of convergence increases with higher bond dimension. We thus assume in the following that any approximation occurring in the \acrshort{bp} framework is purely due to the bond dimension approximation.

We analyze the results as a function of bond dimension and fidelity. The main goal of this analysis is to verify the convergence of the \acrshort{bp} results, so that they can be compared to the results obtained from other classical simulations and from the quantum computer, and highlight where \acrshort{bp} fails as the complexity of the circuits increases.

The quality of the \acrshort{bp} simulation result depends on several model parameters. In general, convergence becomes more challenging for deeper circuits, larger scattering strengths $\eta$, and larger values of $\delta$. We note that circuits up to 4 layers converge for a bond dimension that is just reachable by current computational resources.

\begin{figure}[ht]
\includegraphics[width=0.9\linewidth]{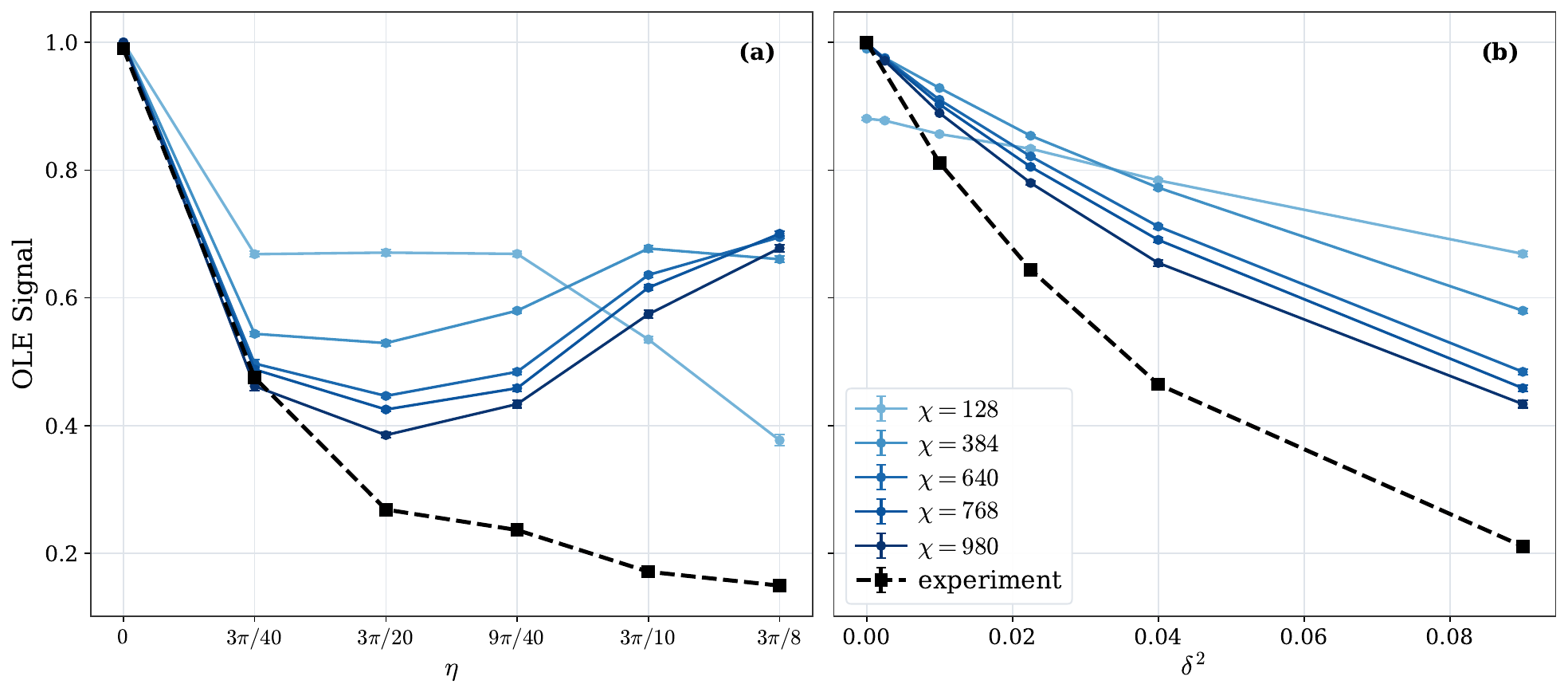}
\caption{
\acrshort{bp} simulation results compared to the experimental data from \texttt{ibm\_boston}. Panel (a): \acrshort{ole} signal as a function of the scattering strength $\eta$ for increasing bond dimension. The computation is performed for the 6-layer circuit at $\delta=0.3$.
Panel (b): signal as a function of $\delta$ for the 6-layer circuit at $\eta=9\pi/40$.}
\label{fig:bp-sc-delta-scan}
\end{figure}

In \cref{fig:bp-sc-delta-scan}(a) we show the \acrshort{ole} signal as a function of the scattering strength $\eta$ for different values of the $\delta$ parameter, for the 6-layer circuit, analogously to \cref{fig:main} (e). Here, we highlight the convergence of the results at increasing bond dimension, starting from $\eta=0$, the trivial case where the \acrshort{bp} computation converges to a value extremely close to $1$ at bond dimension $\chi \gtrsim 256$. As expected, as $\eta$ increases, the \acrshort{bp} computation becomes increasingly hard. For $\eta \gtrsim 0.1$ the agreement with the experimental data breaks down even at the high bond dimensions that we consider.
In panel (b) we show the dependence on $\delta$ for a fixed value of the scattering strength, $\eta=9\pi/40$. Analogously to the dependence on $\eta$, \acrshort{bp} computations are more difficult for larger values of $\delta$, although less drastically than for $\eta$.

We argue that these results show that the \acrshort{bp} simulation is not able to get meaningful and converged results for the 6-layer circuit at $\eta \gtrsim 0.1$. 

\begin{figure}[ht]
\includegraphics[width=.9\textwidth]{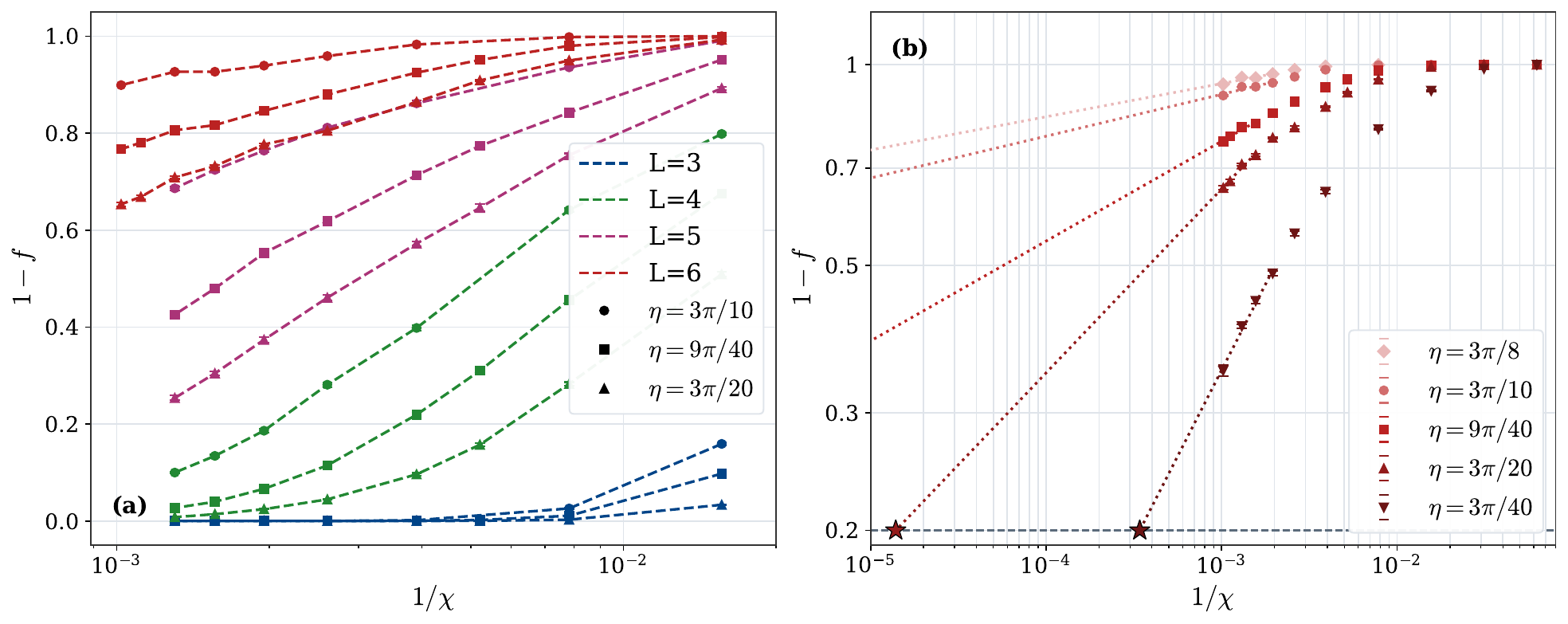}
\caption{
     Panel (a): infidelity  as a function of inverse $\chi$, for different circuits (3, 4, 5 and 6 Floquet layers), $b=0.125$ and $\delta=0.3$. Panel (b): extrapolation of the required bond dimension $\chi$ to obtain fidelity $f=0.8$ for the circuit with $L = 6$.
     }
\label{fig:convergence-results}
\end{figure}

As an indication of the quality of the result, we look at the increase in fidelity, or equivalently the decrease in the infidelity $1-f$, as a function of bond dimension. In \cref{fig:convergence-results} we show that for increasing bond dimension ($1/\chi \to 0$) the infidelity decreases monotonically. In order for the result to be accurate, the fidelity should be sufficiently close to 1, which we define as $f \gtrsim 0.98$. This is achieved for sufficiently high bond dimension for circuits up to $L=4$ and low $\eta$, while for the $L=6$, as well as for high enough $\eta$ for $L=4$, the infidelity is still well above this threshold even at the highest bond dimension considered.

\begin{figure}[ht]
\includegraphics[width=.9\textwidth]{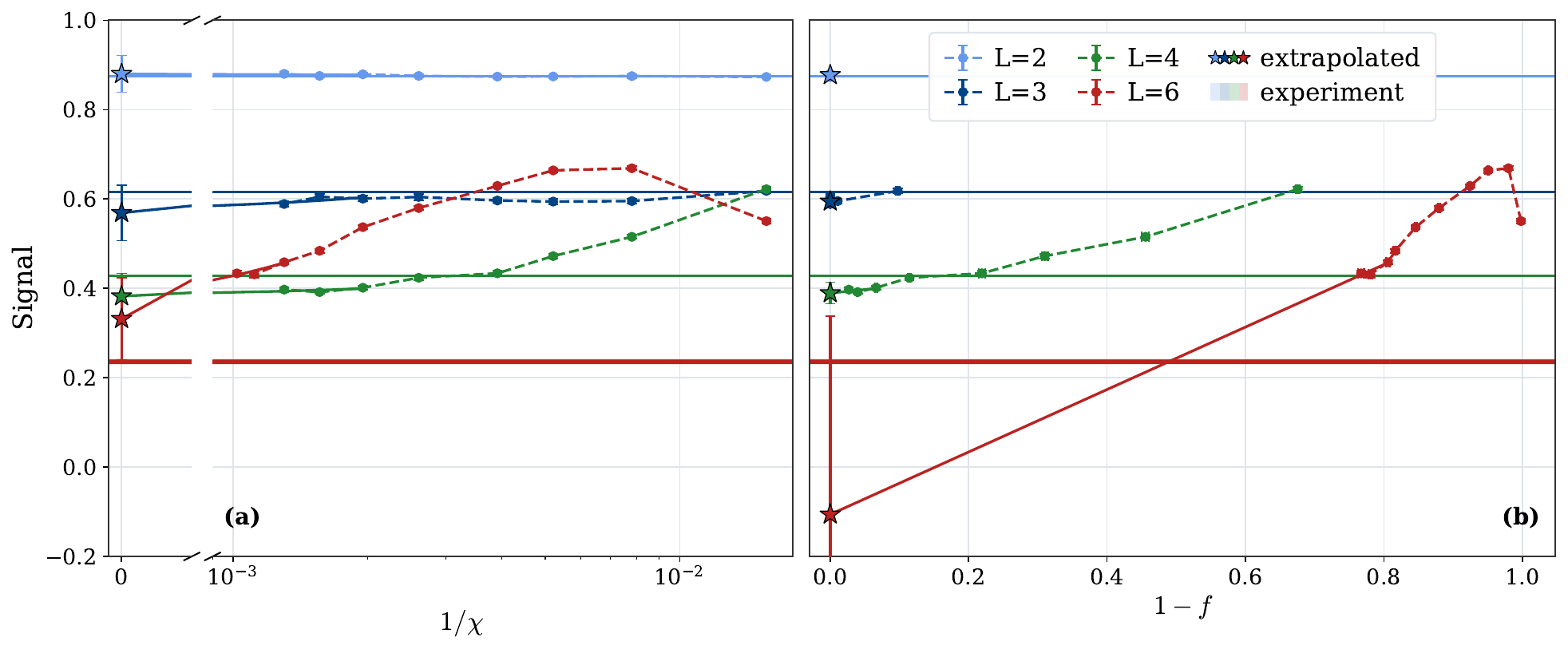}
\caption{
    \acrshort{ole} signal as a function of inverse  bond dimension $1/\chi$ (panel (a)) and infidelity $1-f$ (panel (b)) for different circuits, at fixed $\eta=9\pi/40$. The shaded lines represent the experimental value and their error found with \texttt{ibm\_boston}. The starred value represents the extrapolated value with a linear fit computed with the data points marked with a bold trait.
}
\label{fig:bp-results-fitted}
\end{figure}

With this framework established, we examine the convergence of the results with respect to the inverse bond dimension $1/\chi$ and to the infidelity, shown in \cref{fig:bp-results-fitted}. This convergence behavior provides a basis for comparison against the experimental results, presented both in the figure and in\cref{tab:signal-vs-bd}. For $L=2$ and $L=3$, convergence in bond dimension is achieved at comparatively modest bond dimensions, on the order of 256 or below. The $L=4$ circuit exhibits high fidelity overall, though it falls short of the near-unity values seen for the simpler circuits, remaining approximately at 97\% at $\chi=768$. Nonetheless, in this regime we can reliably extrapolate to infinite bond dimension (discussed below). The $L=6$ circuit, by contrast, exhibits substantially lower fidelity even at $\chi=980$, with the signal continuing to decrease sharply as $\chi$ increases. Consequently, the highest-bond-dimension result cannot be regarded as a faithful proxy for the converged value, and the fidelity is insufficient to support a meaningful extrapolation.

In order to estimate the exact value of the signal in cases where the fidelity is high ($\gtrsim 0.65$) but not sufficiently close to 1, we can reasonably perform an extrapolation. We have compared several extrapolation methods, similarly to~\cite{haghshenas_digital_2026}, and we find that a linear fit of the signal as a function of either inverse bond dimension or infidelity is the most reliable one. The extrapolation is most beneficial for cases such as $L=4$, where the fidelity has not yet fully converged to $1$. The signal is not monotonic in either bond dimension or fidelity, which makes this extrapolation far from trivial. The non-monotonic behavior implies that a linear approximation is only meaningful over a small portion of the convergence curve, namely the region approaching fidelity $1$. Based on findings in the literature, such a region should include points with reasonably high fidelity values ($f \gtrsim 0.8$) \cite{haghshenas_digital_2026}. However, in the case of $L=6$, we consider points at lower fidelity to explore whether loosened extrapolation constraints would still yield a reasonable result.

In both panels of \cref{fig:bp-results-fitted}, the extrapolation is done using only the last three points (bold segments). For both extrapolations, we see that for $L=2, 3$ and $4$ the extrapolated values are compatible (see\cref{tab:signal-vs-bd}). $L=5$ is just below the reasonable fidelity that would be necessary for a good extrapolation, resulting in incompatible values. For $L=6$ the extrapolation clearly fails due to the incorrect fitting window, as results at much higher fidelity would be needed for the $1/\chi$ or $1-f$ curve to saturate to a region where a linear fit would be sensible.

Following this discussion, we argue that the extrapolations shown in \cref{fig:bp-results-fitted} are to be considered a valid approximation of the signal for only up to $L=4$. Our argument for the unreliability of the \acrshort{bp}-derived signal at $L=6$ is twofold; first the low fidelity achieved, and second the high sensitivity of the extrapolated signal to the choice of extrapolation to perform ($\chi\rightarrow \infty$ or $f \to 1$). There's a third, empirical reason to consider the value for $L=6$ unreliable: as discussed in \cref{sec:ole_otoc}, it is expected that the signal should decrease with increasing $L$; however, the signal values obtained by \acrshort{bp} simulations do not exhibit this behaviour. Looking at the values reported in\cref{tab:signal-vs-bd}, we can see how the signal for $L=5$ at $\chi_{\mathrm{max}}=768$ is lower than the value reported at the maximal available $\chi_{\mathrm{max}}=980$ for $L=6$, despite having a higher fidelity. It then follows that the \acrshort{bp} simulations for $L=6$ have not yet reached convergence toward a theoretically reliable result.

\begin{table}[htbp]
\centering
\setlength{\tabcolsep}{5pt}
\small
\begin{tabular}{c c c c c c}
\toprule
$L$ & Experiment & Extrapolation $1/\chi\to0$ & Extrapolation $f\to1$ & OLE$_{\chi_{\max}}$ & $f_{\chi_{\max}}$ \\
\midrule
2 & $0.875 \pm 0.001$ & $0.8800 \pm 0.0410$ & $0.8775 $ & $0.880 \pm 0.004$ & $1.000$ \\
3 & $0.616 \pm 0.002$ & $0.5690 \pm 0.0618$ & $0.5945 \pm 0.0141$ & $0.589 \pm 0.006$ & $1.000$ \\
4 & $0.429 \pm 0.002$ & $0.3824 \pm 0.0507$ & $0.3893 \pm 0.0240$ & $0.397 \pm 0.005$ & $0.974$ \\
5 & -- & $0.2520 \pm 0.0418$ & $0.1864 \pm 0.0642$ & $0.344 \pm 0.004$ & $0.574$ \\
6 & $0.236 \pm 0.005$ & $0.3317 \pm 0.0918$ & $-0.1060 \pm 0.4441$ & $0.434 \pm 0.006$ & $0.233$ \\
\bottomrule
\end{tabular}
\caption{Signal at $\delta=0.3$, $\eta=9\pi/40$ for varying circuit depth $L$: experimental measurement, linear extrapolation to $1/\chi\to0$, linear extrapolation to $f\to1$, and signal with fidelity at the largest bond dimension $\chi_{\max}$ reached for each $L$ ($980$ for $L=6$ and $768$ for the other circuits).}
\label{tab:signal-vs-bd}
\end{table}

The results obtained with \acrshort{bp} are in good agreement with the experimental results for circuits up to $L=4$. As an alternative computation method, we have also tried to perform \acrshort{bp} simulations starting from the central layer of the circuit and contracting outward; this method is discussed in \cref{app:m-o}. In later sections, we will also explore the results of two other standard TN methods (\acrshort{mps} and \acrshort{ttn}). We remark that \acrshort{bp} in the Schr\"odinger evolution is the most reliable tensor network method for simulating circuits considered in this work. 

\subsubsection{Numerical implementation details}
\label{tn-bp-implementation}

We performed the \acrshort{bp} simulations with a modified version of the
open-source Julia package \texttt{TensorNetworkQuantumSimulator.jl}~\cite{TNQS}.
The wavefunction is represented by a tensor-network state on the device graph
and evolved by local gate application with on-the-fly \acrshort{bp}
gauging~\cite{Tindall2023}. We developed a distributed backend in which several
GPUs cooperate on a single \acrshort{bp} simulation. This removes the
single-device memory limit and enabled simulations up to $\chi=980$ on eight
NVIDIA H200 GPUs.

\paragraph{Partitioning.}
For a tensor-network graph $\mathcal{G}=(\mathcal{V},\mathcal{E})$, a map
$\pi:\mathcal{V}\to\{0,\ldots,N_{\mathrm{dev}}-1\}$ assigns each tensor and its
incident messages to a device. We compute the partition with
KaHyPar~\cite{KaHyPar}. Our default objective weights vertices by degree to
balance the memory associated with high-coordination tensors. On nodes with
more available memory, we instead use a uniform-weight min-cut objective to
reduce communication across device boundaries. Each device stores its local
tensors and messages and is responsible for producing the outgoing messages
on cut edges. In particular, the tensors assigned to device $p$ form the set
$\mathcal{V}^{(p)}=\{v:\pi(v)=p\}$. Directed boundary messages are stored with
their source partition, giving every cut edge a unique producer and consumer.

\paragraph{Distributed message updates.}
We divide the graph edges into color classes such that edges of the same color
share no vertices~\cite{Bogle2022}. Messages within one class can then be
updated concurrently without write conflicts. A \acrshort{bp} iteration
processes the color classes sequentially. For each class, every device updates
the messages whose source vertices it owns, after which the newly computed
messages on cut edges are transferred to the neighboring devices. Transfers
use direct NVLink or PCIe peer-to-peer communication when available, with
host-staged copies as a fallback.

Communication is therefore proportional to the number of cut edges rather
than to the total number of tensors or messages. Since each boundary message
has one destination, no global synchronization or collective communication is
required within a color step.

This schedule is a graph-colored block Gauss--Seidel update~\cite{Bertsekas1989,Yang2025}:
it parallelizes independent updates within each color class while making their
results available to subsequent classes in the same sweep. The sequential
feedback also avoids convergence problems that may occur with fully
synchronous updates on loopy graphs~\cite{Elidan2006}.

\paragraph{Gate application.}
Two-qubit gates are applied sequentially. The \acrshort{bp} gauge is refreshed
lazily: a new message-convergence sweep is triggered only when a gate acts on
a vertex modified since the previous sweep. For a gate spanning two devices,
the partner tensor and required environment messages are moved to one device.
The gate is absorbed, the tensor pair is refactorized and truncated by QR/SVD,
and the singular values define the updated message on the gate bond. The
updated tensors and message are then returned to their owners and copied to
any device that tracks the bond. If the temporary factorization exceeds GPU
memory, it is performed on the host.

\subsubsection{Computational cost}
\label{tn-bp-computational-cost}

The leading memory and time costs of the \acrshort{bp} simulations are controlled by the maximal coordination number of the tensor-network graph. For the heavy-hex topology shown in \cref{fig:model}, each tensor has at most three virtual neighbors, with the degree-three vertices located at the corners of the hexagons. Let $z_v \le 3$ denote the number of virtual bonds $\chi$ incident on vertex $v$, and let $d$ be the local physical dimension. The storage for a single tensor scales as $M_v = \mathcal{O}(d\chi^{z_v})$. Because degree-three vertices are never adjacent in the heavy-hex geometry, summing over the network and incorporating the $\mathcal{O}(|\mathcal{E}|\chi^2)$ overhead of directed \acrshort{bp} messages yields a total memory footprint of
\[
    M_{\mathrm{total}} 
    = \mathcal{O}\left( \sum_v d\chi^{z_v} + |\mathcal{E}|\chi^2 \right)
    = \mathcal{O}(N_3 d \chi^3 + N_2 d \chi^2 + |\mathcal{E}|\chi^2) 
    = \mathcal{O}(N_3 \chi^3)
\]
where $N_3$ and $N_2$ are the number of degree-three and degree-two sites, respectively, and the final scaling holds for fixed $d$.

The dominant local compute cost arises during the \acrshort{bp}-gauged Simple Update routine~\cite{Wang2011,Li2012}. Consider a two-qubit gate applied to one bond of a degree-three tensor. Before the QR/SVD truncation, this tensor must be dressed with dense $\chi \times \chi$ \acrshort{bp} environment messages on its two spectator bonds. Since the tensor contains $d\chi^3$ elements, contracting a spectator bond with a message requires $\mathcal{O}(d\chi^4)$ operations. This contraction and the subsequent QR factorizations that scale as $\mathcal{O}(d^2\chi^4)$ dominate the local update. The final reduced-tensor SVD costs \(\mathcal{O}(d^4\chi^3)\) on degree-(3--2) edges and \(\mathcal{O}(d^3\chi^3)\) on degree-(2--2) edges. If $G_3$ is the number of two-qubit gates incident on at least one degree-three vertex and $G$ the total number of gates, the overall gate-application time scales as
\[
    T_{\mathrm{gates}} 
    =\mathcal{O}\left(
    G_3\left(d^2\chi^4+d^4\chi^3\right)
    +
    (G-G_3)d^3\chi^3
    \right)
    = \mathcal{O}(G_3\chi^4)
\]

The environment self-consistency step shares this local exponent at maximally coordinated sites. As reviewed in \cref{app:bp}, updating a directed message $m_{v\to w}$ contracts the local factor at $v$ with all incoming messages except the one from $w$. For a degree-$z_v$ tensor, this operation costs $\mathcal{O}(d\chi^{z_v+1})$. 
A complete \acrshort{bp} sweep updates the $z_v$ outgoing messages of every vertex. Accumulating over the $I_{\mathrm{BP}}$
sweeps executed during the entire simulation gives

\[
    T_{\mathrm{msg}} 
    = \mathcal{O}\left( I_{\mathrm{BP}} \sum_v z_v d\chi^{z_v+1} \right)
    = \mathcal{O}(I_{\mathrm{BP}} d (N_3 \chi^4 + N_2 \chi^3))
    = \mathcal{O}(I_{\mathrm{BP}} N_3 \chi^4)
\]
Summing the gate application and message-passing costs gives the total runtime for a simulation:
\[
    T_{\mathrm{total}} = T_{\mathrm{gates}} + T_{\mathrm{msg}} = \mathcal{O}\big(G_3\chi^4 + I_{\mathrm{BP}} N_3\chi^4\big)
\]
This expression demonstrates that while \acrshort{bp} message passing preserves the leading $\mathcal{O}(\chi^4)$ scaling of the gate updates, the prefactor $I_{\mathrm{BP}} N_3$ can easily dominate the local gate count $G_3$ if the gauge is reconverged too frequently. The lazy \acrshort{bp}-gauging strategy employed here naturally controls this overhead by triggering \acrshort{bp} sweeps only when the current gate sequence reuses a previously perturbed vertex.

\subsubsection{Production performance and computational resource usage}

The multi-GPU backend enabled the highest bond dimensions of this work.
The simulations have been run on two systems, the Leonardo supercomputer nodes with 4 $\times$ NVIDIA A100 (64~GB each), and a high-performance node with 8 $\times$ NVIDIA H200 GPUs (141~GB each). On the Leonardo system, a single-GPU simulation runs out of memory at $\chi \approx 550$, falling back on CPU and showing how essential the multi-GPU implementation is in order to simulate at larger maximum bond dimensions. With all four GPUs, we are able to reach $\chi=768$ in runtimes of the order of one hour per initial state.
The most demanding production simulations were however run on the eight H200 GPUs, reaching $\chi=980$ for the 56-qubit, six-layer circuits, a bond
dimension that exceeds the capacity of any single device by a wide margin, in about one hour and a half per simulation.
\Cref{tab:h200} summarizes representative $\chi=980$ runs on this node, with each
data point being an average over ten random initial states simulations. Aggregated over all instances these campaigns consumed
on the order of $10^{2}$ GPU-hours per averaged data point, and would have been
 infeasible on a single accelerator.

\begin{table}[ht]
    \centering
    \begin{tabular}{c c c c c c}
        \toprule
        $\eta$ & $\delta$ & $t_{sim}$\ (s) & $t_{meas}$\ (s) & fidelity \\
        \midrule
        $3\pi/8$ & $0.0$ & $3556$ & $784$ & $0.975$ \\
        $3\pi/8$ & $0.3$ & $5835$ & $881$ & $0.065$ \\
        $3\pi/10$ & $0.0$ & $3128$ & $744$ & $0.989$ \\
        $3\pi/10$ & $0.3$ & $5717$ & $881$ & $0.101$ \\
        \bottomrule
    \end{tabular}
    \caption{Representative $\chi=980$ multi-GPU production runs on a node of
    eight NVIDIA H200 GPUs, for 56-qubit L=6 circuits. Simulation and
    measurement times are per-run averages over ten random initial state instances; gate error
    and fidelity are the corresponding averages. Such bond dimensions are out of
    reach for any single device.}
    \label{tab:h200}
\end{table}

\subsubsection{Resource estimate for a fidelity-controlled extrapolation.}

A useful figure of merit for the quality of a \acrshort{bp} simulation is the global state fidelity $f$ retained under the successive gate truncations. Following the zero-truncation-error methodology of Ref.~\cite{haghshenas_digital_2026}, we extrapolate the measured observable linearly in $f$ over the largest available bond dimensions. It is therefore instructive to ask what bond dimension would be needed to reach $f\simeq 0.8$ for the circuits studied here, and whether it is within hardware reach. Fitting the measured infidelity $1-f(\chi)$ of the six-layer circuits as a power law in $1/\chi$ over our largest bond dimensions and extrapolating gives the estimates in\cref{tab:fidelity-chi} and is illustrated
in panel (b) of \cref{fig:convergence-results}. For the smallest
case of $\eta=3\pi/40$, where $f\approx0.64$ is already reached at $\chi=980$,
direct convergence to $f=0.8$ requires only $\chi\sim 3\times10^{3}$. For larger values of $\eta$ the fidelity drops sharply and the same extrapolation places the threshold at $\chi\sim10^{6}$--$10^{16}$, which should be understood as lying far beyond any conceivable tensor-network simulation.

\begin{table}[ht]
    \centering
    \begin{tabular}{c c c}
        \toprule
        $\eta$ & $f(980)$ & $\chi_{f=0.8}$ \\
        \midrule
        $3\pi/40$ & $0.64$  & $\sim 3\times10^{3}$ \\
        $3\pi/20$ & $0.35$  & $\sim 7\times10^{4}$ \\
        $9\pi/40$ & $0.23$  & $\sim 8\times10^{6}$ \\
        $3\pi/10$ & $0.10$  & $\sim 10^{13}$ \\
        $3\pi/8$  & $0.065$ & $\sim 10^{16}$ \\
        \bottomrule
    \end{tabular}
    \caption{Bond dimension required to reach a global state fidelity
    $f=0.8$ for the 56-qubit six-layer circuits at $\delta=0.3$, obtained by a
    power-law fit of $1-f(\chi)$ over the inverse of the largest available bond dimensions. Only
    the weakly scattering regime ($\eta\lesssim 3\pi/40$) reaches $f=0.8$ at an
    accessible $\chi$.}
    \label{tab:fidelity-chi}
\end{table}

These bond dimensions must be weighed against the memory ceiling of the
production hardware. On the eight-GPU H200 node the
dominant contribution to the per-device footprint is the set of maximally
coordinated degree-three tensors, each requiring $16\chi^{3}$ bytes in single precision. With the $N_3=12$ degree-three
vertices distributed over eight devices ($\sim 2$ per device) and an additional
factor of $\sim 2$ for the QR/SVD scratch memory and the staged partner and environment
buffers during gate application, the peak per-device memory usage is
$\approx 64\,\chi^{3}$ bytes. Equating this to $141$~GB caps the GPU-resident
regime at $\chi\approx1300$. Reaching $f=0.8$ even for the weakly scrambling case
 therefore already lies above the
single-node memory ceiling.

Beyond the memory and runtime budgets discussed above, the present
implementation faces a more practical limit. The largest bond dimension we can
reach, $\chi=980$, is not set by the hardware but by the dense linear-algebra
backend: \texttt{CUDA.jl} currently dispatches cuSOLVER through its 32-bit
integer API, so the QR/SVD factorizations that dominate gate
application overflow once a single tensor exceeds $\sim 2^{31}$ elements. This
caps the per-tensor dimensions well before the device memory itself is
exhausted. Even if this purely technical ceiling were lifted, the extrapolated
tractable target $\chi\sim4\times10^{3}$ would demand a substantially more
elaborate, multi-node implementation (hybrid CUDA-MPI) whose
engineering cost is considerable.

\subsection{Middle-out evolution}\label{app:m-o}
The structure of Eq.~\eqref{eq:ole_def} suggests a symmetry that could offer real advantages for classically simulating the \acrshort{ole}, as its complex component can be recognized as:
\begin{equation}
    U V_{\delta} U^{\dag}
    \label{eq:m-o}
\end{equation}
where the same unitary $U$ is applied twice to $V_{\delta}$, a perturbation defined via single-qubit generators, once directly and once in mirrored (conjugated) form. In principle, this symmetric structure could let a tensor network simulation apply the unitary gates from the right and the left at the same time, exploiting the mirrored construction directly. To better show the idea, let us rewrite $U$ as $U=\prod_{i} u_i$, with $u_i$ being all the gates defining the scrambling unitary. Let us then rewrite the tensor network at step $i$ as $s_{i+1} = u_i s_i u_{i}^{\dag}$, with $s_0=V_{\delta}$. In this framework, one can argue that each gate application, given a sufficiently small $\delta$, is a simpler task than in the Schr\"odinger evolution, where the analogous operation would be $s_{i+1} = s_i u_{i}$. When $\delta=0$, the operation trivially reduces to the identity regardless of the gate $u_i$; the open question is whether the resulting computational advantage still holds at the larger $\delta$ values considered in the present work.
Building on this idea, we implemented a version of the \acrlong{bp} code that follows the gate-application rule just described. This implementation involves two physical indices per site in the tensor network, one for each direction of evolution. All the necessary ingredients of \acrshort{bp} can still be defined: the two indices of dimension $2$ can be merged into a single index of dimension $4$, which brings the problem back into the standard framework.

This method fails to reproduce results as reliably as the standard Schr\"odinger \acrshort{bp} evolution. Several factors explain why this is the case. First, the tensor network requires twice as much memory to store as its Schr\"odinger counterpart at the same bond dimension, since each site now carries an additional two-dimensional index. Second, although the middle-out computation is verifiably trivial at $\delta=0$, its operator description spans a quadratically larger Hilbert space than the equivalent Schr\"odinger evolution as complexity grows, meaning the same bond dimension captures proportionally less of the relevant information. Taken together, this suggests that although there may be a narrow window of $\delta$ values where the added cost of the operator description is worthwhile, for most values of practical interest this method fails to retrieve the signal correctly. The results of this implementation are shown in \cref{fig:app-messi} in \cref{app:summary}. 

\newpage
\subsection{Matrix product state}\label{app:mps}
As an alternative to a 2D tensor network with loops, we perform classical simulations of the circuits with \acrfull{mps} tensor networks. Instead of constructing a 2D tensor network that directly maps to the heavy-hex topology of the device, we place the qubits on a line according to some ordering. 
The memory and operational cost is then capped at $O(\chi^2)$ for memory occupied by each tensor and $O(\chi^3)$ for gate evolution. 

This cost advantage is however drastically overshadowed by the poor description allowed by the limited connectivity: in order to apply a 2-qubit gate to neighbouring qubits described by two non-neighbouring tensors, the gate effectively becomes a multi-qubit gate at tensor level description, with information needing to pass across all connecting tensors. Because of this, the bond dimension needed by the \acrshort{mps} to encode the same amount of information on the state increases significantly.
Despite this drawback, \acrshort{mps} are routinely used in the literature to simulate circuits on 2D topologies \cite{zhou2020what,mandra2025heuristic} and are worth investigating.

\begin{figure}[t]
    \centering
    \begin{minipage}{0.8\linewidth}
        \centering
        \begin{overpic}[width=\linewidth]{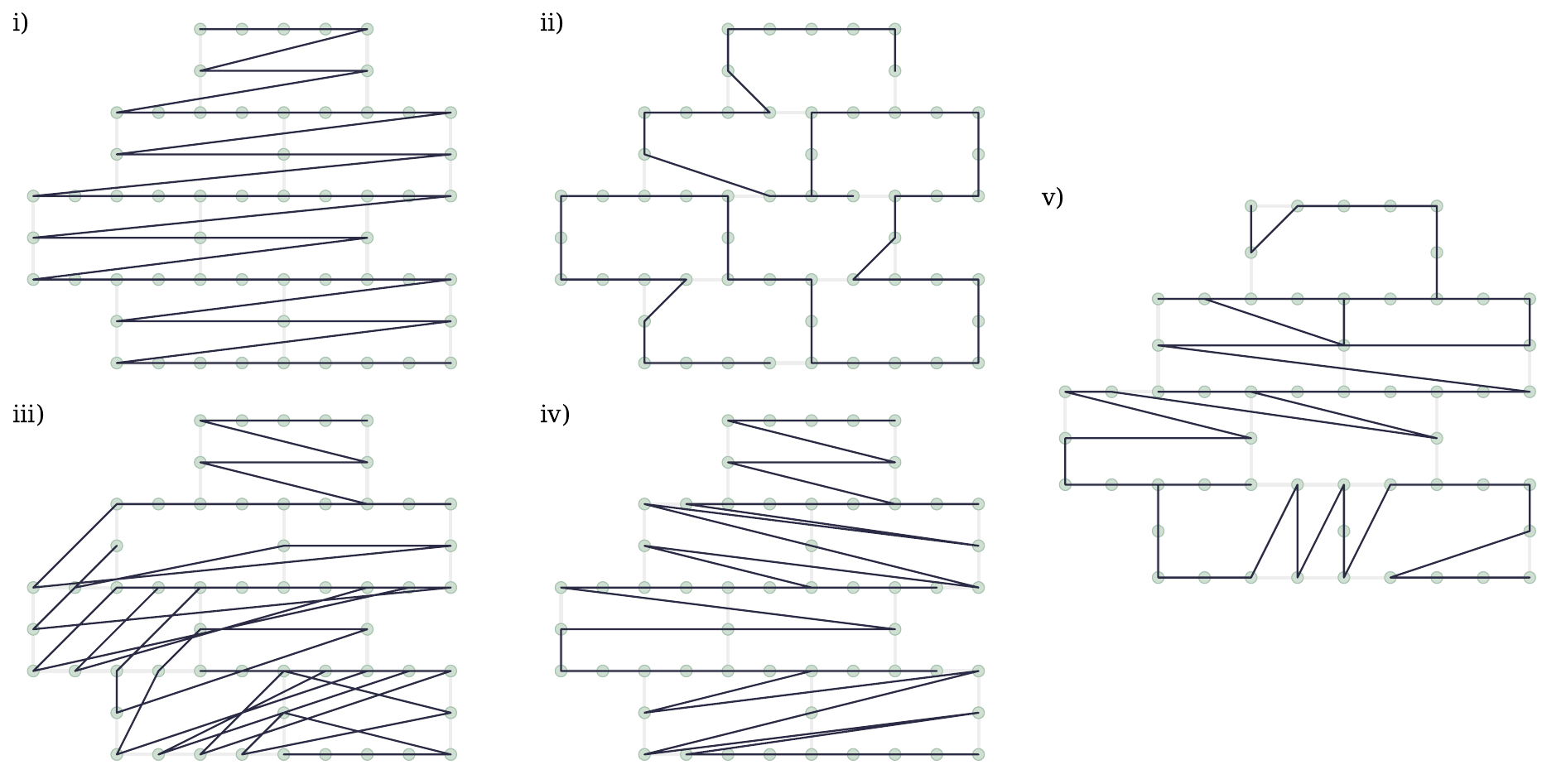}
            \put(2,85){}
        \end{overpic}
    \end{minipage}
    \caption{\acrshort{mps} simulations of the circuit under different qubit orderings. We show the four qubit orderings considered, overlaid on the hardware connectivity layout: (i) the hardware-row ordering, (ii) a snake ordering maximizing short-range interactions, (iii) a reverse Cuthill--McKee ordering, (iv) a spectral (Fiedler-vector) ordering and (v) the annealing optimized ordering.}
    \label{fig:mps}
\end{figure}

\begin{table}[ht]
\centering
        \begin{tabular}{lccc}
            \toprule
            Method & &2q-gate dist.&\\
            & Max & Mean & Median \\
            \midrule
            i) Hardware               & 11 & 2.87 & 1.0 \\
            ii) Snake                   & 32 & 3.21 & 1.0 \\
            iii) Reverse Cuthill--McKee   & 5  & 3.35 & 4.0 \\
            iv) Spectral (Fiedler)       & 6  & 2.92 & 3.0 \\
            v) Annealing optimized       & 20  & 2.60 & 1.0 \\
            \bottomrule
        \end{tabular}
    \caption{Choices of qubit orderings for the \acrshort{mps} simulations, comparing their maximum, mean and median distance in 2-qubit gates. The Cuthill--McKee and spectral orderings minimize the maximum gate distance, while the brute-force annealing optimized ordering minimizes the mean distance. They are expected to be most favorable for \acrshort{mps} simulation.}
\label{tab:mps-orderings}
\end{table}
 
We considered five possible orderings of the qubits, listed in \cref{fig:mps}: i) An \acrshort{mps} that follows the rows of the hardware connectivity layout (used for example in \cite{Tindall2024efficient}); ii) a ``snake" going through the qubits so that the number of short-range interactions is maximized; iii) an ordering computed using a reverse Cuthill-McKee (\acrshort{rcm}) algorithm (to minimize the largest distance between qubits); iv) an ordering using the Fiedler vector of the coupling graph (a technique typically used for \acrshort{dmrg} in quantum chemistry \cite{barcza2011}); v) an annealing optimized (ann-opt) ordering that directly minimizes the distance among the qubits involved in 2-qubit gates.
\Cref{tab:mps-orderings} reports the maximum, mean and median 2-qubit gate distance on the \acrshort{mps} for the different orderings. As suggested by the lower maximum distance, orderings (iii) and (iv) provide the best conditions for the \acrshort{mps} algorithm by maximizing the locality of 2-qubit gates.

As an attempt to improve convergence, in the following discussion we consider both the raw results and a rescaled  version by the signal obtained at $\delta = 0$. Since this value should be trivially 1, any deviation can be used to estimate the error of the \acrshort{mps} simulation and used to rescale the results at larger $\delta$.

In this \acrshort{mps} case, just as in \acrshort{bp}, we  have run simulations across a range of bond dimensions \(\chi\) using at least 20 initial states per circuit setting while running $10$ for the largest bond dimension, \(\chi\) = $4096$.

\begin{figure}[ht]
\includegraphics[width=.8\textwidth]{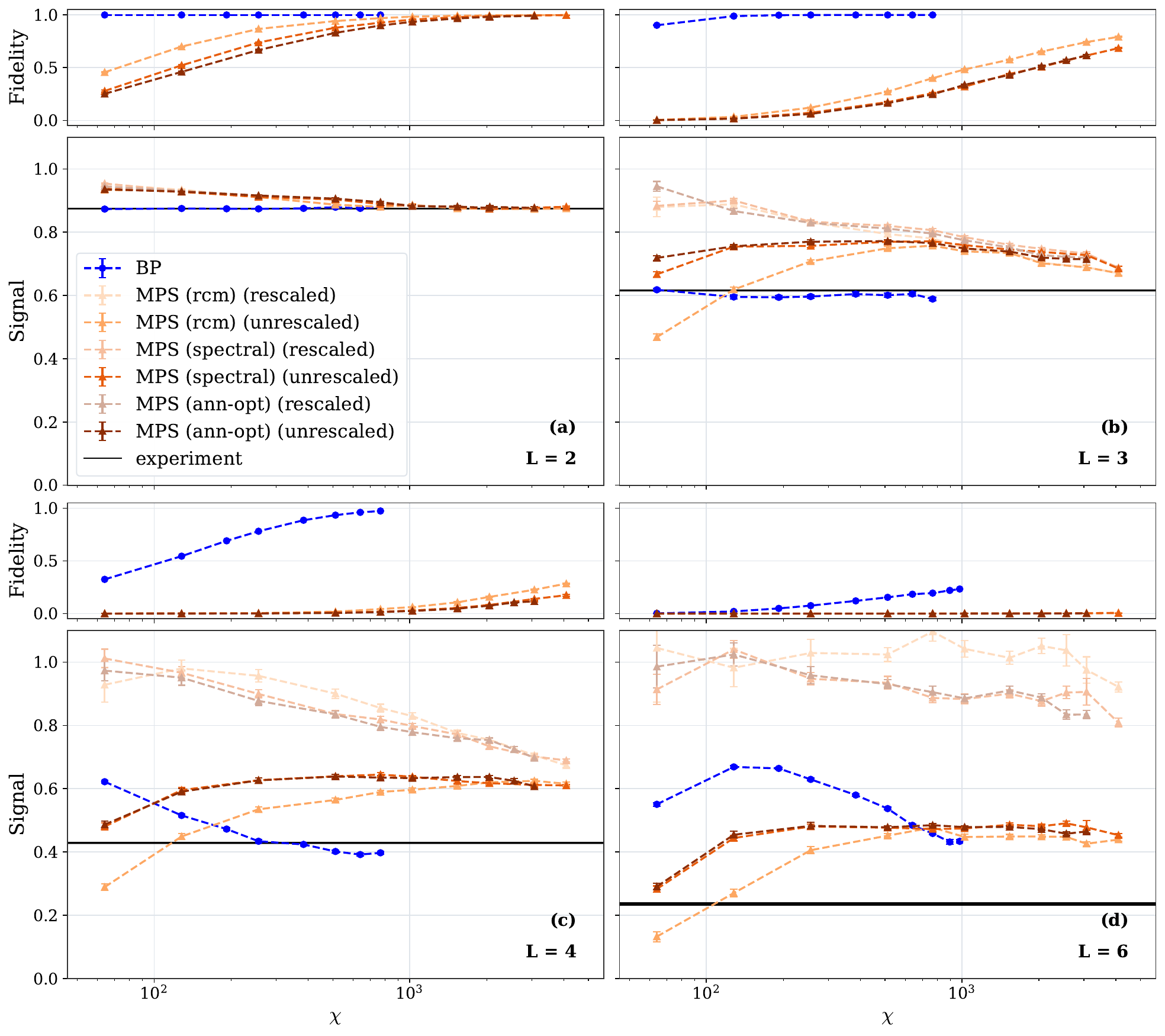}
\caption{
\acrshort{mps} results compared to \acrshort{bp} and the experimental values for the circuits with 2,3,4 and 6 Floquet layers respectively (panels (a) to (d)) and parameters $\eta=9\pi/40$ and $\delta=0.3$. We show the results for spectral, \acrshort{rcm} and ann-opt orderings. The \textit{rescaled} data are rescaled with the corresponding result at $\delta=0$. We show the lack of convergence for L higher than $3$, signaled by the extremely low fidelities.
}
\label{app:mps-verification}
\end{figure}

We test the method over a verifiable regime of lower number of Floquet layers. In \cref{app:mps-verification} we compare the three best orderings of \acrshort{mps} (spectral, \acrshort{rcm}, and ann-opt) with \acrshort{bp} and experimental results as a function of bond dimension. We can derive from this image that \acrshort{mps} has perfectly converged for L=2 (panel a)), whereas for L=3 (panel b)) it is only starting to approach convergence at bond dimension $4096$. For any L higher than 3, both the fidelity and the signal are unconverged and thus unreliable.

\begin{figure}[t]
\includegraphics[width=.98\textwidth]{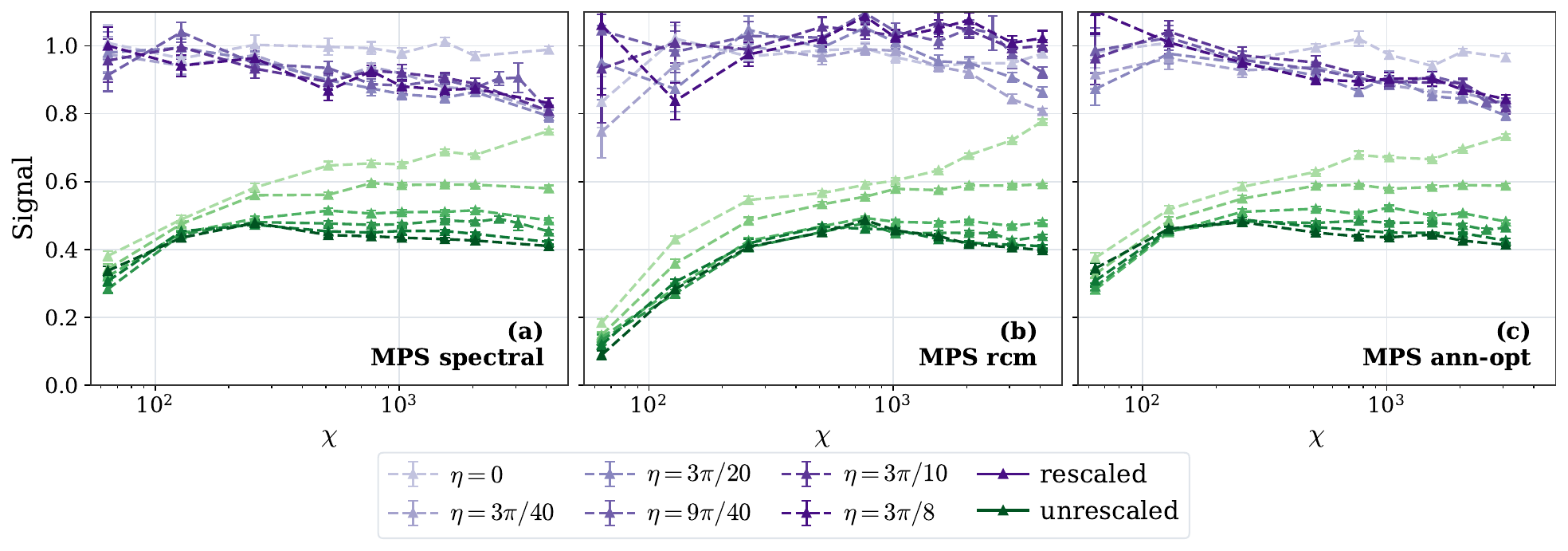}
\caption{
Convergence of the \acrshort{mps} method for the three best orderings, spectral (panel (a)), \acrshort{rcm} (panel (b)), and ann-opt (panel (c)), showing the approximate collapse on a single $\eta$-independent curve. These values are computed for an $L=6$, $\delta=0.3$ circuit. The \textit{rescaled} signal has been divided by the one with equal parameters at $\delta=0$.
}
\label{app:mps-dependence}
\end{figure}

To further this idea, in \cref{app:mps-dependence} we show the result of the \acrshort{mps} simulations for L=6, as a function of the bond dimension, for different values of $\eta$. As the values approximately overlap within their errors at the bond dimensions explored, these results highlight a lack of convergence for almost all $\eta$. At $\eta=0$, where the theoretical value should be trivially $1$, we still see an unconverged result at $\chi=4096$ although the points separate from the other $\eta$ values and approach the expected convergence behavior; we however note that for the same parameters \acrshort{bp} already achieves convergence at bond dimension $64$. 

In panel (d) of \cref{app:mps-verification} we highlight the unconverged results of both \acrshort{mps} and \acrshort{bp} compared to the experimentally obtained value. At L=6, the fidelity remains extremely low across all \acrshort{mps} variants, indicating a lower quality of the result compared to \acrshort{bp}.

These results have been obtained with a GPU code written in Julia on the same large compute node used for the \acrshort{bp} simulations. The time taken to compute the evolution of a single initial state on a single H100 GPU is about 11 hours at bond dimension $4096$ with the most complex circuit settings.

Overall, these results show how \acrshort{mps} fails for simpler circuits than \acrshort{bp}, already at 3 Floquet layers; moreover, at 6 layers it is roughly as expensive as \acrshort{bp} (12 hours vs.\ 16 at respective $\chi_{max}$) while producing less reliable results, which is therefore a preferred tensor-network method to explore this class of circuits.

Note that Heisenberg-picture dynamics, although a candidate method for this computation, would be substantially more computationally heavy.
Other strategies could be considered, involving for example the grouping of qubits into nodes of larger dimensionality, as was done for example in \cite{mandra2025heuristic}. They are left for future work.

\subsection{Tree tensor network}\label{app:ttn}

In addition to the \acrshort{mps} simulations discussed in the previous section, we consider a second, one-dimensional, loopless, tensor network ansatz: the \acrlong{ttn} (\acrshort{ttn})\,\cite{Silvi2019,Felser2026}. A \acrshort{ttn} consists of tensors arranged in a hierarchical binary-tree structure. We exploit this ansatz to simulate the experiment using the QuantumTEA library\,\cite{qtealeaves}. Since the primary limitation of the classical simulations is the available memory\,\cite{Reinic2025}, let us shortly address the memory footprint of the \acrshort{ttn} ansatz. For an \acrshort{mps}, the memory required to store the largest tensor scales as $O(\chi^2)$, whereas for a \acrshort{ttn} it scales as $O(\chi^3)$, since each internal tensor carries three virtual legs rather than two. At a fixed bond dimension $\chi$, a \acrshort{ttn} therefore has a larger memory footprint than an \acrshort{mps}. This additional cost can, however, be compensated by the different connectivity and expressive capabilities of the tree structure. In particular, the hierarchical arrangement provides short paths between distant physical sites and can therefore represent long-range correlations more efficiently\,\cite{Silvi2010}.

We employ the same annealing-based optimization procedure as for the \acrshort{mps} to determine an optimal ordering of the physical qubits\,\cite{cataldi2021,scardicchio2026opt}. More precisely, we minimize
\begin{align*}
    \bar{d}
    =
    \frac{1}{N_{\mathrm{2q}}}
    \sum_{g \in \mathcal{G}_{\mathrm{2q}}}
    d(i_g,j_g),
\end{align*}
where $\mathcal{G}_{\mathrm{2q}}$ denotes the set of all two-qubit gates in the circuit, $N_{\mathrm{2q}} = |\mathcal{G}_{\mathrm{2q}}|$ is the total number of such gates, and $d(i_g,j_g)$ is the chosen distance between the two qubits acted upon by gate $g$. By minimizing $\bar{d}$, we aim to reduce the amount of information that must be propagated through the tensor network during the simulation and thereby mitigate the truncation error introduced by repeated tensor decompositions. Since the number of possible qubit orderings grows factorially with the number of qubits, an exhaustive search rapidly becomes infeasible. Instead, we employ simulated annealing to efficiently search for high-quality orderings. This preprocessing step is computationally inexpensive compared to the subsequent tensor network simulations, typically requiring only a few minutes while significantly improving the quality of the resulting approximations.

For an \acrshort{mps}, the distance between two qubits is uniquely defined by their separation along the one-dimensional chain. For a \acrshort{ttn}, by contrast, different notions of distance can be used. For each two-qubit gate, we consider either the separation between the two qubits along the one-dimensional ordering onto which they are mapped (ann-opt), or the length of the path connecting them within the \acrshort{ttn}, i.e., their graph distance (ann-opt-g). These two distances generally differ because of the hierarchical tree structure.

As a simple baseline, we consider the hardware ordering obtained by following the rows of the heavy-hex hardware layout, see Fig.\,\ref{fig:mps}\,(i). While this ordering preserves some degree of locality in the physical qubit arrangement, it does not take the structure of the quantum circuit into account. The annealing-based optimization of the qubit ordering substantially improves the accuracy of the \acrshort{ttn} simulations compared with the hardware baseline. For certain circuits, the optimized orderings yield results that are converged with bond dimension, whereas convergence is not reached with the hardware ordering. Among the optimization strategies considered in this work, ann-opt-g provides the best-performing qubit ordering for the \acrshort{ttn}.

\begin{figure}[t]
    \centering
    \includegraphics[width=1.0\linewidth]{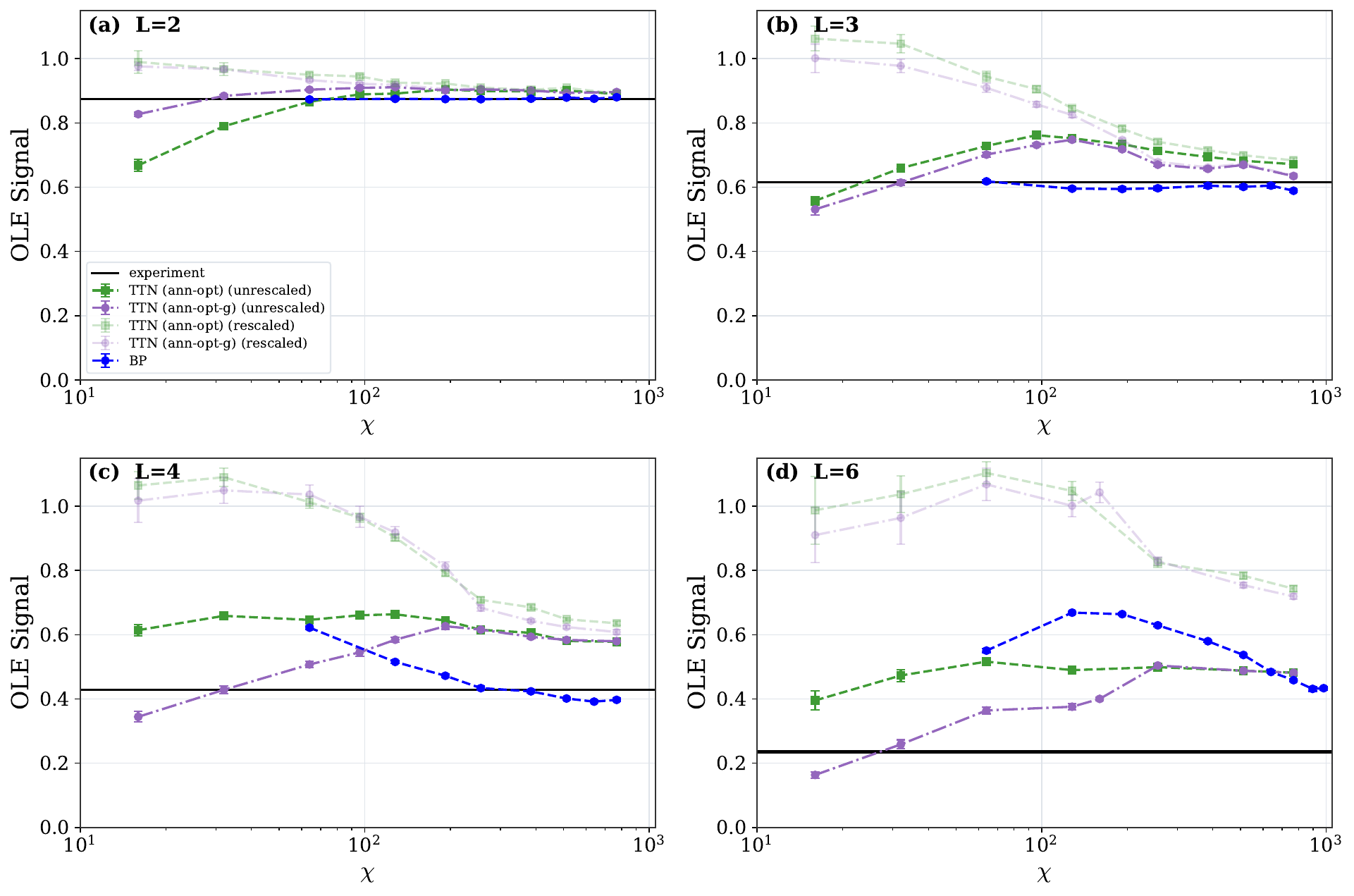}
    
    \caption{ Comparison of annealing-optimized orderings ann-opt and ann-opt-g for the \acrshort{ttn} ansatz with \acrshort{bp} and experimental results for $\eta=9\pi/40$, $\delta=0.3$ and increasing number of layers $L$. The \textit{rescaled} data are rescaled with the corresponding result at $\delta=0$.}
    \label{fig:ttn_numerics2}
\end{figure}

In Fig.\,\ref{fig:ttn_numerics2}, we compare the optimized \acrshort{ttn} simulations with the results obtained using \acrshort{bp} and the experimental data. The optimized \acrshort{ttn} results exhibit the same overall behavior as the \acrshort{mps} results discussed in the previous section. While both the \acrshort{ttn} and \acrshort{bp} simulations agree with the experimental results for $L=2$, the classical simulations increasingly deviate from the experimental data as the number of layers grows. In particular, for $L=4$, the optimized \acrshort{ttn} results already show a sizable deviation from the experimental data, whereas the \acrshort{bp} results remain reasonably close.

This shows that the loss of connectivity inherent in mapping the heavy-hex architecture onto a one-dimensional tensor-network geometry cannot be completely compensated by optimizing the qubit ordering. While the optimized embeddings significantly reduce the approximation error, they do not overcome the fundamental geometric mismatch between the hardware connectivity and the underlying tensor-network ansatz. For this particular benchmark, \acrshort{bp}, which operates directly on a two-dimensional tensor network matching the geometry of the quantum device, therefore provides the most accurate tensor-network description.

The most demanding $L=6$ simulations used NVIDIA H100 GPUs with 80 GB of memory and required approximately 30 minutes of wall time per initial state. We average over 20 initial states per point.

\newpage
\section{Pauli propagation simulation methods}\label{app:pauli_propagation}

Pauli propagation evolves an observable in the Pauli basis by tracking how it spreads under a circuit. Working in the Heisenberg picture, we write
\begin{equation}\label{eq:pp_expansion}
U^\dagger O U = \sum_{P \in \mathcal{P}_n} b_P P,
\qquad b_P = \frac{1}{2^n}\Tr\big(PU^{\dag} O U\big),
\end{equation}
where $\mathcal{P}_n$ denotes the $n$-qubit Pauli strings, with the normalization $\Tr(PP') = 2^n \delta_{P,P'}$. Since in the experimental setting we consider $O$ is a Hermitian Pauli operator with $O^2 = \mathbb{I}$, the coefficients $b_P$ are real and obey $\sum_P b_P^2 = 1$. Thus ${b_P^2}$ defines a probability distribution over Pauli strings (or ``Paulis" for short).

It is often convenient to compile the circuit into Clifford gates and Pauli rotations. A Clifford gate maps a single Pauli string to a single Pauli string, up to a sign, and hence does not increase the number of terms. By contrast, a Pauli rotation $R_Q(\theta) = e^{-i\theta Q/2}$ leaves a Pauli $P$ invariant when $[P,Q]=0$, but branches it into two Paulis when $\{P,Q\}=0$:
\begin{equation}\label{eq:pp_branch}
R_Q^\dagger(\theta) P R_Q(\theta) = \cos(\theta) P + i \sin(\theta) QP,
\qquad \{P,Q\}=0,
\end{equation}
where $i QP$ is again a Pauli. This branching in general drives an exponential growth in the number of Pauli strings with circuit depth.

In conventional deterministic Pauli propagation~\cite{rudolph_pauli_2026, beguvsic2025real, rallSimulationQubitQuantum2019}, truncation schemes are applied to keep the number of Pauli terms with non-zero coefficients in \cref{eq:pp_expansion}, and thus the memory requirements of the algorithm, tractable. Concretely, coefficient truncation discards Pauli strings whose coefficients fall below some threshold~\cite{beguvsic2025real, lerch2024efficient}. Weight truncation instead discards Pauli strings above some maximum Pauli weight, exploiting the fact that low-weight observables and simple input states are often insensitive to very high-weight operators~\cite{rakovszkyDissipationassistedOperatorEvolution2022, angrisani2024classically, aharonov2022polynomial}. Both approaches can be effective when the discarded part of the operator has little subsequent influence on the final observable. However, they also introduce systematic errors which in many regimes can be substantial. 

The operator Loschmidt echo circuit is, by construction, challenging to simulate via Pauli propagation because of operator backflow~\cite{rakovszkyDiffusiveHydrodynamicsOutTimeOrdered2018, vonkeyserlingkOperatorBackflowClassical2021}. That is, the forward part of the circuit can rapidly spread an initially simple observable over many Pauli strings, while the echo structure can subsequently refocus this delocalized operator back into a simpler form. Terms that appear negligible during the outward-spreading stage, and are therefore cut by standard truncation schemes, may therefore become important after the reverse evolution. 

In parallel, the semi-scrambling model is challenging for Pauli propagation because the number of relevant Pauli strings grows rapidly, but the dynamics do not enter a fully scrambling regime in which crude approximations become reliable~\cite{dowling2025bridging, rudolph_pauli_2026, beguvsic2025real}. In a fully scrambling regime, operator weight is pushed into highly non-local strings whose contributions often become negligible or self-averaging allowing weight truncated Pauli propagation or even simple analytic methods to become effective. In the semi-scrambling regime, the operator has spread enough to defeat deterministic truncations, but retains enough structure that the \acrshort{ole} remains sensitive to detailed Pauli coefficients.

The challenge posed by the \acrshort{ole} and semi-scrambling model for Pauli propagation is reflected in the poor estimates of the \acrshort{ole} obtained from standard deterministic Pauli propagation, as shown in Ref.~\cite{advantage_tracker}. For our classical simulation efforts, we therefore develop and apply various approximate variants of Pauli propagation. The variants of Pauli propagation that we will consider differ in how the \acrshort{ole} is approximated, which operator is evolved, how the operator support is truncated and whether sampling is used.

The rest of this appendix is structured as follows. In \cref{sec:OLEexpansion} we outline a number of convenient ways of rewriting the \acrshort{ole} which enable it to be approximately computed more efficiently.
Next, we detail the two heuristic variants of Pauli propagation that we found most tractable for the semi-scrambling \acrshort{ole} circuits: hybrid \acrlong{ppmc} (hybrid \acrshort{ppmc}) in the Heisenberg picture (\cref{sec:PP-MC}) and the quarter-out unitary-expansion method (\cref{app:pp_quarter}). In \cref{sec:TablePPMethods}, we provide a Table summarizing the methods we have explored here. Numerical comparisons across methods and circuit depths are collected at the end of the appendix in \cref{sec:PPResults}.

\subsection{Convenient expansions and approximate expressions for the OLE}\label{sec:OLEexpansion}

While in theory there are many possible ways of computing the \acrshort{ole} via Pauli propagation the method we have found most effective is via its $\delta$ expansion (as introduced in \cref{sec:ole_otoc}). 
Concretely, for small $\delta$ the \acrshort{ole} signal admits the expansion:
\begin{equation}\label{eq:f_delta_expansion}
  S_{\delta} = 1 - \frac{\delta^2}{2}\,  \mathcal{C}_2 + \frac{\delta^4}{4!}\,  \mathcal{C}_4 - \ldots,
  \qquad
   \mathcal{C}_{2m} = \frac{1}{2^n}\,\mathrm{Tr}\Big(\big(\underbrace{[G,[G,\dots[G}_{m},U^{\dag} O U]]]\big)^\dagger
  \big[G,[G,\dots[G,U^{\dag} O U]]\big]\Big),
\end{equation}
where only even powers of $\delta$ contribute and the $2m$-th moment $ \mathcal{C}_{2m}$ involves $m$ nested commutators. 
In the above expansion, the first non-trivial contribution is the standard \acrshort{otoc}
\begin{align}\label{eq:otoc_commu}
    \mathcal{C}_2(t)
    &:= \frac{1}{2^n}
    \Tr\!\left([U^{\dag} O U,G]^\dagger [U^{\dag} O U,G]\right) \, 
\end{align}
a common measure of operator growth and quantum chaos~\cite{abanin_observation_2025, roberts2015diagnosing}. The higher order terms $\mathcal{C}_{2m}$ can be viewed as higher order \acrshort{otoc}s and more fine-grained probes of quantum chaos. Specifically, \refcite{abanin_observation_2025} has extensively studied the $m=1,2$ moments for random circuits.

While for very small values of $\delta$ it is possible to well approximate $S_{\delta}$ by computing just $\mathcal{C}_2$, the standard \acrshort{otoc}, in our experiments we consider the moderate value of $\delta = 0.3$ and therefore high order contributions also become significant and will also need to be estimated. As our methods for higher order contributions are spiritually the same as for \acrshort{otoc}s (but much messier to write down) for pedagogical purposes in the next section we will start by explaining the methods we have explored for estimating \acrshort{otoc}s.

\subsubsection{`Diagonal' and `off-diagonal' OTOC contributions}

\paragraph*{Perturbation $G$ expansion of the OTOC.}\label{sec:diag_offdiag_otoc}
Our methods for estimating \acrshort{otoc}s start with the observation that the contributions to the \acrshort{otoc} in \equref{eq:otoc_commu} can be divided into \textit{diagonal} and \textit{off-diagonal} contributions associated with the terms in $G$. To see this we recall that in our model $G$ takes the form of a sum of local Pauli strings. That is, 
\begin{equation}
    G = \sum_{i\in\mathcal{V}_{\rm P}} G_i ,
\end{equation}
where each $G_i$ is a Pauli string supported on a unique site $i$ and satisfies
\begin{equation}
G_i^2=\mathbb{I}, \qquad [G_i,G_j]=0 \quad (i\neq j).
\end{equation}
Using linearity, the \acrshort{otoc} then decomposes as
\begin{equation}\label{eq:otoc_comm_gi}
    \frac{1}{2^n}
    \Tr\!\left([U^{\dag} O U,G]^\dagger [U^{\dag} O U,G]\right)
    =     \underbrace{\sum_{i}
    \frac{1}{2^n}
    \Tr\!\left([U^{\dag} O U,G_i]^\dagger [U^{\dag} O U,G_i]\right)}_{\text{Diagonal G contribution}} +
    \underbrace{\sum_{i \neq j}
    \frac{1}{2^n}
    \Tr\!\left([U^{\dag} O U,G_i]^\dagger [U^{\dag} O U,G_j]\right)}_{\text{Off-diagonal G contribution}}.
\end{equation}
We further note that because $U^{\dag} O U$ and $G_i$ are Hermitian we have that these terms can be rewritten as: 
\begin{align}
\Tr\!\left([U^{\dag} O U,G_i]^\dagger [U^{\dag} O U,G_j]\right)
&=
2\Tr\!\left((U^{\dag} O U)^2 G_i G_j\right)
-
2\Tr\!\left(U^{\dag} O UG_i U^{\dag} O UG_j\right).
\end{align}

\medskip \paragraph*{Pauli expansion of the OTOC.} Looking ahead to how to estimate these terms using Pauli Propagation based methods, we can alternatively, starting from Eq.~\eqref{eq:otoc_commu}, expand the evolved observable in the $n$-qubit Pauli basis (as in Eq.~\eqref{eq:pp_expansion}) such that \acrshort{otoc} can be decomposed directly as
\begin{equation}\label{eq:c2_diag_offdiag}
\mathcal{C}_2(t)
=
\frac{1}{2^n}
\left(
\underbrace{
\sum_P b_P^2
\Tr([P,G]^\dagger[P,G])
}_{\text{Diagonal Pauli contribution} \, \mathcal{C}_{2, \mathrm{diag}}(t)}
+
\underbrace{
\sum_{P\neq P'}
b_P b_{P'}
\Tr([P,G]^\dagger[P',G])
}_{\text{Off-Diagonal Pauli Contribution}}
\right).
\end{equation}

\medskip \paragraph*{Relationship between the Pauli and Perturbation $G$ expansions.}
For generic dynamics we expect both the diagonal and off diagonal contributions to $\mathcal{C}_2(t)$ to contribute and be different for the Pauli and $G$ expansions. However, numerical evidence from belief propagation methods for the \acrshort{ole} circuits (see \cref{sec:offdiagsmall} just below) suggests that the Pauli off-diagonal contribution is vanishingly small, which also implies that the off-diagonal $G$ contribution is vanishingly small. 
To see this first note that as the $G_i$ terms are themselves Pauli strings it follows from the orthogonality of Pauli strings that:
\begin{align}
\text{if}  \qquad P\neq P'  \qquad \text{then}  \qquad  \Tr([P,G_i]^\dagger[P',G_i]) &=0, \\
\text{and if} \qquad i\neq j \qquad \text{then}  \qquad  \Tr([P,G_i]^\dagger[P,G_j]) &= 0. 
\qquad  \label{eq:offdiag_G_zero}
\end{align}
It therefore follows that if the Pauli off-diagonal contribution vanishes in Eq.~\eqref{eq:c2_diag_offdiag}, i.e.,
\begin{equation}\label{eq:offdiag_P_zero}
    \underbrace{
\sum_{P\neq P'}
b_P b_{P'}
\Tr([P,G]^\dagger[P',G])
}_{\text{Off-Diagonal Pauli Contribution}} = 0,
\end{equation}
then the only remaining contributions are identically the diagonal $G$ contributions
\begin{equation}\label{eq:diag-P-diag-G}
\mathcal{C}_2(t)
=
\frac{1}{2^n}
\underbrace{
\sum_P b_P^2
\Tr([P,G]^\dagger[P,G])
}_{\text{Diagonal Pauli contribution}} = \frac{1}{2^n} 
\underbrace{\sum_{i} \sum_P b_P^2
\Tr([P,G_i]^\dagger[P,G_i])}_{\text{Diagonal G contribution}} := \mathcal{C}_{2, \mathrm{diag}}(t)\, ,
\end{equation}
where for the first equality we have applied \equref{eq:offdiag_P_zero} and for the second we have expanded the generator as $G = \sum_i G_i$ and then canceled cross terms using Eq.~\eqref{eq:offdiag_G_zero}. 

\medskip \paragraph*{Relating the OTOC to the 2-norm distribution of $U^{\dag} O U$.}

Finally, let us consider further rewriting the contributions to the \acrshort{otoc} in Eq.~\eqref{eq:otoc_comm_gi} in terms of the Pauli coefficients of $U^{\dag} O U$. For the diagonal contribution ($i=j$) we start with 
\begin{align}
\Tr\!\left([U^{\dag} O U,G_i]^\dagger [U^{\dag} O U,G_i]\right)
=
2\Tr( (U^{\dag} O U)^2)
-
2\Tr(U^{\dag} O UG_i U^{\dag} O UG_i).
\end{align}
and using the Pauli expansion of $U^{\dag} O U$ obtain 
\begin{align}
\Tr(U^{\dag} O UG_i U^{\dag} O UG_i)
&=
\sum_{P,P'} b_P b_{P'}
\Tr(P G_i P' G_i) \nonumber \\
&=
2^n
\sum_P b_P^2 \,\mathrm{sgn}(P,G_i),
\end{align}
where the sign contribution depends on the commutation relation
\begin{equation}
\mathrm{sgn}(P,G_i)=
\begin{cases}
+1 & [P,G_i]=0, \\
-1 & \{P,G_i\}=0.
\end{cases}
\end{equation}
Therefore, the diagonal contribution can be computed as 
\begin{equation}\label{eq:diag-g}
\mathcal{C}_{2, \mathrm{diag}}(t)
=  2
    \sum_i
    \left(
   \sum_P b_P^2 -\sum_P b_P^2 \mathrm{sgn}(P,G_i)
    \right) =
2
\sum_i
\left(
1-\sum_P b_P^2 \mathrm{sgn}(P,G_i)
\right) \, 
\end{equation}
where for the second equality we use the fact that the initial observable $O$ is a single Pauli string such that $ \frac{1}{2^n} \Tr \left( (U^{\dag} O U )^2 \right) =  \frac{1}{2^n} \Tr \left( O^2 \right) = \sum_P b_P^2 = 1$.
Crucially, this full expression depends only on $|b_P|^2$ (and not on the sign of $b_P$ or cross terms of the form $b_P b_{P'}$) which, as will be made clear below, makes this diagonal contribution easier to compute/approximate. 
Conversely, the off-diagonal contribution contains cross terms of the
form $b_Pb_{P'}$ for distinct Pauli strings $P\neq P'$. It therefore
depends on the relative signs of the Pauli coefficients and cannot, in
general, be estimated from the squared coefficients $b_P^2$ alone.

\subsubsection{Numerical evidence that `off-diagonal' contributions are small}\label{sec:offdiagsmall}

\begin{figure}
    \centering
    \includegraphics[width=0.5\linewidth]{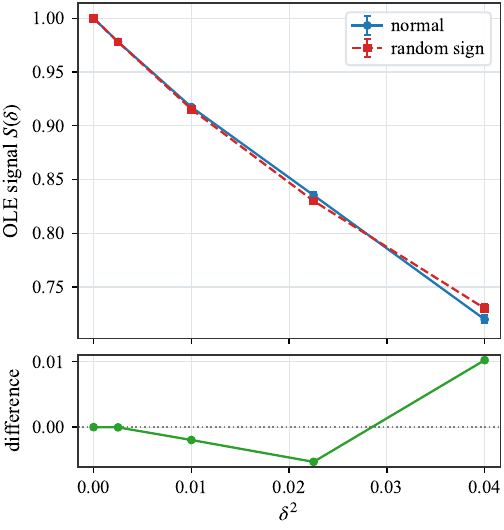}
    \caption{In this picture we numerically demonstrate with TN simulations that the contribution of off diagonal terms to the \acrshort{otoc} is small. We considered a case where \acrshort{bp} simulations converge: $L=3$ and $\chi = 128$, 20 initial bit strings. (top) the \acrshort{ole} signal against the $\delta^2$ for the two settings: in red the signs $s_i$ of the perturbation $G=\sum_{i\in\mathcal{V}_{\rm P}} s_i G_i$ are sampled randomly for each run, in blue they are fixed to one.}
    \label{fig:off_diag_contr}
\end{figure}

Our numerical studies using belief propagation tensor network methods (in regimes where these methods converge) suggest that the off-diagonal contributions to the \acrshort{ole} are insignificant compared to the diagonal contributions. Concretely, in \cref{fig:off_diag_contr} we compare the \acrshort{ole} for { $G = \sum_{i\in\mathcal{V}_{\rm P}} s_i\,G_i$ }, where the signs $s_i$ in front of each perturbation term are chosen randomly in $s_i \in \{-1,+1\}$, and in the other case they are fixed to $1$.
The \acrshort{bp} simulations are repeated for 20 initial bit strings; each run is prepared with a set of random $s_i$. Averaging the signal over perturbations generated by Pauli operators $G$ with random signs effectively converts the perturbation into a Pauli channel. In particular, for a single generator $G_i=X_i$ with a random sign, the resulting channel is a dephasing channel in the $X$ basis.
 This effectively kills off the off-diagonal contributions. Since the simulated signal are similar in the two cases, we concluded that the off-diagonal terms only have a small contribution. This motivates us to develop methods to approximate only the diagonal contribution in subsequent sections. 

\subsubsection{Analytical statistical approach: Full-scrambling assumption}

Assuming the dynamics is fully scrambling ($\eta \neq 0$, $L \rightarrow \infty$), the dynamical observable $U^{\dag} O U$ is composed of random Pauli strings, and the distribution of the squares of the strings' coefficients is uniform. Then the \acrshort{otoc} is given by
\begin{equation}
    C = 2 |\mathscr{V}_P|
\end{equation}
because every Pauli component commutes or anticommutes with the perturbation generator $G$ with probability $\frac{1}{2}$. Each commutator in formula~\eqref{otoc} results in an additional multiplier $2$. Applying the same arguments to the expansion~\eqref{ole-delta-expansion-2}, we obtain the full-scrambling analytical solution $S_{\rm f.s.}(\delta,|\mathscr{V}_P|)$ that depends only on the perturbation strength $\delta$ and the number of qubits involved in perturbation, $|\mathscr{V}_P|$. For the experimental setup in \cref{fig:main}(e), $S_{\rm f.s.}(\delta=0.3,|\mathscr{V}_P|=35) \approx 0.04083$.

\subsubsection{Diagonal approximation of the higher-order OTOCs}
\label{sec:diagonal_higher}

We now extend the diagonal approximation of the previous subsections from the
\acrshort{otoc} $\mathcal{C}_2$ to the higher-order moments $\mathcal{C}_{2m}$. \\

\paragraph*{The diagonal approximation.}
As before, take the perturbation generator $G=\sum_{i}G_i$ to be a sum of single-qubit Pauli
strings on \emph{distinct} sites, so that they mutually commute and square to the identity,
$[G_i,G_j]=0$ and $G_i^2=I$. The higher order \acrshort{otoc} contributions, i.e., the moments of the \acrshort{ole} expansion in \equref{eq:otoc_commu}, can then be written as
\begin{equation}
  \label{eq:moment-def}
  \mathcal{C}_{2m}=\frac{1}{2^n}\Tr\!\Big(\big(\ad_G^{\,m}U^{\dag} O U\big)^{\dagger}\,\ad_G^{\,m}U^{\dag} O U\Big),
\end{equation}
where $\ad_G^{\,m}(X)$ denotes the $m$-fold nested commutator $[G,[G,\dots[G,X]]]$ and
$\mathcal{C}_2$ recovers the standard \acrshort{otoc} of \equref{eq:otoc_commu}.

Expanding $U^{\dag} O U=\sum_P b_P P$ we first note that $\ad_G^{\,m}(\dots)$ is linear and thus $\ad_G^{\,m}U^{\dag} O U  = \ad_G^{\,m} \sum_P b_P P = \sum_P b_P \ad_G^{\,m} P $. Next we recognize that the right hand side of Eq.~\eqref{eq:moment-def} is the Hilbert Schmidt inner product of $\sum_P b_P \ad_G^{\,m} P $ with itself, and so using the linearity of inner products we obtain:
\begin{equation}
    \mathcal{C}_{2m}= \frac{1}{2^n }\sum_{P,P'} b_P b_{P'} \Tr \left( (\ad_G^{\,m}P)^\dagger \, \ad_G^{\,m}P' \right) \, .
\end{equation}
In direct
analogy with \equref{eq:c2_diag_offdiag}, the \emph{diagonal approximation} keeps only the
$P=P'$ terms and discards the cross-correlations between distinct Pauli strings,
\begin{equation}
  \label{eq:diag-def}
  \mathcal{C}_{2m,\mathrm{diag}}=\sum_P b_P^2\,D_{2m}(P),
\end{equation}
where we have defined the norm of the adjoint operator acting on $P$ as
\begin{equation}\label{eq:D2mdef}
    D_{2m}(P) := \frac{1}{2^n} \Tr \left( (\ad_G^{\,m}P )^\dagger \, \ad_G^{\,m}P \right).
\end{equation}
The rest of this subsection computes this term.

\medskip \paragraph*{Preservation Property.} Our derivation below will make use of a useful ``preservation property". Namely, that the set of generators that anticommute with $P$ remain the same at every commutation level. To see this, first note that for two Pauli strings $G_i$ and $P$: $[G_i,P]=0$ if they commute and $[G_i,P]=2G_iP$ if they
anticommute. The commutation sign is multiplicative,
$\mathrm{sgn}(PQ,G_i)=\mathrm{sgn}(P,G_i)\,\mathrm{sgn}(Q,G_i)$, and since the generators
commute with each other, $\mathrm{sgn}(G_j,G_i)=+1$. Hence
$\mathrm{sgn}(G_jP,G_i)=\mathrm{sgn}(P,G_i)$: applying any generator to $P$ leaves the set of
generators that anticommute with it unchanged. This \emph{preservation property} means that
the ``active'' generators, i.e.\ the Paulis that anticommute with $P$, stay fixed at every level
of nesting, and generators that commute with $P$ never contribute. It is useful to define the
\emph{anticommute count}
\begin{equation}
  \label{eq:anti-count}
  n_P:=\#\{\,i:\ \{G_i,P\}=0\,\},
\end{equation}
i.e.\ the number of perturbation terms that anticommute with $P$. Consequently $D_{2m}$
depends on $P$ only through $n_P$. 

\medskip \paragraph*{The first order contribution ($m=1$).}
Let us start by showing for $m=1$ that Eq.~\eqref{eq:diag-def} regains the diagonal approximation to the \acrshort{otoc} that we had earlier (Eq.~\eqref{eq:diag-g}). We first note that the preservation property entails that 
\begin{equation}
  \ad_G^1 P=[G,P]=\sum_{i:\,\text{anti}}2\,G_iP .
\end{equation}
Then from the orthogonality of Pauli strings we have 
\begin{equation}
  \label{eq:D2}
  D_2(P)= \frac{1}{2^n} \Tr \left( (\ad_G^{\,1}P)^\dagger \, \ad_G^{\,1}P \right) =  \frac{1}{2^n}  \sum_{i,j:\,\text{anti}} \, 2^2 \Tr\left(P G_i G_jP \right) = \sum_{i:\,\text{anti}}2^2=4\,n_P,
\end{equation}
so that $\mathcal{C}_{2,\mathrm{diag}}=4\sum_P b_P^2\,n_P$, recovering the diagonal
\acrshort{otoc}.

\medskip \paragraph*{The second order contribution ($m=2$).}
For $m=2$, applying $\ad_G$ twice and using the commutation properties of Pauli operators gives
\begin{equation}
  \ad_G^2 P=4\!\!\sum_{i,j:\,\text{anti}}\!\!G_jG_iP
  =\underbrace{4n_P\, G_i^2 P}_{i=j}\;+\;\underbrace{8\!\!\sum_{\{i,j\}:\,\text{anti}}\!\!G_iG_jP}_{i\neq j},
\end{equation}
where both sums run over the same $n_P$ active generators.
The $i=j$ terms reduce simply to $4 n_P P$ (since $G_i^2=I$), while the $\binom{n_P}{2}$ unordered pairs
$i\neq j$ produce distinct strings $G_iG_jP$. All outputs are mutually orthonormal, so
\begin{equation}
  \label{eq:D4}
  D_4(P )  =  \frac{1}{2^n} \Tr \left( (\ad_G^{\,2}P)^\dagger \, \ad_G^{\,2}P \right)=(4n_P)^2+8^2\binom{n_P}{2}
  =16n_P^2+32n_P(n_P-1)=16\,n_P(3n_P-2).
\end{equation}
Thus we obtain the diagonal approximation to the second order contribution to the \acrshort{ole}: $ \mathcal{C}_{4,\mathrm{diag}}=16 \sum_P b_P^2\,n_P(3n_P-2)$.

\medskip \paragraph*{General order.}
Iterating $\ad_G$ a total of $m$ times, with every index running over the $n_P$
active generators, gives
\begin{equation}
  \label{eq:general-expansion}
  \ad_G^{\,m}P
  =2^m\!\!\sum_{i_1,\dots,i_m:\,\text{anti}}\!\!G_{i_m}\cdots G_{i_1}\,P 
\end{equation}
where the $2^m$ comes from the number of possible orderings of the $m$ commuting $G_i$ contributions. 
Next, because the generators commute and $G_i^2=I$, the product $G_{i_m}\cdots G_{i_1}$ collapses to
$\prod_{i\in S}G_i$, where $S$ is the set of indices occurring an \emph{odd} number of times and
the output string equals $P$ itself only when $S=\varnothing$. Since $G$ is Hermitian, $\ad_G$
is also Hermitian, so the nested commutator norm defined in \equref{eq:D2mdef} can be written as:
\begin{equation}
  \label{eq:kernel-general}
  D_{2m}(P) = \frac{1}{2^n} \Tr \left( P \, \ad_G^{\,2m}P \right)
  =4^{m}\,R_{2m}(n_P),
\end{equation}
where 
\begin{equation}
  \label{eq:R-def}
  R_{2m}(n_P) := \frac{1}{2^n}  \sum_{i_1,\dots,i_m:\,\text{anti}}  \sum_{j_1,\dots,j_m:\,\text{anti}} \!\!  \Tr[ P \, G_{i_m} \dots \,G_{i_1} \, G_{j_1}\, \dots G_{j_m}\,P  ] = \frac{1}{2^n}  \sum_{i_1,\dots,i_{2m}:\,\text{anti}}  \!\!  \Tr[ P \, G_{i_1}\, \dots G_{i_{2m}}\,P  ] \, .
\end{equation}
The factor $4^{m}=(2^{m})^2$ in Eq.~\eqref{eq:kernel-general} collects the $2$ from each of the $2m$ commutators, and only the
sequences whose product returns to $P$ survive the inner product with $P$. These are exactly
the sequences in which every generator is used an even number of times, so
\begin{equation}
  \label{eq:R-count}
  R_{2m}(n_P)=\#\Big\{\text{length-}2m\text{ sequences }(i_1,\dots,i_{2m})\text{ over the }n_P
  \text{ active generators, each used an even number of times}\Big\}.
\end{equation}
For $m=1,2$ this reproduces the exact results $D_2=4n_P$ and $D_4=16\,n_P(3n_P-2)$ found above.

Evaluating $D_{2m}$ thus reduces to the combinatorial problem \equref{eq:R-count}: among the
$n_P^{\,2m}$ ordered length-$2m$ selections of active generators, count the collections in which
every generator that appears is repeated an even number of times,
since only these collapse the product $G_{i_{2m}}\cdots G_{i_1}$ back to $P$ (and so from the orthogonality of Pauli strings, these are the only non-zero contributions to Eq.~\eqref{eq:R-def}). Note that $R_{2m}(n_P)$ is the $2m$-th moment of a sum of $n_P$ symmetric $\pm1$ steps,
i.e.\ of a symmetric Bernoulli random walk, and its closed form is known~\cite{tuenter2006walkingabsolutesum}. In
particular it is a degree-$m$ polynomial in $n_P$, which may be written as the finite sum
\begin{equation}
  \label{eq:R2m-sum}
  R_{2m}(n_P)=\frac{1}{2^{\,n_P}}\sum_{k=0}^{n_P}\binom{n_P}{k}\,(2k-n_P)^{2m}.
\end{equation}
The first few orders read
\begin{equation}
  R_2=n_P,\quad R_4=3n_P^2-2n_P,\quad R_6=15n_P^3-30n_P^2+16n_P,\quad
  R_8=105n_P^4-420n_P^3+588n_P^2-272n_P,\ \dots
\end{equation}
Substituting Eq.~\eqref{eq:R2m-sum} into Eq.~\eqref{eq:kernel-general} and Eq.~\eqref{eq:diag-def} we thus obtain our expression for the diagonal approximation for all the moments in the \acrshort{ole} expansion:
\begin{equation}\label{eq:C2mfinal}
    \mathcal{C}_{2m,\mathrm{diag}}= 4^m \sum_P \sum_{k=0}^{n_P} \, \frac{b_P^2}{2^{\,n_P}}\binom{n_P}{k}\,(2k-n_P)^{2m}.
\end{equation}

Thus we obtain the following procedure for computing the diagonal approximation of any moment $ \mathcal{C}_{2m,\mathrm{diag}}$ in the \acrshort{ole} expansion. 
For a general Pauli sum $U^{\dag} O U$: (i) sweep once over its terms, recording the pair
$(b_P^2,n_P)$ for each Pauli string; (ii) compute $D_{2m}=4^{m}R_{2m}$ via
\equref{eq:R2m-sum} for every count from $0$ to $\max_P n_P$ and the desired orders $m$;
(iii) form each moment as the weighted sum
$\mathcal{C}_{2m,\mathrm{diag}}=\sum_P b_P^2\,D_{2m}(P)$.

\subsection{Hybrid Pauli propagation - Monte Carlo (PP-MC) methods}\label{sec:PP-MC}

\subsubsection{The motivation for 2-norm-based sampling approaches}

In the previous section we derived and justified the diagonal approximations of the standard \acrshort{otoc}. In this section, we show that $\mathcal{C}_{2, \mathrm{diag}}(t)$ in Eq.~\eqref{eq:diag-P-diag-G} can be estimated by Monte Carlo sampling.

To see this, we recall that the squared coefficients  $\{b_P^2\}$ of the Heisenberg-evolved observable ($U^{\dag} O U=\sum_P b_P\,P$) are a valid probability distribution (since $b_P$ are real and obey $\sum_P b_P^2=1$ as $(U^{\dag} O U)^2=\mathbb{I}$). 
Let us start by assuming oracle access to samples from this distribution. That is, we assume we can draw Pauli strings with probabilities
\begin{equation}
q_P := b_P^2 .
\end{equation}
Drawing $P\sim q$ and setting $Y := f(P) = \frac{1}{2^n}\Tr\!\left([P,G]^\dagger[P,G]\right)$, the diagonal contribution to the \acrshort{otoc} (\equref{eq:diag-P-diag-G}) is by construction the mean of $Y$,
\begin{align}\label{eq:otoc_estimator_works}
\mathbb{E}[Y]
&=
\sum_P q_P\, f(P) =
\sum_P b_P^2
\frac{1}{2^n}
\Tr\!\left([P,G]^\dagger[P,G]\right) =
\mathcal{C}_{2, \mathrm{diag}}(t) \, .
\end{align}
The $N$-sample Monte Carlo estimator assembled from independent draws $P_1,\dots,P_N$,
\begin{equation}
\widehat{\mathcal{C}}^{(N)}
=
\frac{1}{N}\sum_{k=1}^N Y_k ,
\end{equation}
is unbiased and its variance is $\Var(\widehat{\mathcal{C}}^{(N)}) = \Var(Y)/N$, with
\begin{equation}
\Var(Y)
=
\sum_P q_P\, f(P)^2
-
\mathcal{C}_{2, \mathrm{diag}}(t)^2 ,
\end{equation}
which is bounded because $f(P)^2 \leq 16 \Tr(G^2)/2^n$ is finite. It follows that
\begin{equation}
\lim_{N\to\infty} \widehat{\mathcal{C}}^{(N)} = \mathcal{C}_{2, \mathrm{diag}}(t) \, .
\end{equation}
That is, the diagonal part of the \acrshort{otoc} is the expectation value of $f(P)$ under the $2$-norm distribution of the Heisenberg-evolved operator $U^{\dag} O U$.

The challenge in practice is to generate samples from this distribution ($q_P := b_P^2$) without explicitly computing all coefficients $b_P$.
In principle, one can obtain the coefficients $b_P$ exactly by deterministic Pauli propagation through the circuit. However, this requires storing and updating a Pauli expansion whose support typically grows exponentially with system size and depth, and is therefore not feasible in general. Consequently, we generally do not have direct access to the exact sampling distribution $\{b_P^2\}$.

This motivates hybrid \acrshort{ppmc} approximations that attempt to generate samples from (or close to) the desired 2-norm distribution by propagating and sampling during the Heisenberg evolution. 
A related approach has been explored in a previous work to simulate the standard \acrshort{otoc}~\cite{abanin_observation_2025}. The difference from our approach is the sampling procedure after exact propagation. The prior approach adopts a random projection procedure for retaining random Pauli strings, whereas our procedure adopts a systematic $2$-norm sampling procedure. We also further apply our method to computing higher order OTOCs and thereby the full OLE.

In the following subsections, we first define the general Monte Carlo procedure via sampling Paulis from an intermediate $2$-norm distribution in \cref{sec:seq_mc}. Then we describe two practical variants: a) single-path MC and b) hybrid \acrshort{ppmc}. Finally, at the end of the propagation of the full circuit, we retain the full set of Pauli strings instead of sampling. We justify this technical detail at the end of the section.

\subsubsection{Description of the PP-MC algorithm}\label{sec:seq_mc}

Here we present a randomized version of Pauli propagation (that we call Pauli propagation Monte Carlo, PP-MC) for approximately producing samples from $q_P := b_P^2$. The algorithm propagates a single Pauli string through the circuit while performing probabilistic sampling at intermediate steps. In particular, we propagate through a gate, or a block of gates, using deterministic Pauli propagation (with light coefficient truncations). Once the number of Pauli strings reaches a memory limit, we randomly sample a single Pauli string and continue Pauli propagation through the next block of gates. 
By changing the block size, we change how much interference from different Pauli paths contributes. At one extreme, \emph{single-path MC} samples one Pauli string after \textit{every elementary gate}; this is computationally cheap, but loses all interference across gate boundaries. More generally, \emph{hybrid \acrshort{ppmc}} groups \textit{several gates} into a block and samples only at the block boundary, thereby interpolating between single-path MC and exact sampling from $q_P = b_P^2$.

To make this precise, let the unitary be decomposed into $n_g$ elementary gates $U = g_{n_g} g_{n_g-1} \cdots g_1 $, grouped into $K$ blocks
\begin{equation}
U = U_K U_{K-1} \cdots U_1,
\end{equation}
each of which is propagated exactly, and let $s \in \frac{1}{\sqrt{2^n}}\{I,X,Y,Z\}^{\otimes n}$ denote normalized Pauli strings. Propagating a Pauli string through the circuit then generates a sum over Pauli trajectories,
\begin{equation}
U^\dagger O U
= 
\sum_{\mathbf{s},s_0}
\Tr(O s_K)
\prod_{\ell=1}^{K}
\Tr(U_\ell^\dagger s_\ell U_\ell s_{\ell-1})
\, s_0, 
\end{equation}
where $\mathbf{s}=(s_K,s_{K-1},\ldots,s_1)$ denotes the intermediate
Pauli strings for a fixed output $s_0$. Equating this expression with
\[
U^\dagger O U
=
\sum_P b_P P
=
\sqrt{2^n}\sum_{s_0}b_{s_0}s_0,
\]
where $s_0=P/\sqrt{2^n}$ and $b_{s_0}\equiv b_P$, gives
\begin{equation}\label{eq:seq_mc_exact}
b_{s_0}
= \frac{1}{\sqrt{2^n}}
\sum_{\mathbf{s}} 
\Tr(O s_K)
\prod_{\ell=1}^{K}
\Tr(U_\ell^\dagger s_\ell U_\ell s_{\ell-1}) .
\end{equation}
Rather than summing over all Pauli paths in Eq.~\eqref{eq:seq_mc_exact} to compute $|b_{s_0}|^2$ (or, equivalently, $|b_{P}|^2$)  exactly, we sample a single Pauli string after each gate (in the case of \textit{Single-path MC}) or after each block (in the case of \textit{Hybrid PP-MC}). 

\medskip \paragraph*{Single-path MC.}

Single-path Monte Carlo is one extreme of the block-size knob, obtained by setting $K = n_g$. Every block then reduces to a single elementary gate, $U_\ell = g_\ell$ with $\ell = 1,\cdots,n_g$, so that no exact propagation is performed anywhere and a Pauli string is resampled after each gate. The path distribution accordingly collapses to
\begin{equation}\label{eq:pp_singlepath}
p_{\text{single}}(\mathbf{s}:s_0)
:=  \frac{1}{2^n}
\Tr(O s_{n_g})^2 \prod_{\ell=1}^{n_g}
\Tr(g_\ell^\dagger s_\ell g_\ell s_{\ell-1})^2 ,
\end{equation}
where $p_{\text{single}}(\mathbf{s}:s_0)$ depends on the set of gates $\{ g_l\}_{l=1}^{n_g}$.
In the infinite-sample limit the estimator converges to the distribution
\begin{equation}
\beta^2_{\text{single}, s_0}  := \sum_{\mathbf{s}} p_{\text{single}}(\mathbf{s}:s_0)
=  \frac{1}{2^n}
\sum_{\mathbf{s}}
\Tr(O s_{n_g})^2
\prod_{\ell=1}^{n_g}
\Tr(g_\ell^\dagger s_\ell g_\ell s_{\ell-1})^2 .
\end{equation}
Because the coherent sum is collapsed after every individual gate, none of the cross terms survive. This makes the scheme extremely cheap, but for circuits in which distinct Pauli paths interfere appreciably it is also systematically biased.

\medskip \paragraph*{Hybrid PP-MC.}

Hybrid \acrshort{ppmc} sits between the two extremes. Exact Pauli propagation with coefficient truncation is carried out within each block and sampling occurs only at the block boundaries, so coherence is preserved across stretches of the circuit rather than destroyed gate by gate. In practice the block size is not fixed in advance but set dynamically: gates are absorbed into the current block until the number of retained Pauli strings reaches the memory limit, a threshold we refer to as the \textit{cache size}. Letting $r_\ell$ denote the number of gates processed after block $\ell$, so that block $\ell$ contains the $r_\ell - r_{\ell - 1}$ elementary gates $U_\ell = g_{r_{\ell}} \cdots g_{r_{\ell-1} + 1}$ and the total gate count is conserved, $n_g = r_K$, the path distribution reads
\begin{equation}
p_{\text{hybrid}}(\mathbf{s}:s_0)
:=  \frac{1}{2^n}
\Tr(O s_K)^2 \prod_{\ell=1}^{K}
\Tr(U_\ell^\dagger s_\ell U_\ell s_{\ell-1})^2 ,
\end{equation}
where $p_{\text{hybrid}}(\mathbf{s}:s_0)$ depends on the set of circuit blocks $\{ U_l\}_{l=1}^{K}$.
In the infinite-sample limit the distribution tends to
\begin{equation}\label{eq:pp_hybrid}
\beta^2_{\text{hybrid}, s_0} := \sum_{\mathbf{s}} p_{\text{hybrid}}(\mathbf{s}:s_0)
=
 \frac{1}{2^n} \sum_{\mathbf{s}}
\Tr(O s_K)^2
\prod_{\ell=1}^{K}
\Tr(U_\ell^\dagger s_\ell U_\ell s_{\ell-1})^2 .
\end{equation}
Written out explicitly in terms of the elementary gates, this (in the limit of no truncations~\footnote{If we apply some weak coefficient truncation during the deterministic propagation through the blocks then there will be an additional bias contribution. In the numerical implementation, the Pauli sum is normalized to $1$ before each 2-norm sampling.}) becomes
\begin{align}
\beta^2_{\text{hybrid}, s_0}
&=  \frac{1}{2^n}
\sum_{s_{r_1},\dots,s_{r_K}}
\left\vert \Tr(O s_{r_K}) \right\vert^2
\prod_{\ell=1}^{K}
\left(
\sum_{s_{r_{\ell-1}+1},\dots,s_{r_\ell-1}}
\prod_{j=r_{\ell-1}+1}^{r_\ell}
\Tr(g_j^\dagger s_j g_j s_{j-1})
\right)^2 ,
\end{align}
where the inner sums run over the Pauli strings internal to each block.
Thus, in contrast to the single-path case, we take a \textit{coherent sum} over the strings internal to each block before squaring the contribution of that block. This retains cross terms, and hence interference, between paths within a block, while eliminating interference between different blocks. In the extreme limit where the block comprises the full circuit, we sum the contributions of all paths before squaring, recovering exact samples from $b_{s_0}$ as in Eq.~\eqref{eq:seq_mc_exact}. In the opposite limit, where each block consists of a single gate, we recover single-path sampling and retain none of the interference between paths encoded in the cross terms. Hybrid PP-MC therefore interpolates between single-path MC and the exact MC calculation enabled by full access to the true propagated-observable distribution. \\

\paragraph*{Using MC-PP to approximate the diagonal contributions to the OLE.} We now show how we can use our samples to approximate $\mathcal{C}_{2, \mathrm{diag}}(t)$ via the replacing $q_P \equiv q_{s_0} = |b_{s_0}|^2$ with $\beta^2_{\text{hybrid}, s_0}$ (or $\beta^2_{\text{single}, s_0}$) in Eq.~\eqref{eq:otoc_estimator_works}. Concretely, each sampled trajectory $(\mathbf{s}, s_0)$ supplies one estimate of the observable,
\begin{equation}\label{eq:est_otoc_deriv}
Y(\mathbf{s}:s_0) =
\Tr([s_0,G]^\dagger [s_0,G]) =  4\,n_{s_0} ,
\end{equation}
where we recall from \eqref{eq:anti-count} that $n_{s_0}$ is the number of generators $G_i$ that anti-commute with the Pauli string corresponding to $s_0$. 
Averaging over $N$ independent trajectories gives the estimator $\widehat{\mathcal{C}}^{(N)} = \frac{1}{N}\sum_{k=1}^{N} Y(\mathbf{s}^{(k)}:s_0^k)$, whose expectation value over the sampled path distribution is then
\begin{equation}\label{eq:exp_otoc_deriv}
\mathbb E[Y]
=
\sum_{s_0}
\beta^2_{\text{hybrid}, s_0}
\Tr([s_0,G]^\dagger [s_0,G]).
\end{equation}
Analogously, higher order \acrshort{otoc} terms can be computed via Eq.~\eqref{eq:C2mfinal}. 

\medskip \paragraph*{The benefits of using the full sum rather than sampling after the final gate.}
Above we supposed that at the end of the circuit we sample a single Pauli string.
Instead, we may choose not to sample the final Pauli string at all. This is more computationally efficient as it throws away less information (and so converges quicker) while also giving an unbiased estimate of the diagonal contributions to the \acrshort{ole}.

The following argument establishes that this ``keep the full sum at the end" approach is also unbiased. In this case, for a given trajectory $\mathbf{s}$ we retain the whole (normalized) Pauli sum $P_{\mathbf{s}} = \sum_{j} a_j\, s^{(j)}_{\mathrm{final}}$ with $\sum_j |a_j|^2 = 1$, and define the estimator $Y_{\mathrm{full}}(\mathbf{s}) = \Tr([P_{\mathbf{s}},G]^\dagger [P_{\mathbf{s}},G])$. Expanding $P_{\mathbf{s}}$ then gives
\begin{align}
Y_{\mathrm{full}}(\mathbf{s})
&=
\sum_{j,k}
a_j^* a_k\,
\Tr\!\left([s^{(j)}_{\mathrm{final}},G]^\dagger [s^{(k)}_{\mathrm{final}},G]\right) \nonumber \\
&=
\sum_j |a_j|^2\, \bar{f}\!\left(s^{(j)}_{\mathrm{final}}\right)
+
\sum_{j \neq k}
a_j^* a_k\,
\bar{K}\!\left(s^{(j)}_{\mathrm{final}}, s^{(k)}_{\mathrm{final}}\right),
\end{align}
where we have separated the diagonal weight $\bar{f}(s_Q) =\Tr([s_Q,G]^\dagger [s_Q,G])$ from the off-diagonal contribution
\begin{equation}
\bar{K}(s_Q,s_Q') =
\Tr\!\left([s_Q,G]^\dagger [s_Q',G]\right).
\end{equation}
Taking the expectation over trajectories, the modified estimator therefore evaluates to
\begin{equation}
\mathbb{E}[Y_{\mathrm{full}}]
=
\sum_{\mathbf{s}} p(\mathbf{s})
\left[
\sum_j |a_j|^2\, f\!\left(s^{(j)}_{\mathrm{final}}\right)
+
\sum_{j \neq k}
a_j^* a_k\,
\bar{K}\!\left(s^{(j)}_{\mathrm{final}}, s^{(k)}_{\mathrm{final}}\right)
\right].
\end{equation}
Comparing the two, the original estimator using the sampled Pauli string $Y_\mathrm{sampled}$ in \equref{eq:exp_otoc_deriv} amounts to sampling a single $s_0$ from the distribution $|a_j|^2$, yielding $\mathbb{E}[Y_\mathrm{sampled}] = \sum_{\mathbf{s}} p(\mathbf{s})\sum_j |a_j|^2 f(s^{(j)}_{\mathrm{final}})$, so that the two estimators differ solely by the off-diagonal cross terms
\begin{equation}
\mathbb{E}[Y_\mathrm{full}] - \mathbb{E}[Y_\mathrm{sampled}] = \sum_{\mathbf{s}} p(\mathbf{s})
\sum_{j \neq k}
a_j^* a_k\,
\bar{K}\!\left(s^{(j)}_{\mathrm{final}}, s^{(k)}_{\mathrm{final}}\right),
\end{equation}
which need not vanish in general.
This term does vanish, however, whenever we only compute the \emph{diagonal} \acrshort{otoc} (c.f.\ \equref{eq:diag-P-diag-G}).
Defining the corresponding diagonal estimators for the single sampled Pauli string and for the full Pauli sum,
\begin{align}
Y_{\mathrm{sampled\, diag}}(\mathbf{s}) &=
\sum_i
\Tr\!\left([s_0,G_i]^\dagger [s_0,G_i]\right),\\
Y_{\mathrm{full\,diag}}(\mathbf{s}) &=
\sum_i
\Tr\!\left([P_{\mathbf{s}},G_i]^\dagger [P_{\mathbf{s}},G_i]\right),
\end{align}
the contribution is now evaluated separately for each perturbation term $G_i$, and the corresponding cross terms vanish identically, $\bar{K}^{(i)}_{\mathrm{diag}}(s_Q,s_Q') = \Tr([s_Q,G_i]^\dagger [s_Q',G_i]) = 0$ for all $Q\neq Q'$ and all $i$. The two estimators therefore agree in expectation, $\mathbb{E}[Y_\mathrm{sampled\,diag}] = \mathbb{E}[Y_\mathrm{full\, diag}]$, and we may safely retain the full Pauli sum rather than sampling at the end of the last block.

\subsection{Unitary `quarter-out' propagation}\label{app:pp_quarter}

We now consider an alternative propagation strategy based on expanding the unitary $U=U_0^\dagger U_\eta $ itself into Pauli strings rather than sequentially propagating Pauli operators.
The label \emph{quarter-out} refers to how the operator propagation is initialized at the seam between the two halves $U_0$ and $U_\eta^\dagger$ of the evolution. At a high level, we propagate $\mathbb{I}$ under the non-physical operation $U_0^\dagger (\dots) U_\eta$, on a gate-by-gate basis, to obtain a Pauli sum approximation of $U_0^\dagger \,  \mathbb{I} \,  U_\eta = U_0^\dagger U_\eta = U$. This Pauli sum approximation for $U$ can then in turn be used to compute $U^\dagger O U$ and thereby other properties of the time evolved observable.  \\

\paragraph*{The core of the algorithm.} This algorithm is tailored to exploiting the fact that simulating $U_0^\dagger U_\eta$ is trivial in the limit that $\eta = 0$ as  $U_0^\dagger U_0 = \mathbb{I}$. Namely, in this limit, $U_0^\dagger \, \mathbb{I} \, U_{\eta=0}$ has an incredibly sparse Pauli sum representation: just $\mathbb{I}$.
Initializing the expansion at the seam is crucial in keeping the Pauli sum sparse. In comparison, propagating the observable inward from the outside would first spread $U_0 O U_0^\dagger$ over exponentially many Pauli strings, only for that growth to be undone again by $U_\eta$ at the very end.

More concretely, following the model convention of \cref{sec:specific-dynamics}, let the circuit given by \equref{UFL}
be described by \(m\) Pauli rotations as
\begin{equation}
    U_\eta
    =
    R_{P_m}\!\left(\theta_m^{(\eta)}\right)
    \cdots
    R_{P_1}\!\left(\theta_1^{(\eta)}\right),
\end{equation}
where \(P_j\) are Pauli strings and \(j=1\) labels the first gate applied. Our
goal is then to propagate \(\mathbb{I}\) according to
\begin{align}
    U_0^\dagger \mathbb{I} U_\eta
    &=
    R_{P_1}^\dagger\!\left(\theta_1^{(0)}\right)
    \cdots
    R_{P_m}^\dagger\!\left(\theta_m^{(0)}\right)
    \,\mathbb{I}\,
    R_{P_m}\!\left(\theta_m^{(\eta)}\right)
    \cdots
    R_{P_1}\!\left(\theta_1^{(\eta)}\right)
    \nonumber\\
    &=
    R_{P_1}\!\left(-\theta_1^{(0)}\right)
    \cdots
    R_{P_m}\!\left(-\theta_m^{(0)}\right)
    \,\mathbb{I}\,
    R_{P_m}\!\left(\theta_m^{(\eta)}\right)
    \cdots
    R_{P_1}\!\left(\theta_1^{(\eta)}\right),
\end{align}
where the second equality uses
\(R_P^\dagger(\theta)=R_P(-\theta)\).
Thus, the two gates meeting at the seam are generated by the same Pauli string
$P_m$ and merge into a single rotation by the difference of their angles,
\begin{equation}\label{eq:seam_merge}
    R_{P_m}(-\theta^{(0)}_{m}) \mathbb{I}
    R_{P_m}(\theta^{(\eta)}_{m})
    =
    R_{P_m}(-\theta^{(0)}_{m}+\theta^{(\eta)}_{m})
    =
    \cos\!\left(\frac{\Delta\theta_m}{2}\right)\mathbb{I}
    -i\sin\!\left(\frac{\Delta\theta_m}{2}\right)P_m,
\end{equation}
where $\Delta\theta_m:=\theta^{(\eta)}_{m}-\theta^{(0)}_{m}$. In the absence
of scattering, the two halves are identical,
$\Delta\theta_m = 0$, so this merged gate is simply
$\mathbb{I}$. Similarly, applying subsequent gates we have $\Delta\theta_j = 0$ for all $j$ and so the evolved operator remains $\mathbb{I}$ as expected.

More generally, for weak scattering the angle differences
$\Delta\theta_j$ are small, and the amplitudes of the non-identity branches are
generally suppressed as
$\sin(\Delta\theta_j/2) \in \mathcal{O}(\Delta\theta_j)$. Thus small angle truncation schemes~\cite{lerch2024efficient} can be applied to maintain a sparser (closer to identity) Pauli sum. More explicitly, we define the sparse approximation of $U$ as 
\begin{equation}
    \tilde{U}_{\epsilon} :=\sum_{a: \, \abs{c_a}>\epsilon} c_{a} P_{a}, \quad c_a = \frac{1}{2^n}\Tr[U P_a],
\end{equation}
where $\epsilon$ is the coefficient truncation threshold for determining which Paulis to retain. If the quarter-out method is applied exactly without truncations then $\tilde{U} = U$ but in general truncations will be essential to make the algorithm tractable and hence  $\tilde{U}$ is only a sparse approximation of $U$. 

\medskip \paragraph*{Computing OTOCs exactly in the `quarter-out' framework.} In the rest of this subsection, we will simply abbreviate the dependence on $\epsilon$, and refer to the truncated unitary expansion as $\tilde{U}$.
The next step is to compute $\tilde{U}^\dagger O \tilde{U}$ which can be expanded as
\begin{equation}\label{eq:uou_expansion}
    \tilde{U}^\dagger O \tilde{U}
    =
    \sum_{a,b}
    c_{a}^\ast c_{b}\,
    P_{a} O P_{b}.
\end{equation}
For each pair \((a,b)\), the product \(P_a O P_b\) is proportional to a unique Pauli string, and we keep track of the relative phase by defining
\begin{equation}
    P_a O P_b := \omega_{ab}(O)\, P_{ab},
\end{equation}
where \(P_{ab}\) denotes the canonical Pauli string obtained from \(P_a O P_b\) after removing its overall phase, and \(\omega_{ab}(O) \in \{\pm 1, \pm i\}\) is the corresponding phase factor. Following \equref{eq:pp_expansion} by writing $\tilde{U}^{\dag} O \tilde{U}=\sum_{P} b_P\,P$, and comparing with Eq.~\eqref{eq:uou_expansion}, we see that the coefficient \(b_P\) is given by
\begin{equation}\label{eq:bP_omega}
    b_P = \sum_{a,b} c_a^* c_b \, \omega_{ab}(O)\, \delta(P_{ab}, P),
\end{equation}
where the delta indicator function enforces grouping of all contributions that yield the same Pauli string \(P\). 

Next, we recall from Eq.~\eqref{eq:diag-g}, that the \acrshort{otoc} diagonal contribution can be written as
\begin{align}\label{eq:diagC_quarter_b}
    \mathcal{C}_{2, \mathrm{diag}}(\tilde{U})
    &= 2
    \sum_i
    \left(
   \mathcal{N}(\tilde{U}^\dagger O \tilde{U})-\sum_P b_P^2 \mathrm{sgn}(P,G_i) \, 
    \right) , 
\end{align}
where 
\begin{equation}\label{eq:norm_def}
\mathcal{N}(\tilde{U}^{\dag} O \tilde{U}) := \frac{1}{2^n}\Tr\!\left( (\tilde{U}^\dagger O \tilde{U}) \tilde{U}^\dagger O \tilde{U} \right) = \sum_P b_P^2.
\end{equation}
is the norm of the propagated observable. In \equref{eq:diag-g}, we were computing the diagonal contribution to the \acrshort{otoc} for a genuine unitary $U$ and $O$ is a Pauli string and hence, by unitary invariance of the Hilbert--Schmidt norm, we had $\mathcal{N}(U^{\dag} O U) = \frac{1}{2^n}\Tr(O^\dagger O) = 1$. Here, however, $\tilde{U}$ is in general a Pauli sum approximation of the true unitary $U$ and thus in general $\mathcal{N}(\tilde{U}^\dagger O \tilde{U}) \neq \frac{1}{2^n}\Tr(O^\dagger O)$.
Using the structure of \(b_P\) given by \equref{eq:bP_omega}, we further expand
\begin{equation}\label{eq:bP_quarter}
    b_P^2
    =
    \sum_{a,b}\sum_{m,n}
    c_a^* c_b\, c_m c_n^*\,
    \omega_{ab}(O)\,\omega_{mn}^*(O)\,
    \delta(P_{ab},P)\,\delta(P_{mn},P).
\end{equation}
Substituting this into \(\mathcal{C}_{2, \mathrm{diag}}\) we have
\begin{align}\label{eq:diagC_quarter}
    \mathcal{C}_{2, \mathrm{diag}}
    &=
    2
    \sum_i
    \left(
    \mathcal{N}(\tilde{U}^\dagger O \tilde{U})-\sum_P \left( \sum_{a,b}\sum_{m,n}
    c_a^* c_b\, c_m c_n^*\,
    \omega_{ab}(O)\,\omega_{mn}^*(O)\,
    \delta(P_{ab},P)\,\delta(P_{mn},P)\right) \mathrm{sgn}(P,G_i)
    \right), \nonumber \\
    &=2\sum_i
    \left[
    \mathcal{N}(\tilde{U}^\dagger O \tilde{U})
    -
    \sum_{a,b,m,n}
    c_a^* c_b\, c_m c_n^*\,
    \omega_{ab}(O)\,\omega_{mn}^*(O)\,
    \delta(P_{ab},P_{mn})\,
    \mathrm{sgn}(P_{ab},G_i)
    \right],
\end{align}
where in the second line the sum over \(P\) collapses the two delta indicator functions into the single constraint $\delta(P_{ab},P_{mn})$.
This is a quartic sum in the number of terms of the unitary expansion, and therefore does not scale favorably.

\medskip \paragraph*{Approximation of the exact evolution.}
Instead of computing the \acrshort{otoc} exactly as in \equref{eq:diagC_quarter}, we can do a partial sum and only retain certain terms. We do this below by approximating $b_P^2$ as the following:
\begin{align}\label{eq:bP_approx}
    b_P^2 
    & \approx \sum_{a,b}\abs{c_a}^2 \abs{c_b}^2 \delta(P_{ab}, P),
\end{align}
where we have only retained contributions from $m=a$ and $n=b$ and used the fact that 
\begin{equation}
\omega_{ab}(O) \omega_{ba}(O) = 1,
\end{equation}
because $(P_{ab})^\dagger = P_{ba}$. Therefore, the approximated quarter \acrshort{otoc} from \equref{eq:diagC_quarter_b} admits an approximation
\begin{align}\label{eq:C2_pair_diag}
\mathcal{C}_{2,\mathrm{diag}}^{\mathrm{approx}}&:= 2
    \sum_i
    \left(
    \mathcal{N}(\tilde{U}^\dagger O \tilde{U})- \sum_{a,b}\abs{c_a}^2 \abs{c_b}^2  \mathrm{sgn}(P_{ab},G_i)
    \right).
\end{align}
In writing the approximate formula, we have carried out the sum over $P$ against the indicator, so that $O$ never needs to be constructed explicitly. Because the cross terms with $(m,n)\neq(a,b)$ have been dropped, each pair $(a,b)$ now contributes independently to $b_P^2$, which reduces the cost from quartic to quadratic in the number of Paulis in the expansion.

\medskip \paragraph*{Approximate normalization for quarter out approach.}
We recall that because truncating the unitary expansion removes unitarity, $\mathcal{N}(\tilde{U}^\dagger O \tilde{U}) = \sum_P b_P^2$ by \equref{eq:norm_def} no longer equals $\frac{1}{2^n}\Tr(O^\dagger O)$, and the norm appearing in \equref{eq:diagC_quarter_b} must be evaluated directly. Using \equref{eq:bP_omega}, we may apply to it exactly the diagonal approximation already made for the estimator in \equref{eq:bP_approx}. Keeping only the diagonal contributions to $b_P^2$ and summing over $P$ to collapse the grouping indicator gives
\begin{align}\label{eq:unitary_norm_approx}
\mathcal{N}_{\mathrm{approx}}
&\approx \sum_P \sum_{a,b} \abs{c_a}^2 \abs{c_b}^2\, \delta(P_{ab}, P) = \sum_{a,b} \abs{c_a}^2 \abs{c_b}^2 = \Big( \sum_a \abs{c_a}^2 \Big)^{2} .
\end{align}
This requires no separate expansion, is evaluated at the same quadratic cost as the estimator, and is approximated consistently with it. The truncation discards interference terms, reducing $\sum_a \abs{c_a}^2$ below one, which amounts to a rescaling of the estimator. 

Given how the truncations reduce the retained unitary norm, directly computing the bare OTOC in \equref{eq:C2_pair_diag} would yield an artificially small value whose scale is largely set by this residual norm. We therefore also consider rescaling the result by the corresponding unitary normalization factor. This procedure is analogous to the rescaling used both experimentally to mitigate noise and in the MPS simulations described in \secref{app:mps}.

The rescaled \acrshort{otoc} is defined by replacing $\tilde{U}^\dagger O \tilde{U}$ with
$\tilde{U}^\dagger O \tilde{U}/\sqrt{\mathcal{N}(\tilde{U}^\dagger O \tilde{U})}$, i.e.\ dividing
\equref{eq:C2_pair_diag} by the norm \equref{eq:norm_def}, to give:
\begin{align}\label{eq:quarter_unitary_C2_diag_app}
    \tilde{\mathcal{C}}_{2,\mathrm{diag}}^{\,\mathrm{approx}}
    &=
    \frac{2 \sum_{a,b} \abs{c_a}^2 \abs{c_b}^2
        \sum_i \big( 1 - \mathrm{sgn}(P_{ab}, G_i) \big)}
        {\mathcal{N}_{\mathrm{approx}}}  \nonumber\\
        & = \frac{ \sum_{a,b} \abs{c_a}^2 \abs{c_b}^2
        (4 n_P)}
        {\mathcal{N}_{\mathrm{approx}}},
\end{align}
where in the first line we used
$\mathcal{N}_{\mathrm{approx}} = \sum_{a,b} \abs{c_a}^2 \abs{c_b}^2$ to
pull the norm inside the pair sum.
To arrive at the second line, the perturbation sum reduces to evaluating the counting of non-commuting Paulis on the support of perturbation $G$ defined in \equref{eq:anti-count}.
The numerator contribution agrees with \equref{eq:D2}, and the denominator renormalizes the \acrshort{otoc}.

\medskip \paragraph*{Higher order OTOC approximations.} As just discussed, an exact computation of the second order \acrshort{otoc} calculation in the quarter-out framework already scales unfavorably with the number of terms in the unitary Pauli expansion, therefore an approximation formula such as \equref{eq:C2_pair_diag} is necessary. The situation is even worse for the higher order contributions to the \acrshort{ole}, hence we here briefly describe how analogous approximations can be used to compute the higher contributions. 

Recall from \equref{eq:diag-def} that to compute the $2m$-th order \acrshort{otoc} we have
\begin{equation}
  \mathcal{C}_{2m,\mathrm{diag}}=\sum_P b_P^2\, \left(\frac{1}{2^n} \Tr \left( (\ad_G^{\,m}P )^\dagger \, \ad_G^{\,m}P \right) \right),
\end{equation}
where we have previously denoted $\ad_G^{\,m}X :=\underbrace{[G,[G,\dots[G}_{m},X]]]$ and defined $D_{2m}(P):=\frac{1}{2^n} \Tr \left( (\ad_G^{\,m}P)^\dagger \, \ad_G^{\,m}P \right)$ for the computation of the trace as in \equref{eq:kernel-general}.
Rather than computing $b_P^2$ exactly, we use the approximation of the coefficients given by \equref{eq:bP_approx} and the approximate normalization in \equref{eq:unitary_norm_approx} to arrive at the $2m$-th order computation using coefficients in the unitary expansion:
\begin{equation}
\tilde{\mathcal{C}}_{2m,\mathrm{diag}}^{\mathrm{approx}}:= 
\frac{ \sum_{a,b} \abs{c_a}^2 \abs{c_b}^2
        D_{2m}(P_{ab})}{\mathcal{N}_{\mathrm{approx}}},
\end{equation}
where the summation only sums over $\abs{c_a}^2$ and never the interference among them. Moreover, the computation of $D_{2m}(P_{ab})$ only requires enumerating through the unitary expansion quadratically $P_{ab}$ and counting the number of anti-commutations. 

\subsection{Summary of Pauli propagation variants explored}\label{sec:TablePPMethods}

\Cref{tab:pp_variants} collects the Pauli-propagation variants we examined, organized along two largely independent axes: how the operator expansion is approximated, and which operator is evolved through the circuit. In addition to previously discussed methods, we have systematically explored other variants that led to less promising results. We do not rule out future improvement built on these methods (or alternatives). For completeness, we describe them below. 

\medskip \paragraph{Schr\"odinger picture evolution.} In this Appendix, we have focused on evolving the observable in the Heisenberg picture $U^\dagger O U$, and compute an \acrshort{otoc} with respect to the static perturbation operator $G$. Similarly, we can perform most of the methods (including deterministic Pauli propagation, single-path MC, and hybrid MC) in the Schr\"odinger picture by evolving the perturbation operator $U G U^\dagger$. 

 \medskip \textit{Schr\"odinger Pauli propagation.} Since $G=\sum_i X_i$ is a sum of one-body operators, it may be tempting to think that one may first evolve the individual $X_i$ to a longer time as it has a smaller lightcone than $O$, which has a support of $12$ Paulis (\cref{sec:graph}). However, the relevant circuit has a depth large enough that the evolved individual operators interfere and it is computationally inefficient to merge their evolved Paulis a posteriori. 

 \medskip \textit{Schr\"odinger Monte Carlo.} An analogous \acrshort{otoc} expression to \equref{eq:otoc_commu} is identically
 \begin{equation}
     \mathcal{C}_2(t) = \frac{1}{2^n}
\Tr([O,G(\bar t)]^\dagger[O,G(\bar t)]),
 \end{equation}
 by the cyclic property of the trace and we write $G(\bar t) = U G U^\dagger$ where the $\bar t$ highlights that we are effectively evolving $G$ under $U^\dagger$ (instead of $U$ in the case of $O$).
 The Monte Carlo approach in the Schr\"odinger picture has the advantage that it is exact using the diagonal formula
 \begin{equation}\label{eq:C2_schro}
\mathcal{C}_2(t)
=
\frac{1}{2^n}
\sum_Q |c_Q|^2
\Tr([O,Q]^\dagger[O,Q]),
\end{equation}
because the observable $O$ is an individual Pauli operator. Therefore, the infinite-cache limit of hybrid PP leads to an exact \acrshort{otoc}. Unfortunately, we have empirically found that the hybrid PP in the Schr\"odinger picture results in an \acrshort{otoc} estimation with a larger variance, therefore requiring more samples for convergence. Thus, we have focused on presenting results in the Heisenberg picture. 

\medskip \paragraph{Quarter-out operator evolution.} Instead of fully evolving in the Heisenberg or Schr\"odinger picture, the quarter-out operator evolution evolves both $O$ and $G$ for a quarter of the circuit each. Following the same notation $U=U^\dagger_0 U_\eta$,
\begin{equation}
O(t/2)= U_0 O U_0^\dagger, \qquad G_S(t/2)=U_\eta G U_\eta^\dagger.
\end{equation}
The \acrshort{otoc} from \equref{eq:otoc_commu} in this picture reads
\begin{equation}\label{eq:c2_quarter_opearator}
\mathcal{C}_2(t) = \frac{1}{2^n}
\Tr([O(t/2),G_S(t/2)]^\dagger[O(t/2),G_S(t/2)]).
\end{equation}
Numerically, we have tried to evolve both using deterministic Pauli propagation. For relatively shallower circuits, this is feasible; however, computing the \acrshort{otoc} is still prohibitively expensive and furthermore it is difficult to do the diagonal approximations as described in \equref{eq:diag-def} because both parts of the commutator are large sums of Pauli operators. 

\begin{table}[h]
\centering
\small
\renewcommand{\arraystretch}{1.35}
\setlength{\tabcolsep}{5pt}
\newcommand{\ppcell}[2]{\parbox[t]{#1}{\raggedright #2\strut}}

\textbf{(a) Approximation of the operator expansion}\\[3pt]
\begin{tabular}{lll l}
\toprule
\textbf{Method} & \textbf{Sampling distribution} & \textbf{Comments} & \textbf{Adopted} \\
\midrule
\ppcell{2.2cm}{Pauli propagation} & $\displaystyle U^{\dag} O U=\sum_P b_P\,P$ & \ppcell{3.3cm}{Deterministic; the number of Paulis grows exponentially.} & \\
\ppcell{2.2cm}{Single-path MC} & $\displaystyle \beta^2_{\text{single}, s_0}= \frac{1}{2^n} \sum_{\mathbf{s}} |\Tr(O s_{n_g})|^2\,\!\!\prod_{\ell=1}^{n_g}\!|\Tr(g_\ell^\dagger s_\ell g_\ell s_{\ell-1})|^2$ & \ppcell{3.3cm}{Sample one Pauli after every gate; no inter-gate interference (c.f. \equref{eq:pp_singlepath}).} & {\cmark~\cref{sec:PP-MC}} \\
\ppcell{2.2cm}{Hybrid PP-MC} & $\displaystyle \beta^2_{\text{hybrid}, s_0}= \frac{1}{2^n} \sum_{\mathbf{s}}|\Tr(O s_K)|^2\!\prod_{\ell=1}^{K}\!|\Tr(U_\ell^\dagger s_\ell U_\ell s_{\ell-1})|^2$ & \ppcell{3.3cm}{Exact within each of $K$ blocks; converges to $ b_P^2$ as the block size increases (c.f. \equref{eq:pp_hybrid}).} & \ppcell{2.2cm}{\cmark~\cref{sec:PP-MC}} \\
\ppcell{2.2cm}{Unitary expansion} & $\displaystyle \tilde{U}=\sum_{a}c_{a}P_{a},\quad b_P^2\!\approx\!\sum_{a,b}|c_a|^2|c_b|^2\,\delta(P_{ab},P)$ & \ppcell{3.3cm}{Diagonal sum is quadratic in the number of terms; the exact sum is quartic (c.f. \equref{eq:bP_approx}).} & \ppcell{2.2cm}{\cmark~\cref{app:pp_quarter}} \\
\bottomrule
\end{tabular}

\vspace{9pt}
\textbf{(b) Choice of evolved operator (picture)}\\[3pt]
\begin{tabular}{lll}
\toprule
\textbf{Picture} & \textbf{Evolve} & \textbf{\acrshort{otoc} estimator (second order)} \\
\midrule
\ppcell{1.7cm}{Heisenberg \\ (\cref{sec:PP-MC})} & $O(t)=U^{\dag} O U$ & \ppcell{8.2cm}{Single-path MC and hybrid PP-MC estimate of diagonal OTOC $\mathcal{C}_{2, \mathrm{diag}}$ via \equref{eq:diag-P-diag-G}.} \\
\ppcell{1.7cm}{Schr\"odinger (\cref{sec:TablePPMethods})} & $G(\bar t)=U G U^\dagger$ & \ppcell{8.2cm}{The diagonal OTOC is exact for Pauli $O$ (c.f. \equref{eq:C2_schro}).} \\
\ppcell{2.1cm}{Quarter-out operator (\cref{sec:TablePPMethods})} & \ppcell{4.5cm}{$O(t/2)=U_O^\dagger O U_O$,\\[2pt] $G(\bar t/2)=U_G G U_G^\dagger$} & \ppcell{7.5cm}{The exact OTOC is given by \equref{eq:c2_quarter_opearator}.} \\
\ppcell{2.1cm}{Quarter-out unitary \\ (\cref{app:pp_quarter})} & \ppcell{3.7cm}{$\tilde{U}=\sum_{a} c_{a} P_{a}$,\\[2pt] $\tilde{U}^{\dag} O \tilde{U}=\sum_{a,b} c_a^* c_b\,P_a O P_b$} & \ppcell{7.5cm}{Diagonal rescaled OTOC via \equref{eq:quarter_unitary_C2_diag_app}.} \\
\bottomrule
\end{tabular}
\caption{Two axes of Pauli-propagation methods for estimating the (second order) \acrshort{otoc} (and hence the \acrshort{ole} signal) of the semi-scrambling circuits. Panel (a): how the evolving operator is approximated; panel (b): which operator is evolved and how the circuit is split. A method from (a) is combined with a method from (b): the two variants we focus on are hybrid \acrshort{ppmc} in the Heisenberg picture (\cref{sec:PP-MC}) and the quarter-out unitary expansion (\cref{app:pp_quarter}). For simplicity, we have shown second order ($m=1$) order OTOC, which generalizes to higher order as discussed in \secref{sec:diagonal_higher}.
We leave notation definitions and details of each method to their corresponding sections.}
\label{tab:pp_variants}
\end{table}

\subsection{Additional numerical studies of Pauli propagation methods}\label{sec:PPResults}

We now report the numerical performance of the two adopted Pauli-propagation variants, hybrid \acrshort{ppmc} (\cref{sec:PP-MC}) and the quarter-out unitary expansion (\cref{app:pp_quarter}), on the target semi-scrambling \acrshort{ole} circuits, and compare them against the \texttt{ibm\_boston} hardware data. 
Unless stated otherwise, all results are for the $56$-qubit heavy-hex circuit at $L=6$ Floquet layers, field $b=0.125$, probe strength $\delta=0.3$, and the weight-$12$ loop observable $O=\bigotimes_{q\in\mathscr{V}_O}Z_q$ of \figref{fig:main}(b), evolved in the Heisenberg picture with perturbation generator $G=\sum_{q\in\mathscr{V}_{\rm P}}X_q$. 
Results are shown as a function of the scattering strength used in the main text [\figref{fig:main}(e)] and throughout \cref{app:bp}, $\eta \in [0,3 \pi / 8]$.
The reported quantity is the \acrshort{ole} signal $S_{\delta}$ of \equref{eq:f_delta_expansion}, reconstructed from the diagonal moments $\mathcal{C}_{2m,\mathrm{diag}}=\sum_P b_P^2\,D_{2m}(n_P)$ of \cref{sec:diagonal_higher} [\equref{eq:diag-def}] rather than truncated at the \acrshort{otoc}; at $\delta=0.3$ the alternating series converges only once moments up to order $2m\simeq 16$--$20$ are retained.

\subsubsection{Hybrid \acrshort{ppmc} in the Heisenberg picture}\label{sec:res-ppmc}

The hybrid \acrshort{ppmc} estimator of \cref{sec:PP-MC} evaluates the diagonal moments $\mathcal{C}_{2m,\mathrm{diag}}$ by Monte-Carlo sampling of the $2$-norm Pauli distribution $\beta^2_{\text{hybrid}, s_0}$, retaining a cache of up to $M=5\times10^8$ Pauli strings per block before resampling back to size 1.
The signal $S_{\delta}$ is assembled per sample and then averaged, with the statistical error taken over the per-sample signals.

\begin{figure}[htbp]
  \centering
  \includegraphics[width=\textwidth]{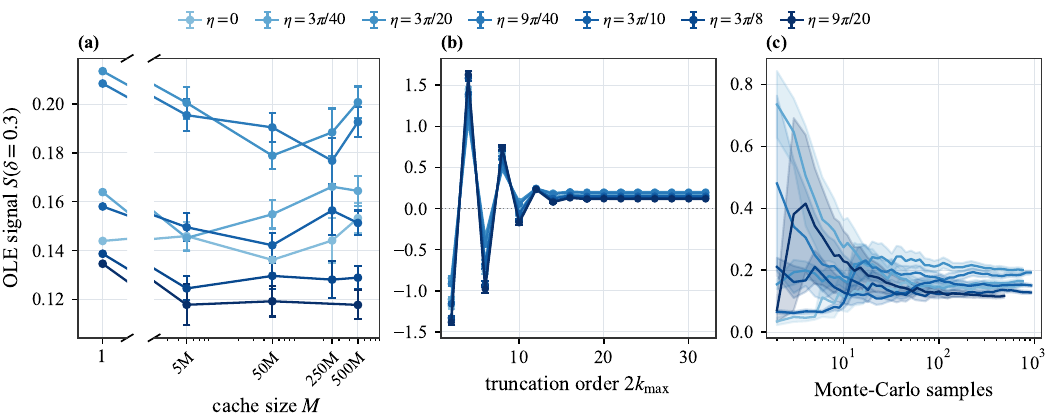}
  \caption{Convergence of the hybrid \acrshort{ppmc} signal $S_{\delta=0.3}$.
    All three panels show the same quantity, on the shared vertical scale, with one curve per scattering strength $\eta$; the colour ramp
    (legend, top) runs light for $\eta=0$ to dark for $\eta=9\pi/20$. (We include one scattering strength beyond the experimental range to make the apparent convergence with cache size more clearly visible.)
    (a)~Versus cache size $M$; $M=1$ is the single-path Monte-Carlo limit, six
    decades below the cached campaign, so the axis is broken. (b)~Versus the
    truncation order $2m$ of the \acrshort{ole} moment series
    [\equref{eq:f_delta_expansion}]. (c)~Versus the number of Monte-Carlo
    samples, with the shaded band giving the running standard error.}
  \label{fig:pp-mcpp-sample}
\end{figure}

\begin{figure}[htbp]
  \centering
  \includegraphics[width=\textwidth]{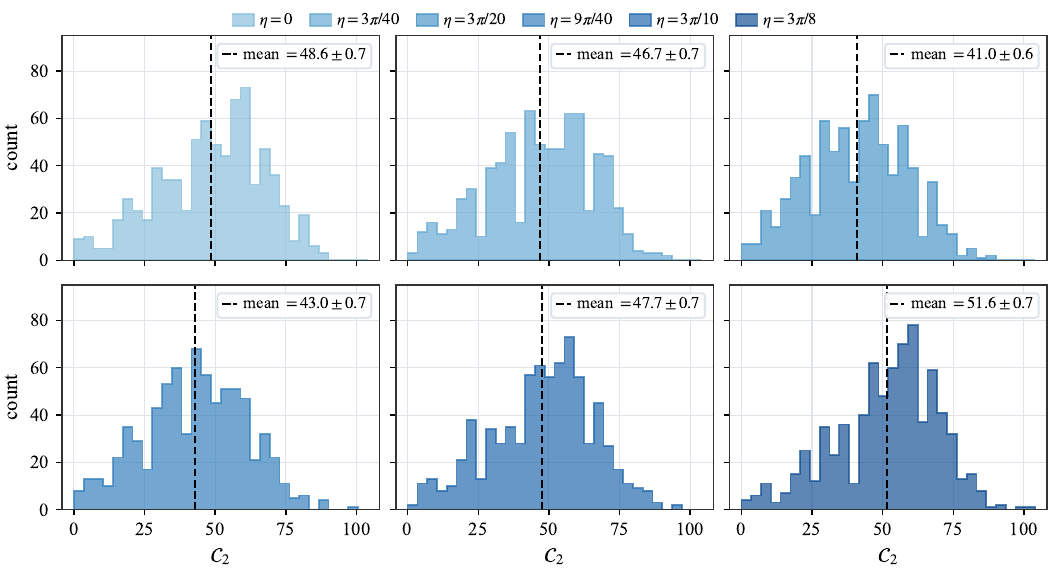}
  \caption{Per-sample \acrshort{otoc} ($\mathcal{C}_2$) distribution from the
    hybrid \acrshort{ppmc} (cache $M=5\times10^8$, Heisenberg picture), one panel
    per scattering strength $\eta$, identified by the shared colour ramp
    (legend, top). Every panel is drawn from the same number of Monte-Carlo
    samples, $N=750$. The dashed line marks the sample mean that enters the signal,
    quoted with its standard error. The distribution is broad but well sampled
    at every $\eta$.}
  \label{fig:pp-mcpp-dist}
\end{figure}

\cref{fig:pp-mcpp-sample} investigates the convergence, one curve per scattering strength $\eta$. The cache-size scan (\cref{fig:pp-mcpp-sample} (a)) is generally non-monotonic and so it does not look straightforward to attempt any extrapolation in the cache size in this context. However, in the case of maximal scattering, $\eta=3\pi/8$, it appears that the estimate is reasonably stable from the single-path limit (cache $M=1$) up to $M=5\times10^8$ with the signal remaining within the statistical error at larger cache sizes.
The moment-order scan (\cref{fig:pp-mcpp-sample} (b)) shows that the second-order (\acrshort{otoc}) approximation of $S_{\delta}$ is qualitatively wrong at $\delta=0.3$  as it can give $S_{\delta}<0$. The full alternating series [\equref{eq:f_delta_expansion}] converges once moments up to order $2m\simeq16$ are included.  
The running \acrshort{ole} estimate (\cref{fig:pp-mcpp-sample} (c)) confirms that the Monte-Carlo error is well controlled with the means stabilizing after a few hundred samples. Any remaining error is due to approximation errors, not statistical uncertainty. \cref{fig:pp-mcpp-dist} shows the corresponding per-sample histograms of the \acrshort{otoc} $\mathcal{C}_2$. The distributions are broad but the means are adequately resolved for every $\eta$.

\subsubsection{Quarter-out unitary expansion}\label{sec:res-quarterout}

The quarter-out unitary expansion of \cref{app:pp_quarter} is, taken exactly, prohibitively costly on these circuits. Building the expansion $\tilde{U}=\sum_a c_a P_a \approx U$ is comparatively cheap, but the OLE is built from the evolved operator, $\tilde{U}^{\dag} O \tilde{U}=\sum_{a,b}c_a^\ast c_b\,P_aOP_b$, whose exact treatment requires \emph{squaring} the number of retained terms of $\tilde{U}$. The coherent expansion of $b_P^2$ is then a quartic sum [\equref{eq:diagC_quarter}], but even assembling just $\tilde{U}^{\dag} O \tilde{U}$ (a quadratic sum) quickly exhausts memory. We therefore use the two approximations introduced in \cref{app:pp_quarter}:
\begin{enumerate}
  \item the \emph{pair-diagonal} approximation of $b_P^2$
    [\equref{eq:bP_approx}], which retains only the $m=a$, $n=b$ contributions
    and lets each pair $(a,b)$ contribute independently, reducing the cost from
    quartic to quadratic and, crucially, letting the estimator be evaluated by
    directly counting the anticommutations $\mathrm{sgn}(P_{ab},G_i)$
    [\equref{eq:C2_pair_diag}] without ever constructing $U^{\dag} O U$ or the individual
    $b_P$; and
  \item the \emph{approximate self-normalization}
    $\mathcal{N}_{\mathrm{approx}}=\big(\sum_a|c_a|^2\big)^2=\|\tilde{U}\|^4$, obtained by
    applying the same pair-diagonal truncation to the norm
    $\mathcal{N}(\tilde{U}^\dagger O \tilde{U})=\sum_P b_P^2$ [\equref{eq:norm_def}].
\end{enumerate}
Since the physical (normalized) estimator divides the raw pair sum by this normalization, $\hat{\mathcal{C}}_{2,\mathrm{diag}}^{\,\mathrm{approx}}$ of \cref{app:pp_quarter}, and likewise every higher moment $\mathcal{C}_{2m}$ that enters $S_{\delta}$, the choice of $\mathcal{N}$ directly sets the reported value. 
The pair-diagonal list moreover gives access to the anticommute counts $n_P$ of every pair, so the higher-order diagonal moments $\mathcal{C}_{2m,\mathrm{diag}}$ of \cref{sec:diagonal_higher} can be accumulated at the same quadratic cost, and the full signal $S_{\delta}$ reconstructed. The size of the retained expansion is controlled by \emph{relative-coefficient (rc) truncation}: after each Pauli rotation, terms whose coefficient magnitude falls below a fraction $\mathrm{rc}$ of the
largest current coefficient are discarded; lowering $\mathrm{rc}$ retains more
terms ($n_{\mathrm{terms}}$ grows) and yields a more faithful $U$ at higher cost.

\begin{figure}[htbp]
  \centering
  \subfloat[\acrshort{otoc} vs $\eta$ ($\mathrm{rc}=5\times10^{-2}$)]{\includegraphics[width=0.48\textwidth]{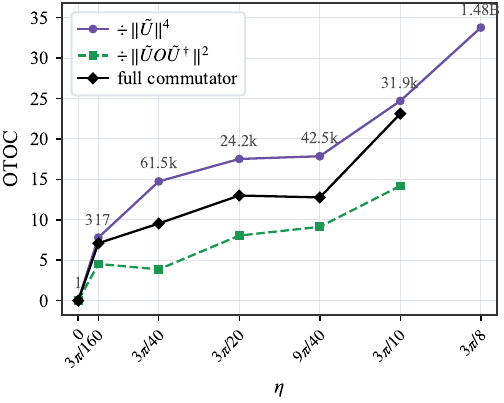}\label{fig:pp-qo-otoc}}\hfill
  \subfloat[signal $S_{\delta}$ vs $\eta$]{\includegraphics[width=0.48\textwidth]{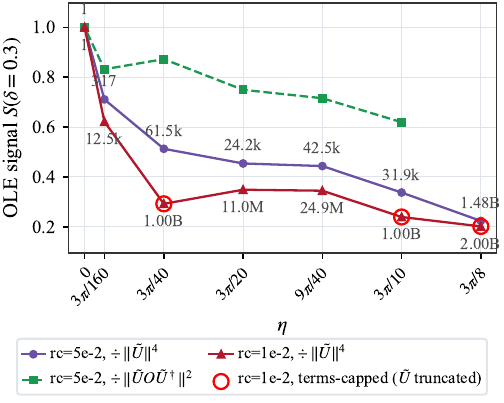}\label{fig:pp-qo-signal}}
  \caption{Quarter-out unitary expansion, normalization comparison. (a)
    \acrshort{otoc} versus $\eta$ at $\mathrm{rc}=5\times10^{-2}$ under the
    self-norm $\|\tilde{U}\|^4$, the exact picture norm $\|\tilde{U}^{\dag} O \tilde{U}\|^2$,
    and the full commutator (exact); the two approximate norms bracket the exact
    result. (b) The corresponding signal $S_{\delta=0.3}$ at
    $\mathrm{rc}=5\times10^{-2}$ and $\mathrm{rc}=10^{-2}$ (self-norm across the
    sweep, picture norm where feasible); point labels give $n_{\mathrm{terms}}$
    in the propagated $U$ and red rings mark memory-capped runs. Note, when we say `exact' here we only mean `exact' in the sense that no diagonal approximations or renormalization methods are applied - there are still, of course, substantial truncation errors.}
  \label{fig:pp-qo-norm}
\end{figure}

\medskip\paragraph*{Coarse truncation ($\mathrm{rc}=5\times10^{-2}$): testing the
approximations.}
At the coarse threshold $\mathrm{rc}=5\times10^{-2}$ the expansion completes across the whole $\eta$ sweep, and even the quadratic computations are tractable, so the two approximations can be tested against more expensive references.\cref{fig:pp-qo-otoc} compares three evaluations of the \acrshort{otoc}: the pair-diagonal estimator normalized by the exact picture norm $\mathcal{N}(\tilde{U}^\dagger O \tilde{U})=\|\tilde{U}^{\dag} O \tilde{U}\|^2$ (dashed; isolates the effect of the pair-diagonal approximation alone), the same estimator normalized by the self-norm $\|\tilde{U}\|^4$ (solid; the fully approximate estimator), and the full commutator evaluated exactly from the truncated $U$ (reference). 
The three disagree: the two normalizations sandwich the exact result, with the self-norm $\|\tilde{U}\|^4$ overestimating and the picture norm $\|\tilde{U}^{\dag} O \tilde{U}\|^2$ underestimating the \acrshort{otoc} (e.g.\ at $\eta=3\pi/40$: $\|\tilde{U}\|^4\!\to\!14.7$, $\|\tilde{U}^{\dag} O \tilde{U}\|^2\!\to\!3.9$, exact $=9.5$), and the same ordering is carried into the signal. 

\begin{figure}[htbp]
  \centering
  \subfloat[signal $S_{\delta}$]{\includegraphics[width=0.48\textwidth]{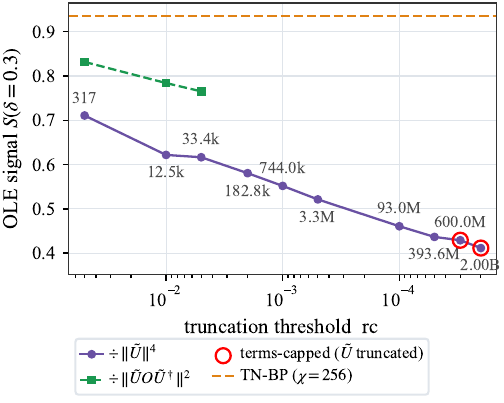}\label{fig:pp-qo-sig-rc}}\hfill
  \subfloat[normalizations]{\includegraphics[width=0.48\textwidth]{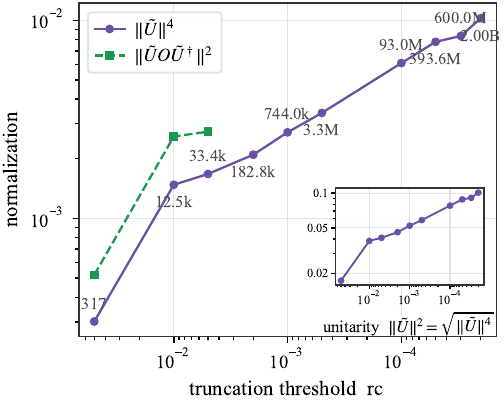}\label{fig:pp-qo-norm-rc}}
  \caption{Truncation dependence of the quarter-out expansion at
    $\mathrm{sc}=0.9$ ($\eta=3\pi/160$): (a) the signal $S_{\delta=0.3}$ and (b)
    the two normalizations $\|\tilde{U}\|^4$  and $\|\tilde{U}^{\dag} O \tilde{U}\|^2$ versus the relative-coefficient threshold $\mathrm{rc}$;
    point labels give $n_{\mathrm{terms}}$ in the propagated $U$ and red rings
    mark memory-capped runs. The self-norm drifts downward without converging,
    while (inner plot) $\|\tilde{U}\|^2=\sqrt{\|\tilde{U}\|^4}\sim10^{-1}\ll1$ shows the propagated $U$ is
    far from unitary.}
  \label{fig:pp-qo-trunc}
\end{figure}

\medskip\paragraph*{Fine truncation and the signal.}
Lowering the threshold to $\mathrm{rc}=10^{-2}$ makes the propagated $U$ far larger ($n_{\mathrm{terms}}$ up to $\sim10^{9}$): the exact picture norm and the full commutator are then no longer computable across the sweep, and the self-normalized signal drops considerably relative to the coarse truncation [\cref{fig:pp-qo-signal}]. To probe convergence we zoom on the weakest scattering, $\eta=3\pi/160$, where $U$ is closest to the
identity and the expansion is smallest, so the truncation can be pushed furthest. \cref{fig:pp-qo-sig-rc} shows the fully-approximate (self-norm) signal decreasing monotonically as $\mathrm{rc}$ is lowered, with no sign of convergence: it keeps drifting downward as more terms are added, all the way to a hard cap on $n_{\mathrm{terms}}$ (red rings). 
This cap is numerically imposed to keep the size bounded and the simulations tractable. Where the exact picture norm is still affordable it lies well above the self-norm curve, so the two normalizations continue to follow each other, but neither the exact norm nor a converged value is reachable. \cref{fig:pp-qo-norm-rc} plots the two normalizations over the same scan: $\|\tilde{U}\|^4$ (and hence $\|\tilde{U}\|^2=\sqrt{\|\tilde{U}\|^4}$) stays an order of magnitude below $1$ throughout ($\|\tilde{U}\|^2\sim10^{-2}$), confirming that only a tiny, strongly non-unitary fraction of $U$ is retained even at the finest truncations. Taken together, the quarter-out expansion is tractable and informative only at very weak scattering, and even there the fully-approximate signal is a moving target that depends sensitively on both the truncation and the choice of normalization.
The verifications above are conducted for $\tilde{U}$ composed of a relatively small number of Paulis. It is possible that the approximations work better for $\tilde{U}$ composed of more Paulis, far from trivial cases ($\eta = 0$). But more studies in this direction have not been conducted.

\subsubsection{Implementation and computational cost}\label{sec:pp-cost}

Both adopted variants are implemented in Julia, extending the \texttt{PauliPropagation.jl} package~\cite{rudolph_pauli_2026}.\\

\paragraph*{Hardware.} Production runs used two AWS systems. The hybrid \acrshort{ppmc} and single-path Monte-Carlo ran on a cluster of single-socket AMD~EPYC~9R14 nodes ($32$ cores, $\sim128$~GB RAM each): each estimate is a sweep over statistically independent samples, dispatched as single-node SLURM array tasks with thread-level parallelism inside Julia, and spread across $15$ distinct compute nodes. The quarter-out expansion is memory-limited, rather than sample-limited, and instead ran on a single large-memory node (Intel~Xeon~Platinum~8375C, $128$ vCPUs, $1$~TB RAM).\\

\paragraph*{Memory.} At the largest cache $M=5\times10^8$ the propagation process peak reaches $65$--$100$~GB, including the transient term growth during each Pauli rotation before truncation. The quarter-out expansion is instead limited purely by the term count of the propagated $U$, which we cap at $6\times10^8$--$2\times10^9$ terms (red rings in \cref{fig:pp-qo-norm,fig:pp-qo-trunc}).\\

\paragraph*{Cost.}\cref{tab:pp-cost} reports the per-job resource usage of single-path MC and the hybrid \acrshort{ppmc} at three cache sizes. For one $\eta$ value, the cumulated time for cache size $5\times10^8$ is $\sim 300$~h.

\begin{table}[ht]
    \centering
    \begin{tabular}{l c c c}
        \toprule
        method & cores/job & peak RSS (GB) & time/job \\
        \midrule
        single path (1000 samples)& $2$ & $0.02$ & $6$~s \\
        hybrid \acrshort{ppmc} ($M = 5\times10^6$) & $3$  & $1.8$--$2.7$ & $\sim2$~min \\
        hybrid \acrshort{ppmc} ($M = 5\times10^7$) & $10$ & $8$--$10$    & $\sim8$~min \\
        hybrid \acrshort{ppmc} ($M = 5\times10^8$) & $32$ & $65$--$102$  & $\sim25$~min \\  
        
        \bottomrule
    \end{tabular}
    \caption{Per-job computational cost of the Monte-Carlo Pauli-propagation runs
    on the AMD~EPYC~9R14 nodes, for the 56-qubit $L=6$ circuits. Each row is a
    single SLURM array task: ``cores/job'' is the allocated core count, peak RSS
    the measured process high-water mark, and ``time/job'' the wall-clock elapsed
    per task. Both are essentially independent of the scattering strength $\eta$.}
    \label{tab:pp-cost}
\end{table}

\subsubsection{Comparison across methods and with experiment}\label{sec:res-comparison}

\begin{figure}[htbp]
  \centering
  \includegraphics[width=0.7\textwidth]{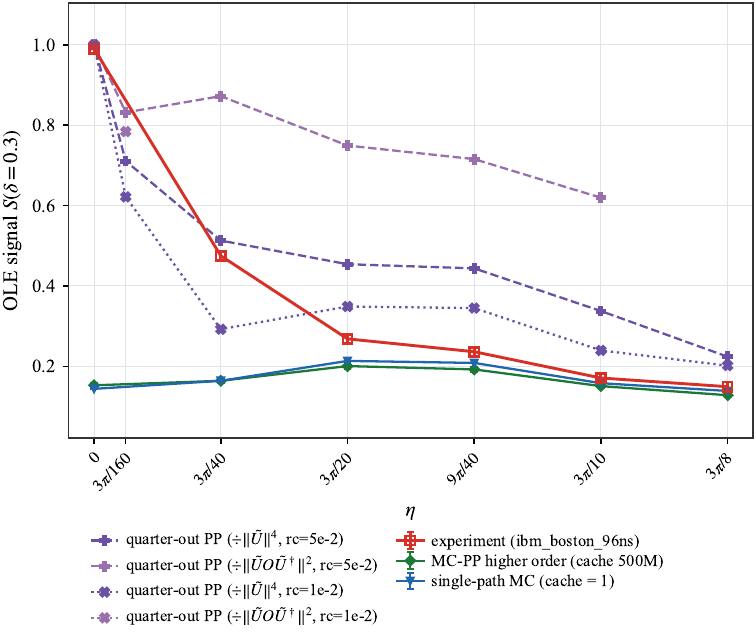}
  \caption{\acrshort{ole} signal $S_{\delta=0.3}$ versus scattering strength
    $\eta$ for the adopted classical methods and the \texttt{ibm\_boston}
    experiment (each rescaled to $f_0=1$): hybrid \acrshort{ppmc}
    (cache $M=5\times10^8$), single-path Monte Carlo (cache $M=1$), and the
    quarter-out expansion (self-norm, $\mathrm{rc}=10^{-2}$; red rings mark
    memory-capped points). The Monte-Carlo methods track the experiment well at large
    $\eta$  and the quarter-out expansion roughly at all $\eta$.
    }
  \label{fig:pp-allmethods}
\end{figure}

\cref{fig:pp-allmethods} overlays the adopted methods on the \texttt{ibm\_boston} data:

\begin{itemize}
    \item The hybrid Monte-Carlo estimator tracks the experiment at strong scattering, consistent with the picture that $2$-norm sampling becomes reliable once the operator has spread over many effectively non-interfering Pauli strings (\cref{sec:PP-MC} and main text). In Fig.~\ref{fig:pp-mcpp-sample} (a), we present evidence that is consistent with our simulations being converged, raising the question of what accounts for the discrepancy with experiment (and how the remaining uncertainties in the two approaches should be assessed).

\item Intriguingly, in this strong scattering limit, the single path Monte Carlo simulation (an exceptionally quick and low memory simulation, see \cref{tab:pp-cost}) also seemingly aligns with the \acrshort{ppmc} and experiment results. This could point towards some deeper structure to the circuit which means that interference effects between paths are small in this regime. Alternatively, it could just be that the single path scheme gets lucky and its various biases cancel out.

\item  The quarter-out signal (self-norm, $\mathrm{rc}=10^{-2}$), by contrast, somewhat tracks the experiment across the full range of scattering strengths, but it is not converged in the truncation (Fig.~\ref{fig:pp-qo-trunc}) and both the renormalization strategy and diagonal approximation lead to substantial errors (Fig.~\ref{fig:pp-qo-norm}). Thus, while this very new method shows some promise, more refinements and validation are needed before it can be viewed as a practical trusted method for large-scale simulations. 
\end{itemize}

\section{Comparison of all classical results and survey of future directions}
\label{app:summary}

\begin{figure}[ht]
\includegraphics[width=0.9\linewidth]{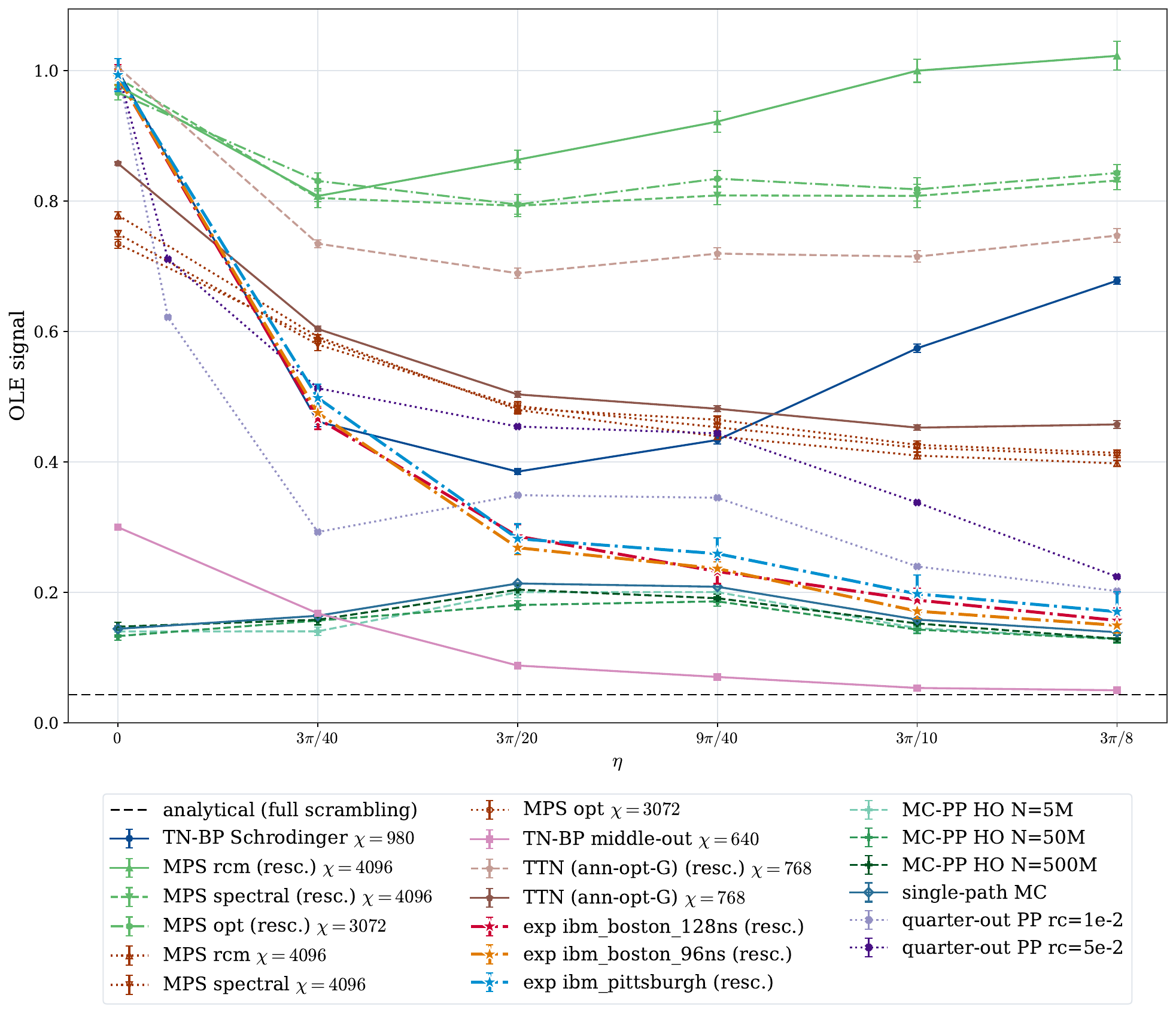}
\caption{
Collection of all methods used to compute the \acrshort{ole} signal for the $L=6$, $\delta=0.3$ circuit.}
\label{fig:app-messi}
\end{figure}

In \cref{fig:app-messi} we collect and compare all data related to the $L=6$ circuit with parameters $b=0.125$ and $\delta=0.3$ for various scattering strengths $\eta$. We have explored several classical methods, some of which are able to correctly simulate the circuit in one of the $\eta$ regimes (close to 0, or large $\eta$). However, we observe that no classical method is reliable across the whole range, making the result from the quantum computer the best choice for a reliable simulation of this system.

The survey of classical simulation methods presented in this paper is necessarily not exhaustive. The difficulty of the circuits considered here arises from several distinct features, including entanglement growth, operator spreading, and interference between Pauli paths. Methods based on different representations may therefore encounter different limitations. We conclude by discussing several possible avenues for future work.

Instead of representing the evolved eigenstates of the observable with tensor networks, one could use neural-network quantum states~\cite{carleo_solving_2017}. These representations are not tied to a fixed tensor-network geometry and may capture correlations that are difficult to encode efficiently with an MPS or a tensor network defined on the heavy-hex graph. Their application to real-time echo dynamics may nevertheless be challenging, since optimization errors accumulated during the forward evolution could be amplified by the reverse evolution.

A second direction is to scale the Pauli-propagation methods developed here to much larger computational resources. Distributing the retained Pauli strings across a large cluster would allow longer periods of exact propagation between sampling steps and could preserve more of the interference generated by the circuit. Sparse Pauli-propagation methods have already been deployed on the Fugaku supercomputer at RIKEN, retaining up to one trillion Pauli strings~\cite{broers2025scalable}. A similar distributed implementation could be combined with the sampling and truncation strategies introduced here.
Increasing the cache size would not by itself guarantee convergence, since the number of relevant Pauli strings may continue to grow faster than the available memory. If converged results required leadership-scale classical resources, it would then be interesting to compare the classical and quantum approaches in terms of wall-clock time, energy consumption, and total computational cost.

Methods based on proximity to a Clifford circuit provide a complementary approach. Clifford perturbation theory~\cite{Begusic2023} or stabilizer decompositions organize the computational cost according to the non-Clifford content of the circuit rather than its entanglement or Pauli support. They may be useful near Clifford parameter values or after decomposing the circuit into regions of different effective complexity. Their accuracy could be tested by increasing the perturbative order or stabilizer rank.

It could also be possible to estimate the \acrshort{ole} signal via a plausible hydrodynamics argument of the operator spreading.
Or if the majority of an evolved observable is supported on a smaller lattice, a simulation of the dynamics restricted to such a lattice may provide a good approximation to the \acrshort{ole} signal. Such `effective light cone' methods have found success when applied to some other large-scale quantum circuits~\cite{Tindall2024efficient, torre2023dissipative, kechedzhi2024effective}. 

As shown in subsections~\ref{app:mps} and \ref{app:ttn}, \acrshort{mps} and \acrshort{ttn} do not qualify as competitive methods for predicting the \acrshort{ole} signal, even when the site ordering is optimized by minimizing the two-qubit gate count. An alternative, however, would be to optimize the ordering with respect to circuit-specific quantities, such as $\eta, \delta$, and the measured plaquette, as in Fig.~\ref{fig:main}. Since these settings determine the complexity of the simulation, they could all be incorporated into a tailored optimization of the ordering.

Another potential avenue is to simulate the noisy circuit rather than the ideal one. Noise suppresses coherence, entanglement, and high-weight operator components, potentially making the evolution easier to represent classically. Error mitigation could then be applied to infer the ideal \acrshort{ole}, as previously explored using MPS simulations~\cite{rakovszkyDissipationassistedOperatorEvolution2022}. The reduced entanglement must be weighed against the cost of representing a mixed state. Quantum trajectories may provide a useful alternative to explicit density-matrix evolution by replacing it with an ensemble of stochastic pure-state simulations, at the expense of an additional sampling cost.

Another possible approach within the heavy-hex tensor network framework is to trade spatial entanglement for temporal
entanglement using influence functionals, in which the environment of a subsystem is encoded as
a matrix product state in the time direction~\cite{Park2025IFBP}. The influence functional
belief propagation (IF-BP) algorithm extends this construction to arbitrary graphs and has been
applied to the kicked Ising model on the heavy-hex lattice, where the temporal entanglement
entropy grows only logarithmically in time; the locally tree-like geometry keeps the
belief-propagation error small, and loop corrections can be added systematically through a
cluster expansion. Whether the temporal entanglement remains as benign for echo dynamics, where
the backward evolution doubles the extent of the influence functional in time, is an open
question.

{\color{blue} And finally, one could apply the tensor network Monte Carlo (TNMC) approach used in Ref.~\cite{abanin_observation_2025} with a BP tensor network backend instead instead of an MPS backend. The approach can be understood as a marriage between the hybrid MC-PP and TN-BP methods employed in this work. By applying random Pauli operators or sampling from intermediate representations in the Pauli basis throughout the TN-BP simulation, one projects out non-diagonal terms in the \acrshort{ole} and potentially reduces the maximal bond dimension. By averaging over several randomized instances, one achieves a result with tunable bias depending on the amount of projections and the maximal bond dimension.}

Regardless of the chosen method, the same validation principles addressed in this work should be applied before its prediction is treated as reliable. These include convergence with respect to controllable numerical parameters, agreement with exact results at smaller depths, recovery of analytically tractable limits, and consistency among independent methods.

\newcommand{\gates}[1]{\ensuremath{\mathsf {#1}}}
\newcommand{\TODO}[1]{(\textcolor{magenta}{TODO:#1})}
\renewcommand{\trace}[1]{\text{Tr}\left(#1\right)}
\newcommand{\xslow}{\gates{X}$_{\text{slow}}$}

\section{Experimental methods and calibration details} \label{app:exp}

\subsection{Superconducting Quantum Hardware}\label{app:hardware}

\begin{figure}[h]
\includegraphics[width=0.55\textwidth]{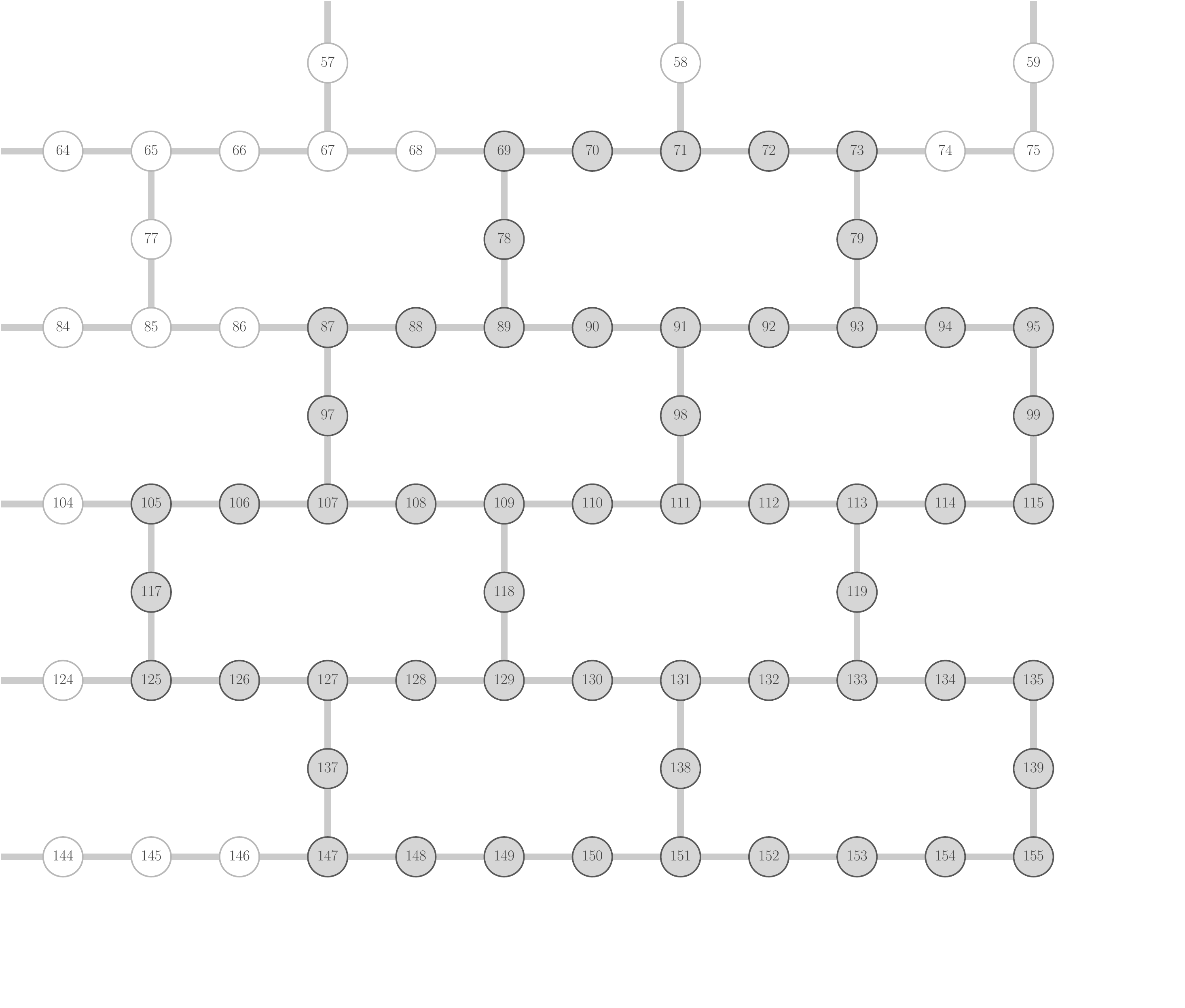}
\caption{
The 56 qubits of $\texttt{ibm\_boston}$ (gray nodes) used in our experiments.}
\label{fig:device_label}
\end{figure}

\begin{figure}[h]
\includegraphics[width=0.75\textwidth]{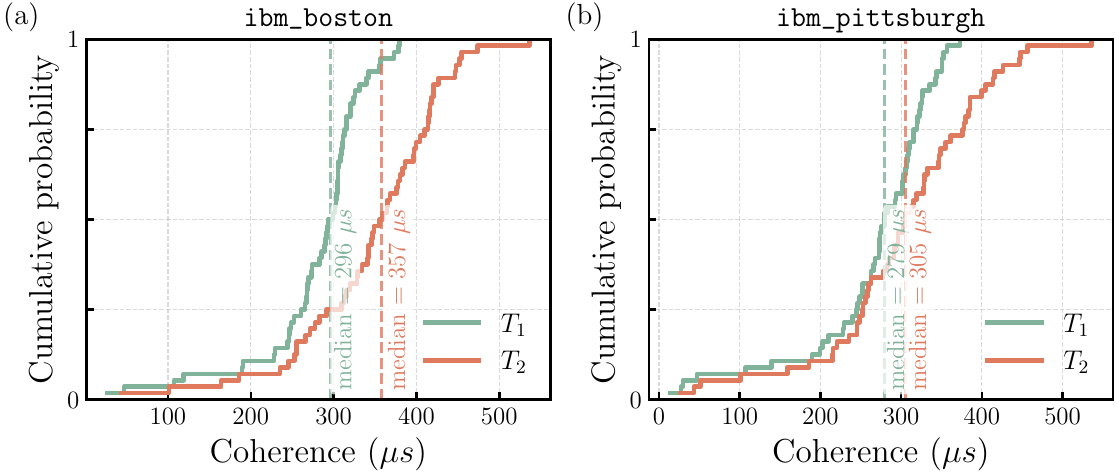}
\caption{
Cumulative distribution functions of the coherence times of the 56 qubit sections on (a) $\texttt{ibm\_boston}$ and (b) $\texttt{ibm\_pittsburgh}$.}
\label{fig:coherences}
\end{figure}

Experiments were performed on $\texttt{ibm\_boston}$ and $\texttt{ibm\_pittsburgh}$, which are superconducting quantum processors available through the IBM Quantum Platform. Both devices are Heron (R3) architectures comprising 156 fixed-frequency transmon qubits arranged in a heavy-hexagonal connectivity. Qubit coherence times are shown in Fig.~\ref{fig:coherences}. Two-qubit gates are enabled by flux-tunable couplers, which provide native CZ two-qubit gate instructions.
The 56 qubits on $\texttt{ibm\_boston}$ are shown in \cref{fig:device_label}.
We use the same labeling convention to refer to qubits throughout the manuscript.

For rescaling experiments (data shown in \cref{fig:main} of the main text), arbitrary single-qubit operations are decomposed into R$_X$ and virtual R$_Z$ gate instructions (one R$_X(\theta)$ gate per single-qubit gate with variable angle $\theta \in [0,\pi]$) using \SI{32}{ns} single-qubit gates, and \SI{96}{ns} CZ gates. For PEC data on $\texttt{ibm\_boston}$ (data shown in \cref{fig:pec_results} of the main text), arbitrary single-qubit operations are decomposed into $\sqrt{X}$ (SX) and virtual R$_Z$ gate instructions (two SX gates per single-qubit gate). 
To suppress classical single-qubit drive crosstalk, the drive phase of each qubit is randomly selected for each shot \cite{eddins2026computing}.

Qubits are initialized using two conditional reset instructions. The repetition rate is one sample per \SI{250}{\micro s}. Terminal measurements comprise \SI{2.2}{\micro s} ($\texttt{ibm\_boston}$) and \SI{2.58}{\micro s} ($\texttt{ibm\_pittsburgh}$) readout instructions. 

\subsection{Two-Qubit Gate Calibrations}

The OLE circuit consists of repetitions of three unique layers of CZ gates. Therefore, the CZ gates were calibrated with identical pulse duration and parallelized using the three sets of matching batches of gates~\cite{kim_evidence_2023}. Calibration included optimization of the $ZZ$, $IZ$, and $ZI$ rotation angles effectuated by the gate pulse, as determined by error-amplification sequences~\cite{wei2024heat}. As any classical (signal) crosstalk present in the CZ layers is replicated during these batched calibrations, gate calibration implicitly compensates for potential excess $ZZ$, $IZ$, and $ZI$ over-rotations due to crosstalk between simultaneous CZ gates. 

Gate error rates are characterized using the $\texttt{LayerFidelity}$ experiment in the $\texttt{Qiskit-Experiments}$ library to provide simultaneous, direct randomized benchmarking of CZ gates twirled by random single-qubit Cliffords~\cite{proctor2019rb, mckay2023lf,Kanazawa2023}. 
    Single-qubit Cliffords are transpiled into a device-native instruction set that matches the single-qubit-gate decomposition used for the corresponding target circuit.
We report the error per CZ gate defined as $1-F_{ij}$ where $F_{ij}$ is the process fidelity measured for the CZ gate on adjacent qubits $i$ and $j$. The error per CZ gate is inclusive of errors during single-qubit gates. Benchmarks are measured simultaneously in three groups matching the three CZ layers that appear in the target circuit.

\subsection{Gate Transient Errors}\label{app:transients}

\begin{figure}[t]
\includegraphics[width=0.35\textwidth]{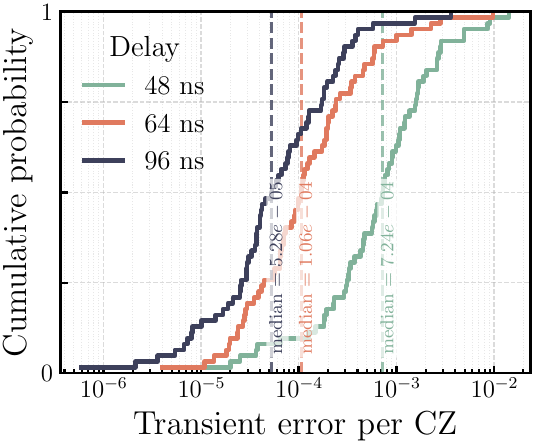}
\caption{Cumulative distribution function of CZ gate errors due to pulse transients as a function of delay time between CZ gates. At each delay value, gate error rates are converted from the measured $IZ$, $ZI$, and $ZZ$ over-rotation angles, referenced to that at \SI{400}{ns} delay.}
\label{fig:transient_errors}
\end{figure}

Pulse transients due to signal distortions in flux control lines may impart coherent errors to CZ gates and are a source of non-Markovianity that is particularly a challenge for noise learning and PEC. 
Figure~\ref{fig:transient_errors} demonstrates transient gate errors as a function of delay times between CZ gates. 
This implies that the noise model learned at a particular separation between CZ pulses may not be representative of the noise in a target circuit with different durations between CZ repetitions. 
We take particular care to mitigate the effect of such transient errors through advanced Finite Impulse Response (FIR) filter calibrations, decomposing single-qubit operations into SX and R${}_Z$ instructions so that CZ layers are separated by \SI{64}{ns}, and/or targeted transpiler passes. To construct a targeted transpiler pass,
we measure the $IZ$ and $ZI$ over-rotation errors of each CZ gate as a function of the time between that CZ gate and its previous occurrence. 
Over-rotation error measurements employ error-amplification sequences having variable delays added between CZ gate repetitions. 
A transpiler pass then applies corrections to each CZ gate in the OLE circuit. 
For each CZ gate, the transpiler determines the time between that CZ gate and its prior occurrence in the scheduled circuit. 
The transpiler then inserts corrective virtual R${}_Z$ gates adjacent to the CZ gate. 
The angles of the R$_Z$ gates are the inverse of the corresponding measured $IZ$ and $ZI$ over-rotation errors.

\subsection{Noise stabilization}\label{app:TLSmod}
For PEC experiments, drift in the landscape of two-level system defects (TLS) between noise model learning and target-circuit execution could cause changes in the noise channel and therefore invalidation of the learned model. To stabilize the noise channel, electrodes placed above each qubit that tune the qubit-TLS interaction are modulated at a low frequency (0.1-1 Hz) throughout the entire PEC campaign~\cite{kim2025error}. Modulation amplitudes and offsets are chosen for each qubit to avoid collision between the qubit and any strongly coupled TLS. Importantly, the modulation frequency is low compared to the sampling rate so that the electrode bias is quasi-static during each measurement sample, yet shot repetitions execute outside looping over Pauli twirling instantiations so that, for each Pauli twirl, data is sampled at many electrode bias values. As such, the noise model is sampled across an ensemble of TLS landscapes throughout both learning and target circuit execution.

For rescaling experiments, the electrode biases are fixed at values selected to maximize the qubit coherence times~\cite{dane2025performance}.

\subsection{Post-selection of non-Markovian leakage errors}\label{app:postselect}

\begin{figure}
    \centering
    \includegraphics[width=0.75\linewidth]{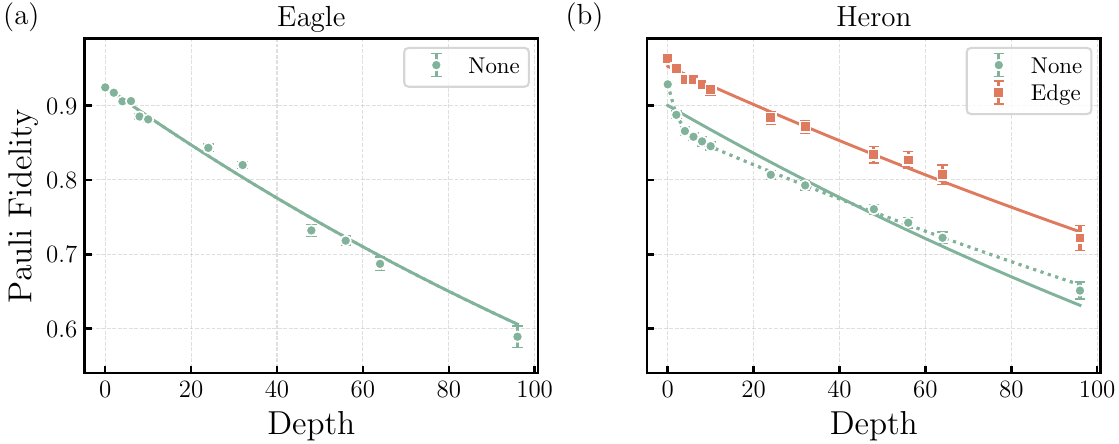}
    \caption{\textbf{Non-Markovian errors in Pauli noise learning.} \textbf{(a-b)} Pauli fidelity after repeated identity gates, measured on an Eagle (a) and a Heron (b) device for 40 qubits. The Pauli fidelity is extracted by measuring local Pauli Transfer Matrix elements of circuits with increasing net-identity repetitions of the twirled gate.  On Heron, we have observed decay curves with a fast initial decay at short depths (green markers, no PS), which the single-exponential fit (solid line) fails to capture. A double-exponential fit (dotted line) better matches the data. After using the edge-based post-selection (PS) protocol, the data better follows the expected exponential decay. This is quantified by a smaller reduced-$\chi^2$ as a goodness-of-fit metric, as shown in~\ref{fig:pec_results}d.}
    \label{fig:quasistatic_freq_shifts}
\end{figure}

Error mitigation protocols often rely on having a representative model of the device noise that is informed by an experimental noise learning step. An efficient Pauli noise-learning protocol based on the sparse Pauli-Lindblad noise model was first proposed and demonstrated at scale on fixed-frequency Eagle processors~\cite{van_den_berg_probabilistic_2023,kim_evidence_2023} with fixed qubit-qubit coupling. These noise learning protocols~\cite{Erhard2019characterizing,Flammia_2020} often measure the decay of local Pauli expectation values from circuits consisting of repeated, Pauli-twirled, net-identity blocks (such as pairs of CZ gates), enabling the extraction of Pauli fidelity decay parameters in a SPAM-free manner. An example Pauli fidelity noise learning curve for an identity gate layer on an Eagle device is shown in Fig.~\ref{fig:quasistatic_freq_shifts}a, which shows good agreement with the exponential decay model. On Heron processors, however, we observe non-exponential features in the noise learning decays that are often signatures of non-Markovian device errors~\cite{Fogarty2015,fong2017randomizedbenchmarkingcorrelatednoise,Figueroa_Romero_2021}. Leakage out of the computational subspace is a common source of such non-Markovian errors. These can arise for a number of reasons, including interactions with TLS and/or other couplings on the device and the weaker anharmonicity of transmons on the Heron devices.

\begin{figure}[h]
\includegraphics[width=0.65\textwidth]{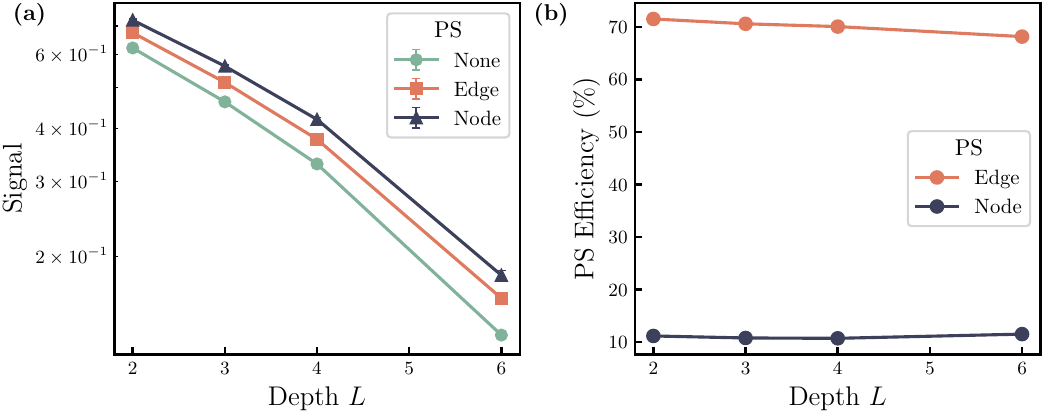}
\caption{(a) The unmitigated signal of the weight-12 observable at $\delta=0$ as a function of the OLE depth parameter $L$ without PS and with edge/node PS.
(b) Efficiency of different PS strategies. 
The signal is boosted the most with node-based PS, however the reduction in efficiency is substantial.}
\label{fig:ps_survival}
\end{figure}

Non-Markovian noise sources are a potential source of systematic error in error-mitigated expectation values, as our error mitigation approach assumes Markovian noise. We therefore develop a protocol to detect and mitigate the impact of such errors on a shot-by-shot basis. We check if the qubits are in their computational state space by testing if a $\pi$ pulse (X) successfully flips the qubit state. These checks can be inserted at the start (pre-check) and/or the end of the circuit, using pulse-sequences described in Table~\ref{table:ps}. Since the bandwidth of a typical $\pi$ pulse (X) spans approximately $\sim 14$ MHz, we enhance the detection sensitivity of our protocols by engineering a narrow-band $\pi$ pulse (\xslow) with a $\sim1~\mu$s duration and sub-MHz spectral width. We mitigate the effect of these non-Markovian errors by post-selecting on the efficacy of these (\xslow) gates on individual qubits, based on the logic described in Table~\ref{table:ps}, and refer to this as ``node-based" post-selection. We note that the post-selection circuits may also be affected by other errors such as readout errors from qubit decay~\cite{thorbeck2024readout} and back-actions~\cite{gambetta2007protocols}, and imperfect calibration of the \xslow~and X gates. As a result, the post-selection survival rate on circuits with many qubits can be prohibitively low.
The post-selection efficiency can be improved though, albeit at the cost of detecting only a subset of the possible non-Markovian errors, by post-selecting on correlated errors. Here, we only disregard shots that involve failures on connected qubits, and we refer to this as ``edge-based" post-selection and employ it across all experiments reported in the main text as a terminal check. Pre-check circuits were not used in any experiments in this work to optimize for post-selection efficiency.

Post-selection survival rates are shown for ``edge-based" and ``node-based" post-selection in Fig.~\ref{fig:ps_survival} as the depth of the OLE circuit is increased, as well as corresponding increase in OLE signal.
The post-selection capability is available in $\texttt{qiskit-addon-utils}$~\cite{qiskit-addon-utils}.

\begin{table}[t!] 
\centering
\begin{tabular}{||c | c | c | c||} 
 \hline
 check name & circuit & bitstring & decision \\ [0.5ex] 
 \hline\hline
 \multirow{2}{5em}{pre-check} & \multirow{2}{8em}{[\xslow-\gates{X}-\gates{M}$_\text{pre}$]} & $M_\text{pre}=0$ & keep \\ 
  & & $M_\text{pre}=1$ & discard \\ 
 \hline
 \multirow{2}{5em}{post-check} & \multirow{2}{8em}{\gates{M}$_\text{t}$-[\xslow-\gates{M}$_\text{post}$]} & $M_\text{t}M_\text{post}=01/10$ & keep \\ 
  & & $M_\text{t}M_\text{post}=00/11$ & discard \\ 
 \hline
\end{tabular} 
\caption{\textbf{Table for post-selection logic.} Post-selection logic is described for pre-check and post-check circuits. $M_\text{pre}$ represents a readout bitstring from pre-check circuit. $M_\text{t}$ stands for the bitstring obtained from terminal measurement. $M_\text{post}$ is a readout bitstring from post-check circuit.}
\label{table:ps}
\end{table}

\subsection{Cliffordizations}\label{app:cliffordization}
In our experiments, we use \emph{Cliffordizations} of our target circuit to calibrate and validate the noise models~\cite{merkel_when_2025}.
Recall that the OLE circuits are decomposed into CZ, R$_Z(2h)$, and R$_X(2b)$ where $h = \pi/8$ and $b \in \{b_1,b_2,b_1-\eta\}$.
A Cliffordization of the OLE circuit is constructed by replacing the values of $h$ and $b$ by either $0$ or $\pi/4$. These angles result in Clifford circuits that can be efficiently simulated.
We repeatedly use two particular Cliffordizations, which share
$h = b_1 = \eta = \pi/4$ and only differ in $\theta = 2b_2 \in\{0,\pi/2\}$. 

\subsection{Maintaining an Accurate Noise Model}\label{app:filtering}

In the idealized limit where the learned noise model coincides with the true device noise, probabilistic error cancellation (PEC) yields unbiased expectation values with rigorous error bounds~\cite{temme_error_2017}. In practice, however, device noise may have components that are not well described by a sparse Pauli-stochastic noise channel, such as non-Markovian errors, non-local errors, and non-zero off-diagonal matrix elements in the Pauli Transfer Matrix. These out-of-model errors cannot be captured by the learned model, and therefore lead to model disagreement and residual bias~\cite{govia2024bounding}. As discussed above and in the main text, these effects are largely mitigated through optimized gate calibrations, Pauli twirling, post-selection, and electrode bias modulation. Despite these mitigation strategies, small out-of-model errors and temporal drift can remain, leading to residual biases in measured PEC expectation values.

To build confidence in the PEC results, we carry out several tests for the quality of the learned model as a representation of the device noise.

\textbf{Pauli channels:} 
We rely on Pauli twirling (implemented with \texttt{samplomatic}~\cite{samplomatic}) to effectively eliminate the off-diagonal components of the Pauli Transfer Matrix (PTM) of the noise channel. In Fig.~\ref{fig:pec_results}(b), we measure the off-diagonal components of a Cliffordization at $\theta = \pi/2$. Specifically, we measure $(Z,X)$ PTM transition elements, over the set of qubits involved in the A loops of the device layout. We find that these off-diagonal elements are reliably zero, indicating little evidence of non-Pauli terms in the PTM. 

\textbf{Sparsity:} 
The next assumption about the noise model is that it is local, i.e.\,, we restrict the noise generator coefficients to at most weight 2 on connected qubits on the device topology. To test this assumption, in Fig.~\ref{fig:pec_results}(c), we evaluate expectation values of higher weight observables from the noise learning data set, and compare them to the values predicted by the learned sparse Pauli-Lindblad noise model (x axis). We observe good agreement between the data and model for these higher weight observables, validating the sparsity assumption of the noise model.

\textbf{Noise learning fit quality:} 
In Fig.~\ref{fig:pec_results}(d) we probe the assumption that the device noise is Markovian, by analyzing how well the Pauli fidelities follow an exponential decay. We use the reduced-chi-squared $\chi_\nu^2$ as a measure of goodness-of-fit: a smaller $\chi_\nu^2$ indicates better agreement with an exponential decay. We find that using edge-based post-selection following the protocol described above improves the $\chi_\nu^2$ measure of exponential decay.

\textbf{Model agreement on Cliffordizations:} 
During the noise learning stage, we perform additional checks that the noise models are accurate via a model-agreement (MA) experiment, in which Clifford circuits with known noiseless answers are measured on hardware and compared against the noise model's own prediction of that same noisy circuit. In particular, we focus on circuits that are structurally identical to the OLE circuits of interest, and choose the two Cliffordizations mentioned in \cref{app:cliffordization}. These Cliffordized versions of the OLE circuits are thus entangling mirror circuits. In \cref{fig:pec_results}(e), the unmitigated expectation values of a Cliffordized OLE model agreement circuit are compared against the predictions of a simulation of the circuit with the learned noise models. Agreement between the experiment and the noisy simulations provides additional confidence that the noise model is accurate.

\begin{figure}[t]
    \centering
    \includegraphics[width=1.0\linewidth]{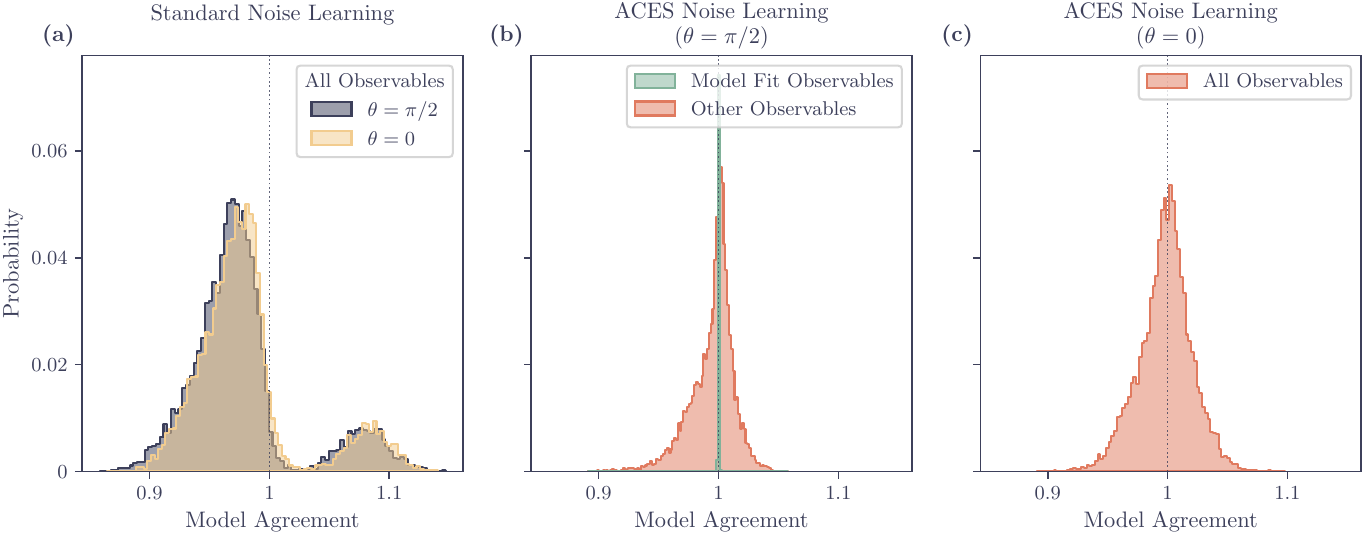}
    \caption{Histograms of model agreement for stabilizers for the Cliffordization circuits for $L=4$. The dashed, vertical lines are the ideal model agreement. (a) The model agreement when using the standard noise model learning method. The bi-modality of these distributions is due to specific qubits which skew the distribution. (b) The model agreement of the observables for the $\theta=\pi/2$ Cliffordization circuit. The ``Model Fit Observables'' were used in fitting the model. The ``Model Fit Observables'' distribution contains $55$ observables. (c) The model agreement of the observables for the $\theta=0$ Cliffordization circuit. None of the observables for $\theta=0$ were used in the ACES Noise Learning. The fact that the distribution here is centered around $1$ suggests that we do not overfit to the Cliffordization circuit that we chose.}
    \label{fig:aces_model_agreement_hist}
\end{figure}

\textbf{Noise model improvement using ACES:} The noise model learning procedure need not strictly include the circuits used in standard cycle benchmarking. The ACES method~\cite{flammia2022aces} generalizes this, allowing us to incorporate the Cliffordizations described above into our estimate of the noise model. To do this, we select a subset of the stabilizers of the Cliffordization at $\theta = \pi/2$, which correspond to new rows in the design matrix used in learning. The improvement in model agreement compared to the standard noise learning method is shown in Fig.~\ref{fig:aces_model_agreement_hist}.

\textbf{Validation at $\delta = 0$:} 
To validate on a circuit that is even closer to the target circuit than its Cliffordizations, we perform PEC on the OLE circuit at $\delta = 0$.
Except for the one single-qubit perturbation $V_\delta$, this circuit is identical to the target circuit.
The exact OLE signal at $\delta = 0$ is 1, providing a reliable point for validation. 
In \cref{fig:pec_results}(f), we show that PEC mitigated estimates are consistent with 1, up to the PEC error bounds.

\begin{figure}
    \centering
    \includegraphics[width=1.0\linewidth]{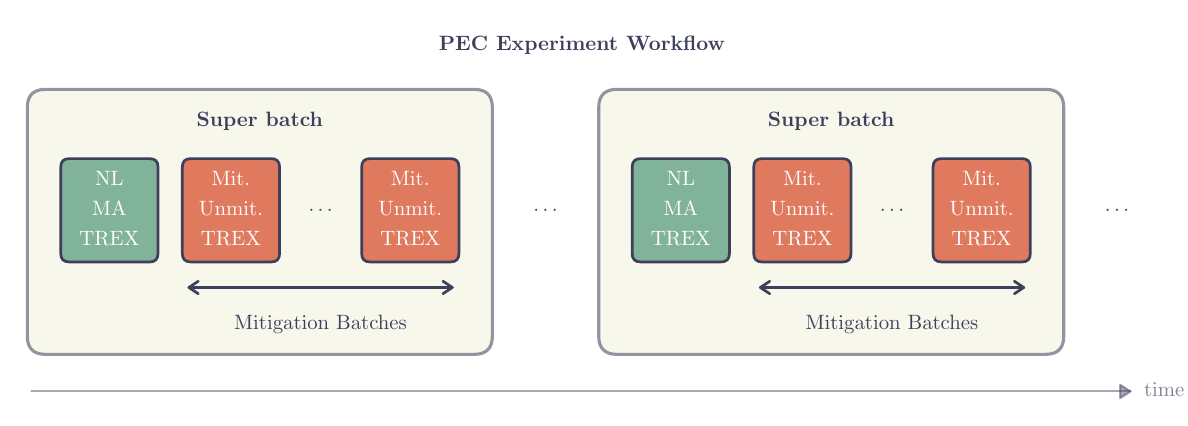}
    \caption{PEC Experiment execution workflow. Data is repeatedly collected in ``super-batches", which comprise of a noise-learning step that includes model agreement (MA) circuits, and then repeated batches of PEC-mitigated circuits, as well as unmitigated OLE circuits, to monitor device stability.}
    \label{fig:pec_workflow}
\end{figure}

\begin{figure}[h]
    \centering
    \includegraphics[width=\linewidth]{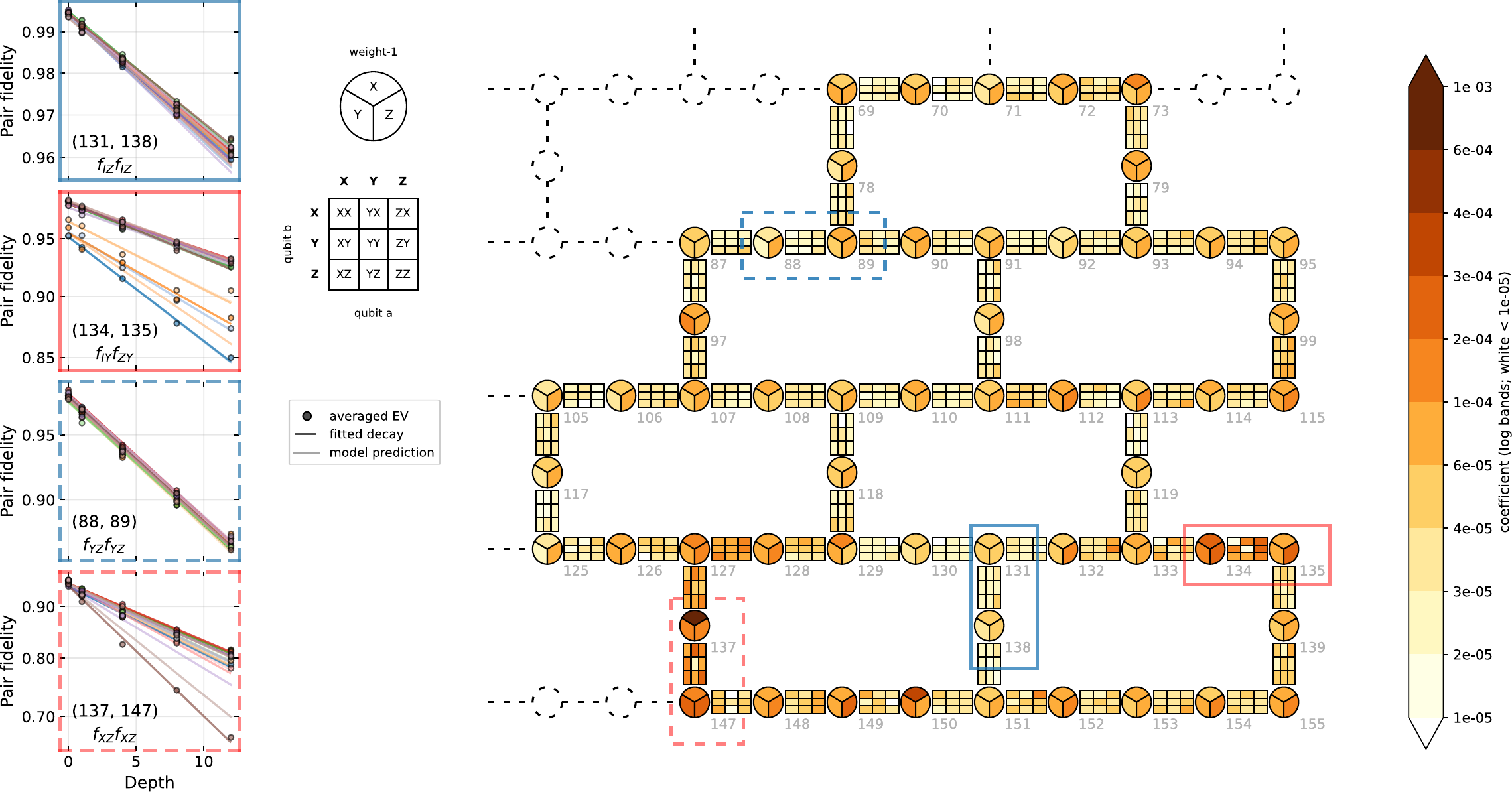}
    \caption{Left panels: Noise learning traces over the course of an example experiment. Different colors correspond to different super-batches. Learning curves boxed in blue indicate more stable examples, and boxed in red indicate areas with greater variations over super-batches.
    Right panel: The standard deviation of noise rates over 12 super-batches. Darker regions indicate higher fluctuations in the noise rates during the experiment.
    }
    \label{fig:noise_stability}
\end{figure}

\begin{figure}
    \centering
    \includegraphics[width=0.99\linewidth]{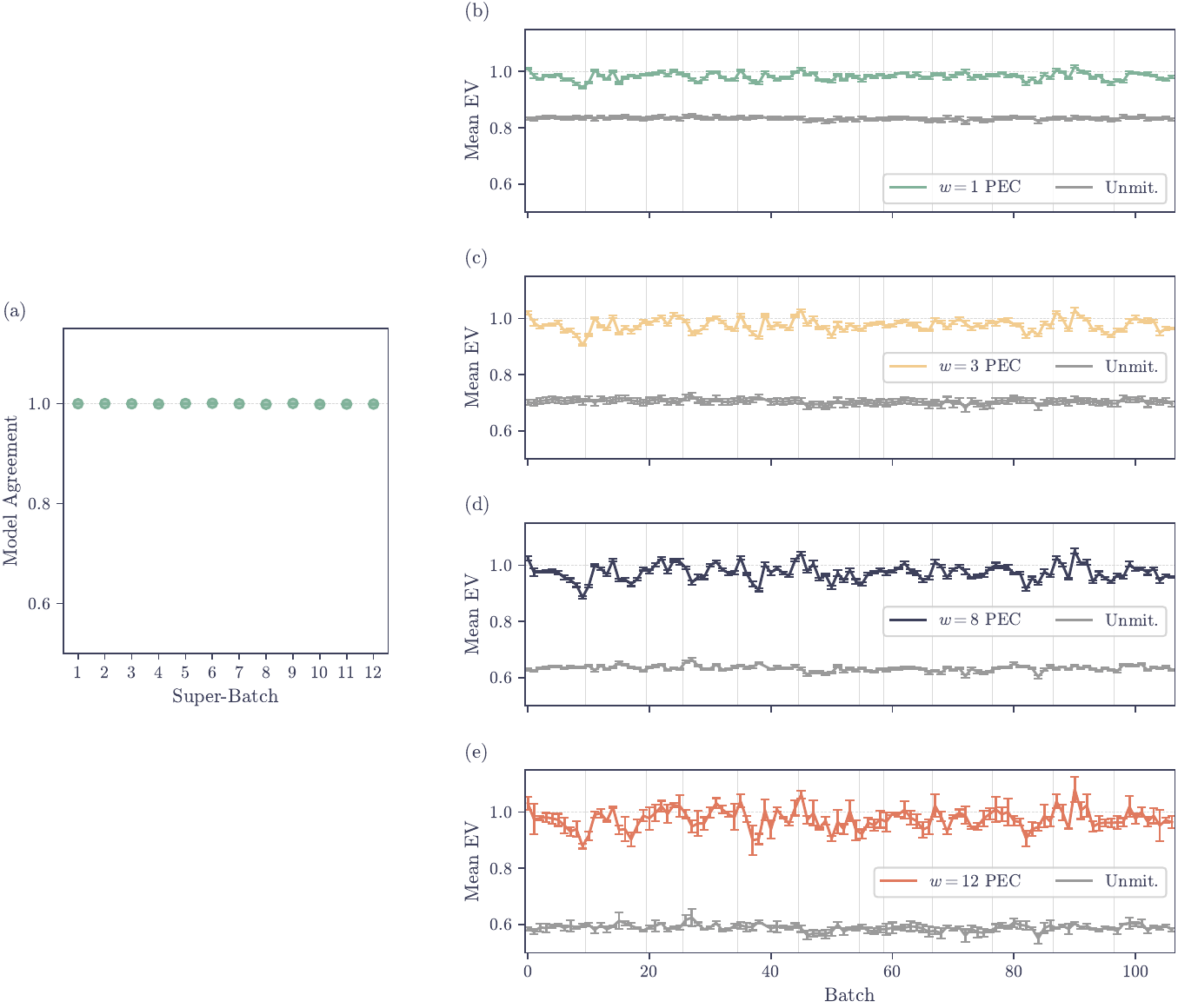}
    \caption{Model Agreement and PEC stability. (a) Model agreement for the $L=2$ PEC experiment over the course of the experiment, which includes a total of 12 super-batches. Each data point is averaged over all weight-one observables. Error bars correspond to the standard error of the mean (SEM) over the observables. (b)-(d) Unmitigated (grey) and PEC mitigated expectation values (EVs) for observable weights 1, 3, 8, and 12, respectively. The EVs are averaged over all observables for a given weight. Across all weights, both the unmitigated and mitigated EVs remain stable across the course of the experimental run. Error bars correspond to the SEM over the observables. The vertical, grey lines denote the borders between different super-batches. The horizontal, dashed grey line is at $1.0$ for visual clarity.}
    \label{fig:pec_stability}
\end{figure}

\textbf{PEC Experiment workflow and stability:}
To monitor the experiment for device instabilities and enable filtering of data for noise model compliance, if needed, we run PEC experiments in a workflow according to Fig.~\ref{fig:pec_workflow}: the experiment is organized into independent ``super-batches'' that are run repeatedly. Each super-batch begins with a noise-learning (NL) step that fits the Pauli--Lindblad noise models of the CZ gate layers. In addition to the noise learning circuits, we run Cliffordized OLE circuits that are used to measure model agreement (MA) on circuits that are structurally similar to the target circuit. In Fig.~\ref{fig:pec_stability}(a), we plot the MA, defined as the ratio of the unmitigated expectation values of the Cliffordized OLE circuit to the classically simulated prediction, for each of the 12 super-batches run for the $L=2$ PEC experiment (Fig.~\ref{fig:pec_results} of the main text). Across all super-batches, the MA remains approximately equal to one, even as learning traces across super-batches may display fluctuations in decay rates (Fig.~\ref{fig:noise_stability}). 
This emphasizes the importance of repeated and interleaved noise learning in our workflow and its ability to mitigate the effects of noise model drift during the experiment.

In the mitigation stage of the PEC workflow, a series of PEC circuit ``batches'' are submitted, where each batch corresponds to a new set of randomly sampled PEC circuits. Circuits without PEC mitigation are also run in this step to monitor for device drift. Fig.~\ref{fig:pec_stability}(b)-(d) shows both the unmitigated and mitigated expectation values (EVs) at $\delta=0$ for each circuit batch run across the experiment. Across all observable weights, both the unmitigated and mitigated EVs remain stable across the experiment. Overall, these results highlight the stability of the experiment over time and do not show evidence of significant performance degradation due to resonant interaction with two-level systems (TLSs). 

Lastly, we note that readout error-mitigation (TREX~\cite{van_den_berg_model-free_2022}) circuits are executed in both the noise learning stage and the mitigation stage.

\section{Probabilistic Error Cancellation}\label{app:pec}

In this section, we provide a review of the Probabilistic Error Cancellation (PEC) method used in the main text. 
PEC is a technique to mitigate the effect of noise in quantum computations by injecting random gates, carefully chosen to cancel the effect of noise on average~\cite{temme_error_2017,van_den_berg_probabilistic_2023}.

Given a noisy layer of noisy quantum circuit $\tilde{\mathcal{U}}$, we can express it as a composition of the ideal channel $\mathcal{U}$ and a noise channel $\Lambda$: $\tilde{\mathcal{U}} = \Lambda \circ \mathcal{U}$.
If $\mathcal{U}$ is a Clifford channel, we apply Pauli twirling to transform $\Lambda$ into a Pauli-Lindblad form:
\begin{align}
    \Lambda (\rho) = \prod_{P \in \mathcal P} \left[w_P \rho + (1-w_P) P \rho P \right],
\end{align}
where $\mathcal P$ is a set of Pauli strings, $w_P = (1 + e^{-2\lambda_P})/2$, and $\lambda_P$ is the error rate.
In our experiment, we assume that $\Lambda$ is sparse: each $P \in \mathcal P$ is supported on only nearest-neighboring qubits on the hardware.
Due to fundamental learnability limits, $\lambda_P$s are only learnable up to gauge degrees of freedom~\cite{Chen2023learnability,chen2026gateset}. We fix this gauge by assuming that the rates are invariant under the conjugation by $U$, i.e. $\lambda_P = \lambda_{P'}$ if $P = U^\dag P' U$, which is consistent with the physics of our hardware~\cite{malekakhlagh2026symmetriespaulinoiselindbladian}.
We validate the Pauli-ness and sparsity assumptions in Fig.~3 of the main text.
Under these assumptions, the rates $\lambda_P$ can be efficiently learned via cycle benchmarking~\cite{van_den_berg_probabilistic_2023}.

Once we have learned the noise model, PEC implements the inverse map $\Lambda^{-1}$ to cancel the effect of noise:
\begin{align}
    \Lambda^{-1}(\rho) = \gamma \prod_{P \in \mathcal P} \left[w_P \rho - (1-w_P) P \rho P \right],
\end{align}
where $\gamma = \exp\left(2\sum_{P\in\mathcal P} \lambda_P\right)$.
Operationally, PEC applies a Pauli operator $P$ with probability $1-w_P$.
Depending on the number of $P$ that were applied, the measurement outcome in each randomization is multiplied by a global phase $\pm1$.
Averaging these outcomes and multiplying by $\gamma$  recovers the signal from the ideal, noise-free channel.

If noise learning is exact, the error of PEC is purely statistical.
With probability $1-2/e^2\approx 73\%$, the distance from a PEC mitigated estimate to the noiseless value is upper bounded by $\sigma = 2\gamma/\sqrt{N}$, where $N$ is the number of samples~\cite{van_den_berg_probabilistic_2023}.
Therefore, $\gamma^2$ is typically used as a proxy for the sampling overhead of PEC.
In \cref{sec:slc}, we discuss a postprocessing technique that leverages classical simulation to reduce the sampling overhead of PEC.

In practice, noise learning is not perfect. There are different effects that lead to statistical and systematic errors in the noise models. We discuss these errors in \cref{sec:noise-prop} and detail how we can propagate them to error bounds on the PEC mitigated estimates.

\subsection{Shaded lightcones}\label{sec:slc}
\begin{figure*}[t]
    \centering
    \begin{minipage}{0.70\textwidth}
        \raggedright \textbf{(a)}\par
        \includegraphics[width=\linewidth]{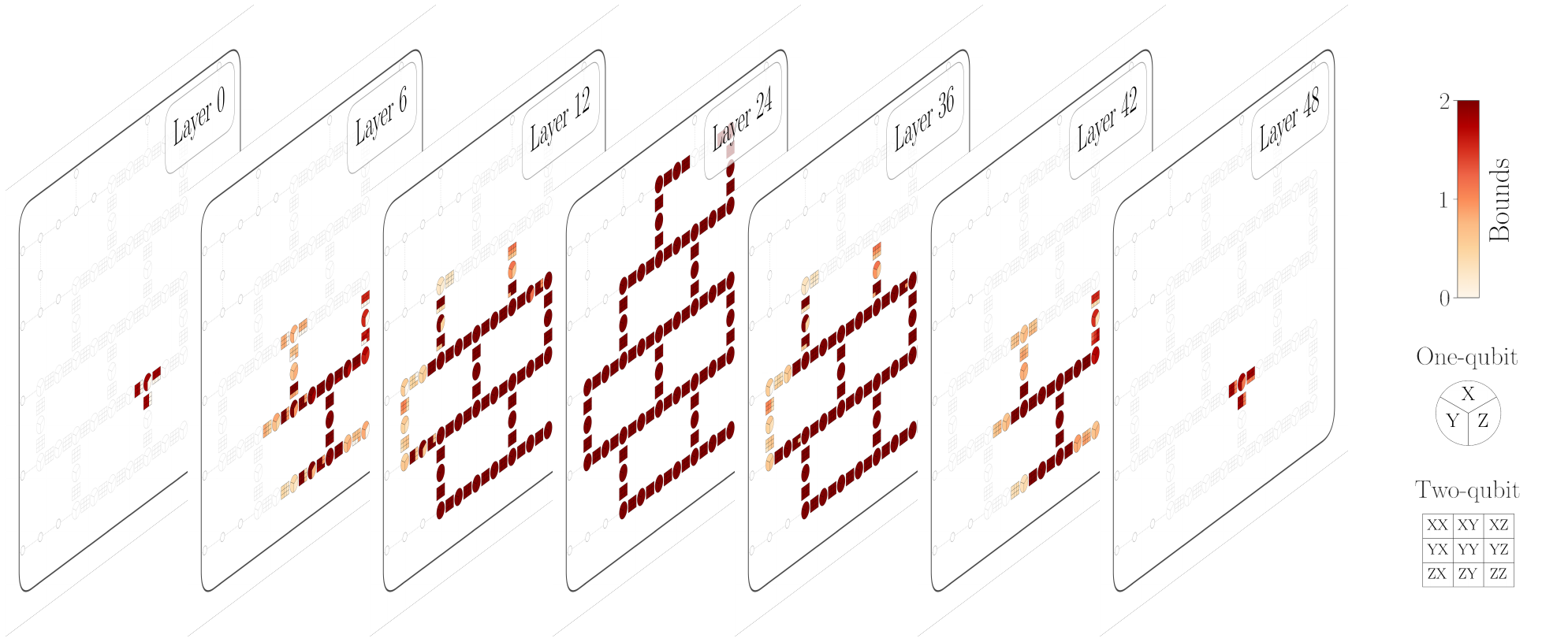}
    \end{minipage}\hspace{0.01\textwidth}%
    \begin{minipage}{0.28\textwidth}
        \raggedright \textbf{(b)}\par
        \includegraphics[width=\linewidth]{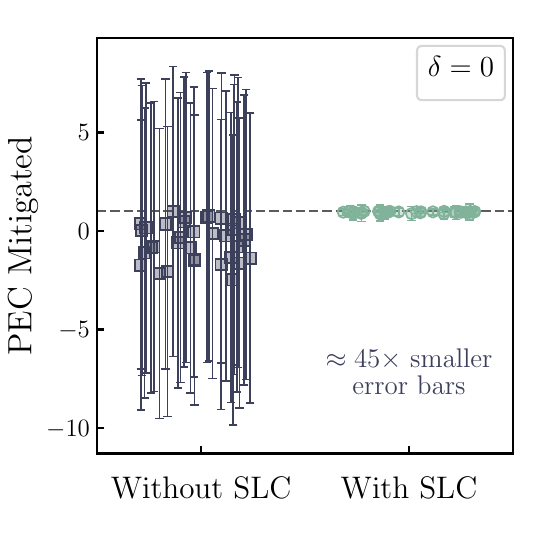}
    \end{minipage}
    \caption{Shaded lightcone (SLC) for PEC. (a) The SLC of a representative weight-one observable shown at several CZ layers of the depth-$48$ ($L=4$) OLE circuit at $\delta=0$; each one- and two-qubit error channel is shaded by the classical bound on its influence on the observable. (b) PEC-mitigated signal at $L=4$, $\delta=0$ (ideal echo value $1$, dashed).
    Without SLC, the estimates scatter broadly due to the large sampling overhead.
    With SLC they form a tight cluster near $1$.}
    \label{fig:app_slc_overview}
\end{figure*}

Standard PEC~\cite{temme_error_2017,van_den_berg_probabilistic_2023} inverts every error channel in the circuit, including channels that commute with the observable and therefore leave its expectation value unchanged. Mitigating these channels wastes sampling overhead. The Shaded Lightcone (SLC)~\cite{tran_locality_2023,eddins_lightcone_2024} supplies classical upper bounds on how much each error channel can influence a given observable. PEC-with-SLC inverts only the channels whose bounds are large and leaves the remainder unmitigated, incurring a small, bounded bias in exchange for a substantial reduction in the sampling overhead $\gamma^2$. 
Throughout the paper, we allow this bias to be at most $0.01$. 
Panel (a) of \cref{fig:app_slc_overview} shows the SLC of a representative weight-one observable across the CZ layers of the $L=4$ OLE circuit: its support first expands and then contracts under the mirror-like structure of the circuit. Panel (b) shows the effect at $\delta=0$, where the ideal echo value is $1$: full PEC scatters broadly about $1$ because of its large sampling overhead, whereas PEC with SLC yields a tight cluster near the exact value.

\begin{figure*}[h]
    \centering
    \begin{minipage}{0.32\textwidth}
        \raggedright \textbf{(a)}\par
        \includegraphics[width=\linewidth]{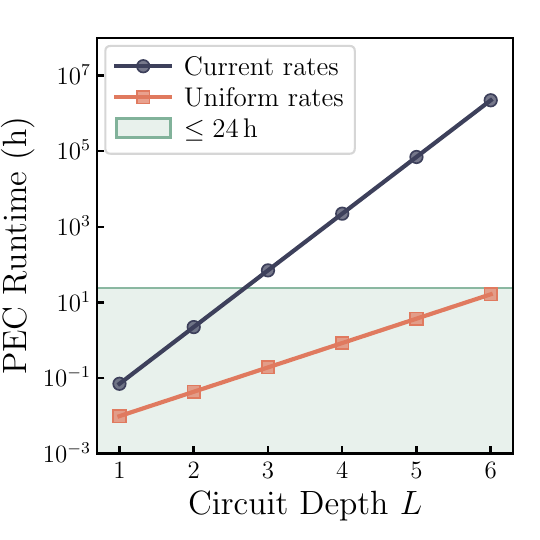}
    \end{minipage}\hfill
    \begin{minipage}{0.32\textwidth}
        \raggedright \textbf{(b)}\par
        \includegraphics[width=\linewidth]{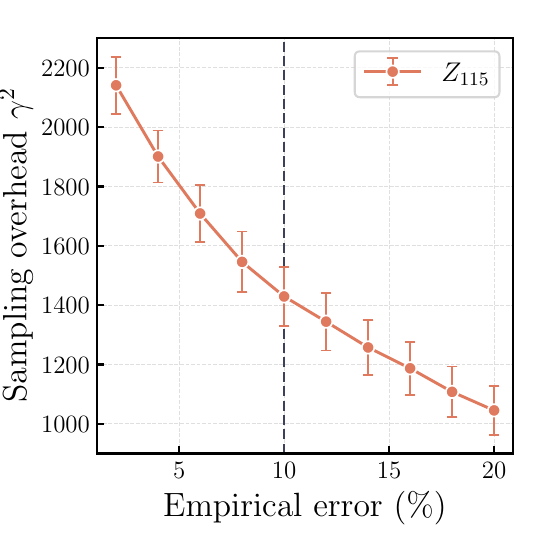}
    \end{minipage}\hfill
    \begin{minipage}{0.32\textwidth}
        \raggedright \textbf{(c)}\par
        \includegraphics[width=\linewidth]{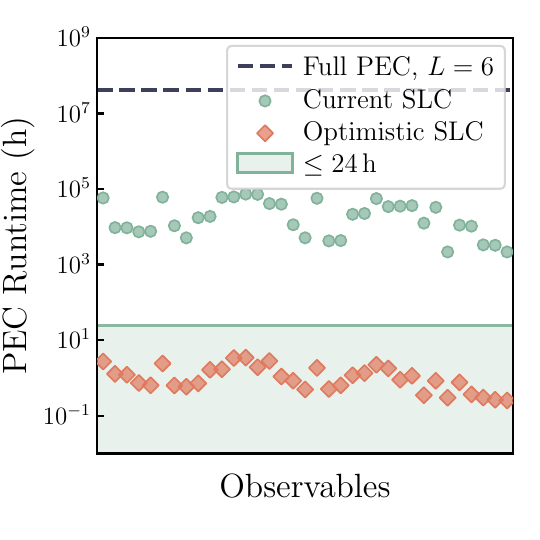}
    \end{minipage}
    \caption{(a) Projected quantum-processor runtime for PEC without SLC to reach a target standard error on the OLE observable versus circuit depth $L$.
    The ``Current rates'' curve uses the learned Pauli--Lindblad error rates and the ``Uniform rates'' curve sets every channel rate to the current median, collapsing the heavy tail. 
    (b) The sampling overhead $\gamma^2$ versus deviation in the mitigated expectation values obtained by skipping PEC applications in postprocessing for some noise channels inside the current SLC of a representative weight-one observable $Z_{115}$ at $L=4$.
    The sampling overhead $\gamma^2$ can be reduced by about $50\%$ while the expectation value only changes by 10\%, indicating that many noise channels have negligible influence on the observable despite being inside the current SLC.  
    (c) Projected PEC runtime at $L=6$ for weight-one observables under full PEC (dashed), the current SLC bounds (circles), and an optimistic, tighter SLC (diamonds). Shaded bands mark the $24$-hour feasibility window.}
    \label{fig:app_pec_runtime_depth}
    \label{fig:app_slc_runtime_obs}
\end{figure*}

\textbf{Projected PEC runtime:}
\Cref{fig:app_pec_runtime_depth}(a) projects the quantum runtime required for PEC without SLC to reach a target standard error of $0.1$ on the OLE observable, as a function of circuit depth $L$ (from $L=1$ to $L=6$). The runtime is $T = 4\gamma^2 / (\epsilon^2 R)$ with a sampling rate $R = 50$~shots/s and $\epsilon = 0.1$, where the sampling overhead $\gamma^2$ grows with the total circuit noise as $\gamma^2 = \gamma_{1}^{2L}$ and $\gamma_1$ is the sampling overhead per $L$. 
The ``Current rates'' curve uses the device's learned Pauli-Lindblad error rates. 
The ``Uniform rates'' curve is a hypothetical scenario in which the rate distribution is tightened so that every Pauli-Lindblad channel rate is equal to the current median rate, collapsing the heavy tail of the rate distribution.
This improvement would lower $\gamma^2$ and bring the projected runtime below the $24$-hour feasibility band (shaded) through $L=6$.

A complementary approach to reducing the PEC runtime is to improve the classical computation of the SLC bounds.
The bound the SLC computation puts on the importance of each Pauli noise channel can be loose; a channel can have a large SLC bound despite having almost no influence on the observable. 
If this is indeed the case, skipping the mitigation of such channels in the PEC postprocessing, which would reduce the overhead $\gamma^2$, should have minimal effects on the mitigated expectation value.
In \cref{fig:app_pec_runtime_depth}(b), we plot this reduced $\gamma^2$ against deviation of the expectation value when some noise channels are excluded from the PEC postprocessing. The figure shows that $\gamma^2$ can be reduced by $50\%$ while the mitigated expectation value stays within 10\% of the current SLC implementation. 
This robustness suggests that the SLC bounds can be significantly tightened. 

For an estimate of how much $\gamma^2$ may be further reduced, \cref{fig:app_slc_runtime_obs}(c) projects the PEC runtime at $L=6$ for several weight-one observables to reach a $0.1$ standard error, under three scenarios. 
Full PEC (dashed line) is standard PEC without SLC---a single large overhead $\gamma \sim 1.2\times 10^5$, out of reach on current hardware. Current SLC (circles) uses the SLC bounds that our current classical simulation produces, reducing the runtime by three to four orders of magnitude relative to full PEC but still costly on current devices. Optimistic SLC (diamonds) is the best-case estimate in which the SLC overhead scales inversely quadratically with the decay factor $\tilde S_0$, an expected feature of a typical circuit~\cite{filippov2023scalable}.
Whether and how such a best-case scenario can be realized are likely contingent on substantial progress in classical methods for computing the SLC bounds. We leave the answers to these questions to future works.

\subsection{Error propagation}\label{sec:noise-prop}

As mentioned earlier, if noise learning were exact, the error of a PEC-mitigated estimate would be purely statistical. 
In practice the learned rates $\lambda_P$ carry both statistical uncertainty, from the finite number of cycle-benchmarking shots, and systematic error, from the assumptions built into the noise model itself. 
Here, we consider several such sources and estimate their magnitudes.

\begin{figure}
\includegraphics[width = 0.45\textwidth]{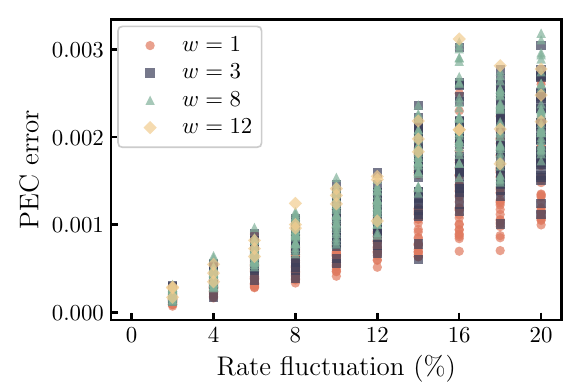}
\includegraphics[width = 0.45\textwidth]{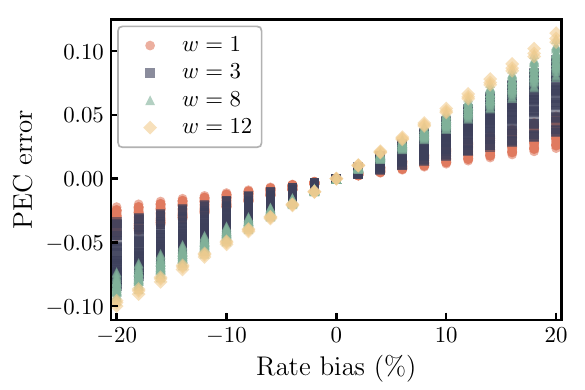}    
\caption{
Propagation of statistical fluctuation and bias from the noise model to error bounds on the PEC mitigated estimates at $L = 2, \delta = 0,$ and $ \eta = 9\pi/40$. 
Left panel: Each individual rate is assumed to be sampled from a normal distribution with the mean $\lambda$ being the learned rate and the standard deviation $\sigma$ being $x\%$ of $\lambda$.
The panel plots the standard deviation in the PEC estimate as the result of fluctuation of strength $x$\%.
Different colors correspond to different observable weights $w$.
Right panel: Each individual rate is assumed to be exactly $x\%$ different from its learned value. This systematic bias in the rate leads to a shift in the PEC estimate.
The PEC error can be computed efficiently for each observable using retroactive resampling.      
}
\label{fig:reweight}
\end{figure}

These errors are then propagated to error bounds on the mitigated estimate by retroactive resampling~\cite{aharonov2026reliablehighaccuracyerrormitigation}, a data processing procedure to shape the randomizations drawn from a distribution to resemble a different distribution. In \cref{fig:reweight}, we provide examples of propagating two types of deviations in the learned model to errors in the PEC estimate at $L = 2$, $\delta = 0$, and $\eta = 9\pi/40$. 
The first kind is the statistical shot noise, leading to an approximately unbiased fluctuation in the learned rates. 
The second is a systematic bias in the learned rates, caused by non-Markovian noise for example. 
Below, we estimate the contribution of these two sources of errors, which we expect to be the dominant in our PEC experiments.

\textbf{Shot noise:} 
In cycle benchmarking, we measure the fidelity $f_d$ of an observable as it evolves under $d$ repetitions of the layer pair $U^2$ the noise of which is being characterized.
The learning trace---the decay of the fidelity $f_d$ at different $d \in \{0,1,4,18,24\}$---is fitted to an exponential decay to find the fidelity $f_1$ of the observable after each layer pair.
From these layer fidelities $\vec f$ for different observables, we obtain the rates $\vec \lambda$ by minimizing $\Vert M \vec \lambda + \frac{1}{2}\ln \vec{f}\Vert_2^2$, where $M$ is the \emph{design matrix}~\cite{van_den_berg_probabilistic_2023}. This is accomplished with \texttt{qiskit-noise-learning}~\cite{qiskit-noise-learning}.

In our experiment, each $f_d$ is measured using a finite number $N_{\text{CB}}$ of shots.
The resulting shot noise in $f_d$ leads to a statistical fluctuation in the learned rates $\vec \lambda$.
We bootstrap the fidelity $f_d$ at each depth to estimate this fluctuation.
At $N_{\text{CB}} = 20000$ as in our experiment, we find that the largest 50\% of noise rates, which are responsible for more than 90\% of the overhead $\gamma^2$, fluctuate by roughly $17\%$ between bootstrapped samples.
This rate fluctuation propagates to PEC errors at $L = 2, \delta = 0$ between $7.7 \times 10^{-4}$ and $2.8 \times 10^{-3}$ for the considered observables of weights between 1 and 12 [\cref{fig:reweight}].
We repeat the error propagation for $L \in \{2,4\}$ and $\delta \in \{0,0.3\}$.
We note that the dependence of the learned rates on the fidelities $f_d$ is nonlinear and in principle also leads to a bias in the learned rates.
However, this nonlinearity bias should be negligible at $N_{\text{CB}} = 20000$. 

\begin{figure}
\includegraphics[width = 0.35\textwidth]{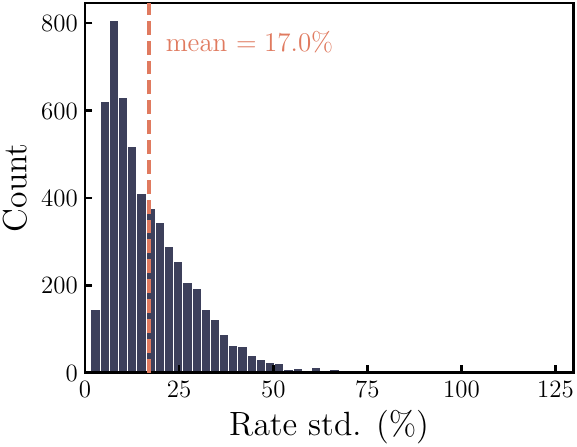}
\caption{A histogram of the bootstrapped standard deviations due to shot noise of the largest $50\%$ of the learned rates. The standard deviation for each rate is computed over 20 bootstrap repetitions.}
\label{fig:rate-std-bootstrap}
\end{figure}

\textbf{Non-Markovianity:} The effect of non-Markovianity on the learned noise model is harder to quantify directly. 
Rather than a statistical fluctuation, its signature is a deviation of the fidelity-versus-depth data from a single exponential decay~\cite{kim2025error}. 
To quantify its effect, we model the non-Markovianity as follows.

We assume that, in each experiment run, the fidelity for each layer pair is $e^{-\tilde \lambda}$, where the layer-pair decay rate $\tilde \lambda \sim \mathcal N(\lambda,\sigma^2)$ is independently sampled from a normal distribution with mean $\lambda$ and standard deviation $\sigma$.
However, the $d$ repetitions in the same depth-$d$ experiment are perfectly correlated and share the same sampled $\tilde \lambda$, leading to the overall fidelity $\tilde f_d = e^{- \tilde \lambda d}$.
The fidelity at depth $d$ averaged over experiment runs is 
$$
\mathbb E[\tilde f_d] = \exp(-\lambda d + \sigma^2 d^2/2).
$$
Therefore, under this model, the non-Markovianity biases the fidelity up at large depths and rates obtained from a simple exponential fit would underestimate $\lambda$.

To quantify this bias, we refit the learning traces to an exponential decay with the additional correction $\exp(\sigma^2 d^2/2)$, where $\sigma$ is an independent parameter.
We observe that the simple exponential fit results in a median underestimation of $16\%$ for the corresponding rates.
From \cref{fig:reweight}, this underestimation would cause a bias between $1.8\times 10^{-2}$ and $8.1\times 10^{-2}$ at $\delta = 0$ for the PEC estimates of the considered observables.
Again, we repeat the bias propagation for $L \in \{2,4\}$ and $\delta \in \{0,0.3\}$. 
The bias tolerance in the SLC calculation and the propagated error in the PEC estimate due to model bias and fluctuation form the confidence band illustrated for each observable in Fig.~3 (f) and (g) of the main text.

Here, we use a simplified model of quasistatic noise. We consider a limiting case in which the noise is perfectly correlated across all layers within a sample but independent between samples. The present bounds should therefore be viewed as loose estimates. A better understanding of the physical noise processes and their correlation times would allow for a more realistic model and more accurate bounds. Improvements in hardware stability and mitigation protocols designed specifically for non-Markovian errors could also reduce the bias itself and improve the accuracy of PEC.

\section{Experiments on truncated lattices}
\label{sec:finite-size-scaling}

\begin{figure}[h]
    \includegraphics[width = 0.48 \textwidth]{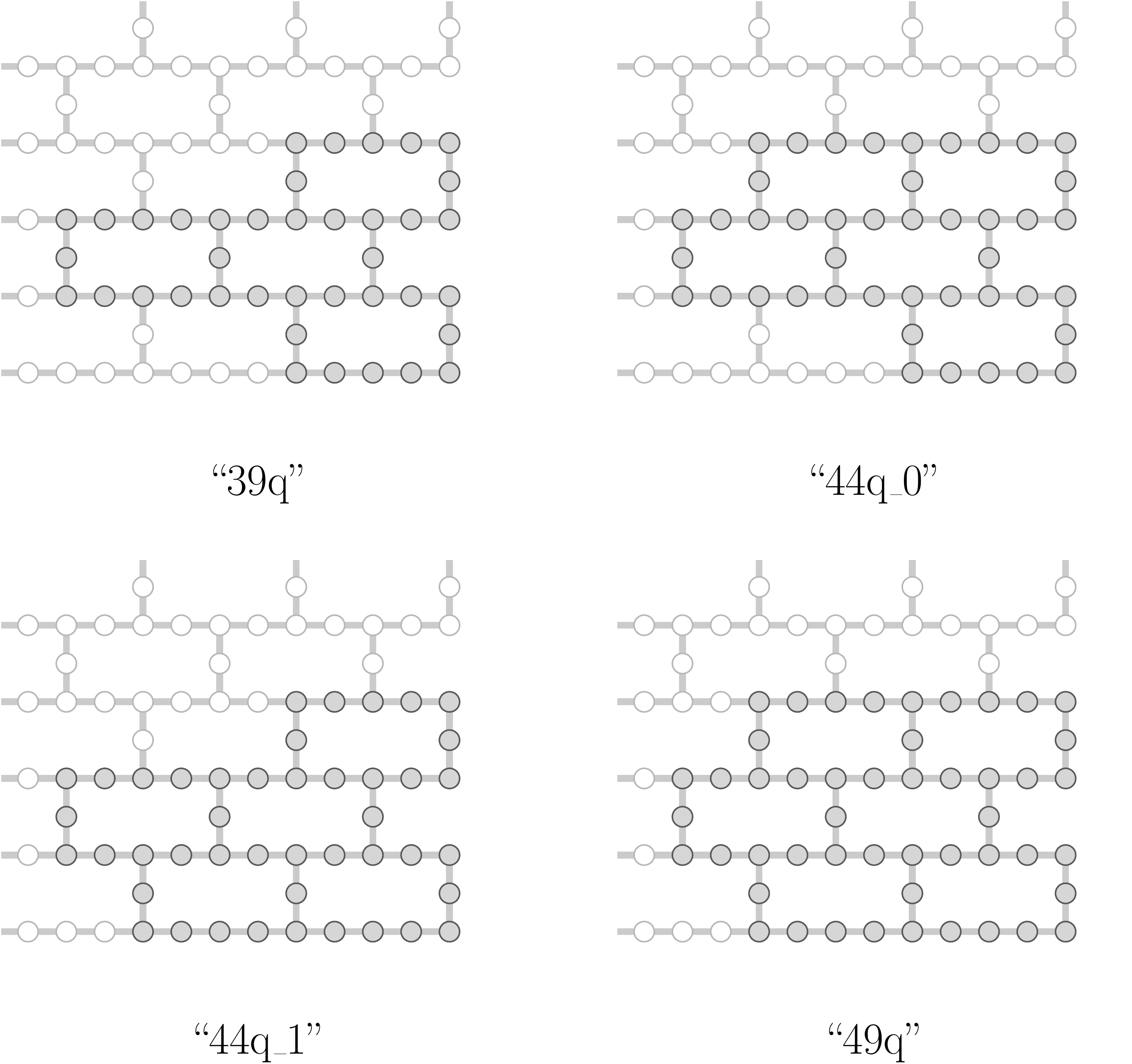}
    \caption{Sub-lattice variants of the full 56-qubit OLE circuit. Gray circles indicate qubits that remain in the sub-lattices.}
    \label{fig:variants}
\end{figure}

In this section, we explore whether the OLE signal on the 56-qubit circuit can be inferred from those of smaller lattices.
If that is the case, the convergence of the signal with increasing lattice sizes would enable a reliable path for validating the OLE signal in large systems.
Here, we compute the globally rescaled OLE signals on the four circuit variants supported on 39, 44, and 49 qubits (\cref{fig:variants}) on $\texttt{ibm\_boston}$ and compare the estimates with that of the 56-qubit circuit.

\begin{figure}[h]
\centering
    \includegraphics[width = 0.48\textwidth]{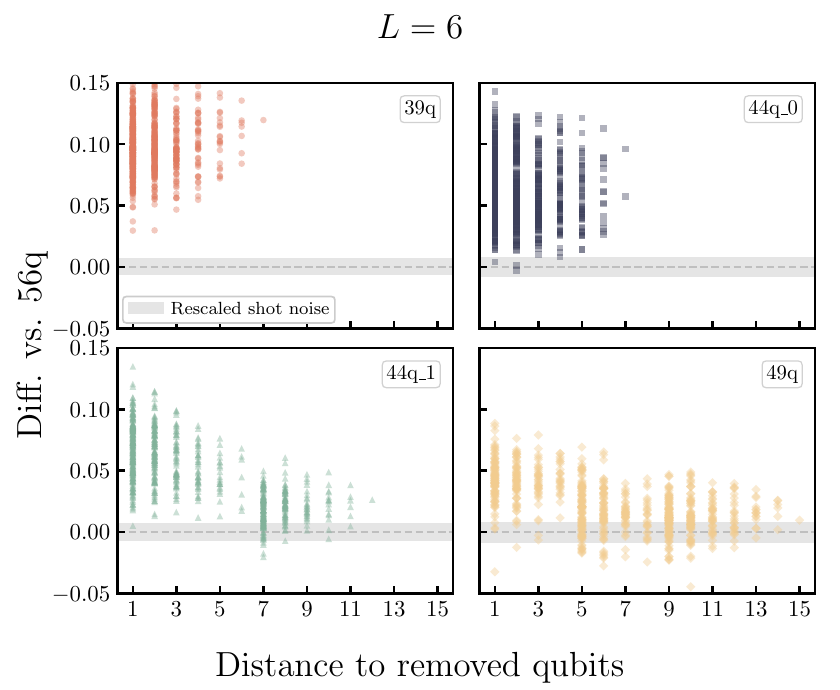}
\includegraphics[width = 0.48\textwidth]{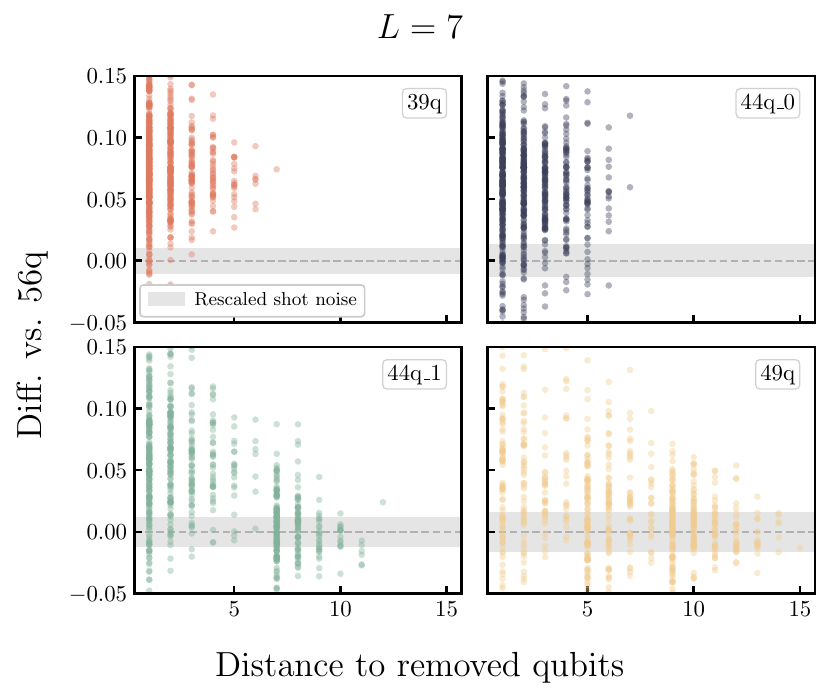}
\caption{Difference in the globally rescaled signal at $\delta = 0.3, \eta = 3\pi/8$ between the 56-qubit circuit and its sub-lattice variants for several observables, sorted by the distance from the support of the observables to the nearest removed qubits, for $L = 6$ (left) and $L = 7$ (right). 
The gray band around 0 is the typical rescaled shot noise, computed by rescaling the standard error over $614400$ shots by the median of the global rescaling factors over the considered observables.
The estimates from the sub-lattice variants broadly disagree with the estimates from the full 56-qubit circuit. 
}
\label{fig:finite-size-scaling}
\end{figure}

In \cref{fig:finite-size-scaling}, we plot the difference between the globally rescaled OLE signal of several observables in the 56-qubit circuit and the four sub-lattice variants at $L \in [6, 7]$, $\delta = 0.3$, and $\eta = 3\pi/8$.
The scatter points are sorted by the distance from the support of the observables to the nearest removed qubits. 
Generally, at a fixed depth, we expect the difference to approach $0$ as the observable moves further away from the removed qubits. 
This is roughly the trend observed in \cref{fig:finite-size-scaling}.
However, except for the few observables at the furthest distance, the majority of estimates are sensitive to the removal of qubits from the 56-qubit lattice, indicating that the evolved observables have propagated to the removed qubits at $L \geq 6$.

While the ``49q'' circuit's estimates are the closest to the full ``56q'' circuit among the variants, they still broadly disagree at both $L = 6$ and $L = 7$.
We also note that some estimates from the ``49q'' variant for observables at relatively short distance appear to match that of ``56q'', e.g. those that fall within the shot-noise band at distance $5$ on the ``49q'' plot.
However, observables much further away, such as those at distance 12, are still sensitive to the missing 7 qubits in ``49q'', suggesting that the agreement for those at distance 5 above cannot be explained by the lack of operator spreading. 
Taken together, these observations cast doubt on the reliability of predicting the estimates of the full 56-qubit circuit in our experiment by simulating smaller systems.
We note that one can also attempt to fit the estimates at small system sizes to a theory-derived functional and extrapolate to the 56-qubit lattice. This heuristic approach may provide additional validations for our experiments. 
However, studying the accuracy and reliability of such extrapolations is beyond the scope of the current manuscript.

\end{document}